\documentclass[a4paper,11pt]{article}

\usepackage[margin=2.5cm]{geometry}

\usepackage[utf8]{inputenc}
\usepackage[final]{microtype}
\usepackage{lmodern}
\usepackage{exscale}
\usepackage[T1]{fontenc}
\usepackage{microtype}

\usepackage{amsmath}
\usepackage{amssymb}
\usepackage{mathtools}

\numberwithin{equation}{section}

\allowdisplaybreaks

\usepackage[linktocpage]{hyperref}
\usepackage[dvipsnames]{xcolor}
\hypersetup{
    colorlinks,
    linkcolor={RoyalBlue},
    citecolor={RoyalBlue},
    urlcolor={RoyalBlue}
}

\usepackage{cite}

\usepackage{mathrsfs}

\usepackage{relsize,exscale}
\usepackage{tensor}

\usepackage[low-sup]{subdepth}

\newcommand{\scr}{\scriptscriptstyle}

\newcommand{\ch}{ \, {\rm ch} }
\newcommand{\sh}{ \, {\rm sh} }

\newcommand{\longsim}{\scalebox{1.8}[1]{$\sim$}}

\usepackage[labelfont=bf]{caption}

\makeatletter
\newcommand{\dalembertian}{\mathop{\mathpalette\dalembertian@\relax}}
\newcommand{\dalembertian@}[2]{%
  \begingroup
  \sbox\z@{$\m@th#1\square$}%
  \dimen0=\fontdimen8
    \ifx#1\displaystyle\textfont\else
    \ifx#1\textstyle\textfont\else
    \ifx#1\scriptstyle\scriptfont\else
    \scriptscriptfont\fi\fi\fi3
  \makebox[\wd\z@]{%
    \hbox to \ht\z@{%
      \vrule width \dimen0
      \kern-\dimen0
      \vbox to \ht\z@{
        \hrule height \dimen0 width \ht\z@
        \vss
        \hrule height 2\dimen0
      }%
      \kern-2.5\dimen0
      \vrule width 2.5\dimen0
    }%
  }%
  \endgroup
}
\makeatother

\usepackage{orcidlink}

\begin{document}

\begin{center}
{\bf \Large Graviton propagator in de Sitter space\\ in a simple one-parameter gauge}

\

\renewcommand{\thefootnote}{\fnsymbol{footnote}}

{Dra\v{z}en Glavan}\,\orcidlink{0000-0002-1983-0448}\,\footnote{email: \href{mailto:glavan@fzu.cz}{\tt glavan@fzu.cz}}

\setcounter{footnote}{0} 

\medskip

{\it CEICO, FZU --- Institute of Physics of the Czech Academy of Sciences,}
\\
{\it Na Slovance 1999/2, 182 21 Prague 8, Czech Republic}

\medskip

\medskip

\parbox{0.86\linewidth}{
We construct the graviton propagator in de Sitter space in a one-parameter 
family of non-covariant gauges. This family generalizes the simple gauge in which 
most graviton loop computations in de Sitter space have been performed. The 
resulting propagator has a relatively simple form and will facilitate checks of 
the gauge dependence of one-loop computations and proposed observables.

}

\end{center}

\medskip

\hrule

\tableofcontents



\

\section{Introduction}
\label{sec: Introduction}

Primordial inflation stands out in the history of our Universe as a period of extreme
conditions during which quantum gravitational effects, although still small, need not
be prohibitively suppressed. This is due to gravitational particle
production~\cite{Parker:1968mv,Parker:1969au,Parker:1971pt}: a vast ensemble of long-wavelength
inflationary gravitons is produced because of their non-conformal coupling to the
rapidly expanding and accelerating background~\cite{Starobinsky:1979ty}. Since gravitons couple universally to all matter, this ensemble enhances graviton loop corrections.
The magnitude of this enhancement is controlled by the loop-counting parameter 
that can be as large as~$\kappa^2 H^2 \!\sim\! 10^{-10}$ for some models of
inflation~\cite{Planck:2018jri}. Although this effective coupling remains small enough for 
loop corrections to be computed within the effective theory of quantum
gravity~\cite{Donoghue:1994dn,Donoghue:1995cz,Burgess:2003jk}, it is nevertheless
significantly larger than in post-inflationary epochs.

While realistic inflationary dynamics involves a slowly rolling scalar field that leads to a
quasi-exponential expansion, the idealization to exact exponential expansion usually provides
an excellent approximation. This corresponds to the de Sitter limit, which offers further
technical simplifications because it is a maximally symmetric space. The infrared effects
relevant for inflationary gravitons are well captured in this limit, which underscores the
importance of perturbative quantum gravity in de Sitter space. Here we compute the graviton
propagator (two-point function) in de Sitter space in a particularly simple one-parameter gauge, which can
be used as a basic ingredient in perturbative loop computations.

Long-wavelength inflationary gravitons, enhanced by tree-level gravitational particle
production, can communicate information about the accelerating background to the matter
fields with which they interact, and which would otherwise be insensitive to the expansion.
The effects of this interaction are captured perturbatively by one-loop self-masses and
self-energies~\cite{Tsamis:1996qm,Tsamis:1996qk,Tsamis:2005je,Miao:2005am,Kahya:2007bc,Miao:2012bj,Leonard:2013xsa,Glavan:2015ura,Miao:2017vly,Glavan:2020gal,Glavan:2021adm},\footnote{One-graviton-loop correction to the self-mass of a massless, conformally coupled scalar was also reported
in~\cite{Boran:2014xpa,Boran:2017fsx} using the simple graviton gauge, but that result disagrees 
with~\cite{Glavan:2020gal}.} 
which have thus far been computed mostly in the simple graviton gauge~\cite{Tsamis:1992xa,Woodard:2004ut}.
These 1-particle-irreducible two-point functions appear in the quantum-corrected equations of
motion for matter fields. Solving these effective equations yields quantum corrections to
the classical behaviour of fields in de Sitter
space~\cite{Miao:2006gj,Glavan:2013jca,Glavan:2016bvp,Wang:2014tza,Miao:2018bol,Glavan:2021adm,Glavan:2023tet,Tsamis:1996qq,Mora:2013ypa,Glavan:2020ccz,Tan:2021ibs,Tan:2021lza,Tan:2022xpn,Katuwal:2023wtl} (see also \cite{Janssen:2008dw,Frob:2018tdd,Lima:2020prb} for works on graviton loop corrections in power-law inflation and slow-roll inflation). 
These corrections can exhibit secular, i.e.~time-dependent, behaviour, which can increase
their magnitude relative to naive dimensional analysis by up to two orders of magnitude for
minimal inflation, and even more for longer periods of inflation. This secular growth is
relevant to assessing the potential observability of quantum-gravitational effects generated
during inflation.

However, the reality of secular corrections has been debated in the 
literature~\cite{Garriga:2007zk,Tsamis:2007is,Higuchi:2011vw,Miao:2011ng,Morrison:2013rqa,Miao:2013isa,Frob:2014fqa,Woodard:2015kqa}, 
with the main question being whether they are a gauge artifact. Although this issue has not yet been 
settled by explicit computations, the only available one-loop 
computation~\cite{Glavan:2016bvp,Glavan:2015ura} in the exact generally covariant graviton 
gauge~\cite{Miao:2011fc,Kahya:2011sy,Mora:2012zi} has demonstrated that the leading secular 
correction is gauge dependent. When compared to the result~\cite{Wang:2014tza} obtained in the simple 
gauge, the two computations exhibit the same secular behaviour of the one-graviton-loop
correction to the dynamical photons in de Sitter, but the coefficients of the leading 
secular logarithms differ. We note that this does {\it not} imply that the gauge-independent part of 
the secular corrections must vanish. It is therefore important to understand how to determine the 
physical corrections reliably.

In flat space quantum field theory, the question of gauge-independent observables is resolved 
by the construction of the S-matrix, which is guaranteed to be gauge independent. However, the 
S-matrix is not available in cosmology, and, moreover, 
it does not correspond to the initial-value 
formulation of cosmological quantum field theory. Therefore, it is necessary to construct 
one-loop quantum-gravitational observables in inflation and to test their gauge dependence.

A program to address the gauge dependence of secular corrections to de Sitter–space
quantum-gravitational observables was initiated in~\cite{Miao:2017feh}. It is based on the
earlier works~\cite{Donoghue:1993eb,Donoghue:1994dn,Bjerrum-Bohr:2002fji,Bjerrum-Bohr:2002gqz},
which showed how to isolate from the S-matrix the leading infrared graviton-loop corrections
to exchange potentials in flat space. These corrections become gauge independent once the
graviton-loop contributions to the source and the observer are included, as they are part of
the flat-space S-matrix. In~\cite{Miao:2017feh}, it was shown that one can infer the
gauge-independent part of the self-mass by incorporating these corrections without taking
the asymptotic-time limits that enter the S-matrix definition. This approach allowed the
derivation of the gauge-independent one-loop–corrected effective field equation for the
self-mass of the minimally coupled massless scalar~\cite{Miao:2017feh}, and likewise for
electromagnetism~\cite{Katuwal:2021thy}.

Demonstrating that this program also works in de Sitter space requires first proving that 
the secular correction to the scalar exchange potential~\cite{Glavan:2024elz} obtained in the simple 
gauge is gauge independent. This requires recalculating the correction in different 
graviton gauges with arbitrary gauge-fixing parameters, in order to show that any dependence 
on these parameters drops out from the final result. One possibility is to use the propagator 
from~\cite{Glavan:2019msf}, which was obtained in a generalization of the simple gauge 
containing two infinitesimal gauge-fixing parameters. In this paper, we construct an 
alternative propagator suitable for this purpose in a one-parameter family of non-covariant 
gauges, characterized by the following gauge-fixing action:
\begin{equation}
S_{\rm gf}[h_{\mu\nu}] = \int\! d^{D\!}x \, \sqrt{-g} \,
	\biggl[ - \frac{ g^{\mu\nu} \mathscr{F}_\mu \mathscr{F}_\nu }{2\alpha} \biggr]
	\, ,
\quad \
\mathscr{F}_\mu = g^{\rho\sigma}
	\biggl[ \bigl( \nabla_\rho \!-\! 2n_\rho \bigr) \delta g_{\sigma\mu}
		\!- \frac{1}{2} \bigl( \nabla_\mu \!-\! 2n_\mu \bigr) \delta g_{\rho\sigma} \biggr]
		\, ,
\label{GaugeIntro}
\end{equation}
which contains an arbitrary gauge-fixing parameter~$\alpha$. Here the graviton 
field is defined as the perturbation of the metric,~$g_{\mu\nu} \!=\! \overline{g}_{\mu\nu} \!+\! \kappa \delta g_{\mu\nu}$, around the de Sitter 
background,~$\overline{g}_{\mu\nu} \!=\! a^2 \eta_{\mu\nu}$,
with~$\kappa \!=\! \sqrt{16\pi G_{\scr N}}$ denoting the loop-counting 
parameter, and where~$n_\mu \!=\! aH \delta_\mu^0$ is a time-like vector,~$n^\mu n_\mu\!=\!-H^2$, with~$H$ the constant Hubble rate of de Sitter space. This gauge 
constitutes a one-parameter generalization of the simple gauge, which corresponds to 
$\alpha\!=\!1$.

One could also consider redoing the computation in generally covariant gauges. However, the 
loop computation in the one-parameter exact gauge~\cite{Glavan:2015ura} proved to be 
considerably more difficult than the corresponding computation in the simple 
gauge~\cite{Leonard:2013xsa}. Moreover, the most general two-parameter generally covariant 
gauge propagator reported in~\cite{Frob:2016hkx} does not reproduce the exact gauge 
propagator~\cite{Miao:2011fc,Kahya:2011sy,Mora:2012zi} in the appropriate limit, a problem 
that must be resolved before further computations can be attempted. Experience with photon 
gauges~\cite{Glavan:2022dwb,Glavan:2022nrd,Domazet:2024dil,Glavan:2025iuw} also shows that 
appropriately chosen non-covariant gauges tend to be considerably simpler than covariant 
ones. Since gauge choices are ultimately a matter of convenience, it is advantageous to 
employ simpler non-covariant gauges. Indeed, the propagator constructed here is 
considerably simpler than its generally covariant counterparts.

The purpose of this work is to construct the graviton two-point function in this 
one-parameter extension of the simple gauge, providing the input required for 
subsequent loop computations to be addressed in future work. The construction is 
presented in seven sections. Section~\ref{sec: Graviton in de Sitter} gives the details 
of the canonical formulation of linearized gravitons in de Sitter space, in both the 
gauge-invariant and gauge-fixed formulations. The latter formulation serves as the starting 
point for quantization on an indefinite metric space of states appropriate for multiplier 
gauges in Section~\ref{sec: Graviton quantization}. The dynamics of the graviton field 
operator is solved in Section~\ref{sec: Dynamics and state}, relying on the results for 
scalar two-point functions collected in Section~\ref{sec: Scalar mode functions and 
two-point functions}. Section~\ref{sec: Two-point function} presents the graviton 
two-point function, which is the main result of the paper. A brief summary and discussion 
of the results are given in Section~\ref{sec: Discussion}, and further technical details 
required to verify the main result are collected in the Appendix.

\section{Graviton in de Sitter}
\label{sec: Graviton in de Sitter}

In this section we introduce the linearized graviton system in de Sitter and develop its canonical (Hamiltonian) formulation. We first analyze the gauge-invariant theory, 
from which we obtain 
the first-class constraints associated with linearized diffeomorphisms. We then introduce a 
one-parameter family of non-covariant multiplier gauges and derive the corresponding 
gauge-fixed canonical formulation, which will provide the basis for the quantization in the 
next section.

The dynamics of general relativity with a cosmological constant is encoded
in the action
\begin{equation}
S_{\scr \rm EH}[g_{\mu\nu}] = \int\! d^{D\!}x \, \sqrt{-g} \,
	\frac{1}{\kappa^2} \Bigl[ R - (D\!-\!2) \Lambda \Bigr] \, ,
\label{EHaction}
\end{equation}
where~$g$ is the determinant of the positive-signature
metric~$g_{\mu\nu}$,~$\Lambda$ is the positive cosmological constant, 
and~$\kappa^2\!=\!16 \pi G_{\scr N}$ is the rescaled Newton constant.
The Ricci scalar $R\!=\!g^{\mu\nu}R_{\mu\nu}$ is obtained by contracting the Ricci tensor
$R_{\mu\nu} \!=\! {R^\rho}_{\mu\rho\nu}$, which in turn is obtained by contracting the Riemann 
tensor $R^\rho{}_{\mu\sigma\nu} \!=\! \partial_\sigma \Gamma^\rho_{\mu\nu}
	\!-\! \partial_\nu \Gamma^\rho_{\mu\sigma}
	\!+\! \Gamma_{\mu\nu}^\rho \Gamma^\alpha_{\sigma\alpha}
	\!-\! \Gamma_{\mu\sigma}^\alpha \Gamma^\rho_{\nu\alpha}$,
defined in terms of the Christoffel symbols $\Gamma^\alpha_{\mu\nu} \!=\! \frac{1}{2} g^{\alpha\beta} \bigl( \partial_\mu g_{\nu\beta} \!+\! \partial_\nu g_{\mu\beta} 
	\!-\! \partial_\beta g_{\mu\nu} \bigr)$.

The vacuum solution of this theory is the maximally symmetric de Sitter space, for which
all curvature tensors can be expressed in terms of the metric,
\begin{equation}
R_{\mu\nu\rho\sigma} = \frac{2\Lambda}{D \!-\! 1} g_{\mu[\rho} g_{\sigma]\nu}
\, , \qquad \quad
R_{\mu\nu} = \Lambda g_{\mu\nu} \, ,
\qquad \quad
R = D \Lambda \, .
\end{equation}
This spacetime is of particular interest in cosmology, since its exponentially
expanding Poincar\'{e} patch provides a good leading-order model for slow-roll 
inflation. The metric on this patch is conveniently written in conformally
flat form,
\begin{equation}
g_{\mu\nu} = a^2(\eta) \eta_{\mu\nu} \, ,
\end{equation}
where $\eta_{\mu\nu}$ is the $D$-dimensional Minkowski metric, and the
scale factor
\begin{equation}
a(\eta) = \bigl[ 1 \!-\! H(\eta \!-\!\eta_0) \bigr]^{-1} \, ,
\end{equation}
is a function of conformal time~$\eta$, which ranges over the interval
$\eta\!\in\!(-\infty,\eta_0\!+\!1/H)$, with~$\eta_0$ denoting the time at which
$a(\eta_0)\!=\!1$. Spatial slices are~$(D\!-\!1)$-dimensional Euclidean spaces.
The Hubble rate is related to the cosmological constant as 
$H^2\!=\!\Lambda/(D\!-\!1)$, and we often express the expansion in terms of the
conformal Hubble parameter~$\mathcal{H}\!=\!H a$.

For our purposes it is convenient to define the linearized graviton field
$h_{\mu\nu}$ as the perturbation of a conformally rescaled de Sitter metric,
\begin{equation}
g_{\mu\nu} = a^2 \bigl( \eta_{\mu\nu} \!+\! \kappa h_{\mu\nu} \bigr) \, ,
\label{MetricSplit}
\end{equation}
where, henceforth, indices on $h_{\mu\nu}$ are raised and lowered with the Minkowski metric.
This definition is just a conformal rescaling of the metric perturbation introduced 
in~(\ref{GaugeIntro}),~$h_{\mu\nu} \!=\! a^{-2} \delta g_{\mu\nu}$.
Expanding the action~(\ref{EHaction}) to second order in $h_{\mu\nu}$ yields
the quadratic action for the linearized theory (see e.g.~\cite{Glavan:2020gal}),
\begin{align}
\MoveEqLeft[12]
S[h_{\mu\nu}] 
	= \int\! d^{D\!}x \, a^{D-2} \, \biggl[
	\frac{1}{2} (\partial_\mu h_{\nu\rho}) (\partial^\nu h^{\mu\rho})
	- \frac{1}{2} (\partial_\mu h^{\mu\nu}) (\partial_\nu h)
	+ \frac{1}{4} (\partial_\mu h) (\partial^\mu h)
\nonumber \\
&
	- \frac{1}{4} (\partial_\rho h_{\mu\nu}) (\partial^\rho h^{\mu\nu}) 
	+ \frac{1}{2} (D\!-\!2) \mathcal{H}  h_{0\mu} \partial^\mu h
	\biggr] \, .
\label{InvariantAction}
\end{align}
This action is invariant under the linearized gauge transformation
\begin{equation}
h_{\mu\nu}
	\longrightarrow
	h_{\mu\nu} 
	+
	\frac{1}{a^2} \nabla_{(\mu} \xi_{\nu)}
	=
	h_{\mu\nu}
	+
	\Bigl(
	\delta^\rho_{(\mu} \partial_{\nu)}
	-
	\eta_{\mu\nu} \delta^\rho_0 \mathcal{H}
	\Bigr)
	\frac{\xi_\rho}{a^2}
	\, ,
\label{GaugeTransformation}
\end{equation}
for an arbitrary vector field $\xi_\mu$. The nonstandard form of the
transformation reflects the metric split in~(\ref{MetricSplit}), in which two
powers of the scale factor have been factored out.

The equation of motion that follows from the gauge-invariant action~(\ref{InvariantAction})
can be written compactly as
\begin{equation}
\boldsymbol{L}^{\mu\nu\rho\sigma} h_{\rho\sigma} = 0 \, ,
\end{equation}
with the help of the modified Lichnerowicz operator,
\begin{equation}
\boldsymbol{L}^{\mu\nu\rho\sigma} 
	=
	V^{\mu\nu\rho\sigma\omega\lambda} \partial_\omega a^{D-2} \partial_\lambda
	-
	\frac{1}{2} (D\!-\!2) a^{D-2} \mathcal{H} 
		\Bigl[
		\eta^{\mu\nu} \delta_0^{(\rho} \partial^{\sigma)}
		- \eta^{\rho\sigma} \delta_0^{(\mu} \partial^{\nu)}
		- (D\!-\!1) \mathcal{H} \eta^{\mu\nu} \delta_0^\rho \delta_0^\sigma \Bigr]
		\, ,
\label{LichnerowiczDefinition}
\end{equation}
whose double-derivative part inherits the tensor structure from its flat-space counterpart,
\begin{equation}
V^{\mu\nu\rho\sigma\omega\lambda}
	=
	\frac{1}{2} \eta^{\rho\sigma} \eta^{\omega(\mu} \eta^{\nu)\lambda}
	-
	\eta^{\omega(\rho} \eta^{\sigma)(\mu}  \eta^{\nu) \lambda}
	+
	\frac{1}{2} \eta^{\mu(\rho} \eta^{\sigma)\nu} \eta^{\omega\lambda}
	+
	\frac{1}{2} \eta^{\mu\nu} \bigl( \eta^{\omega(\rho} \eta^{\sigma)\lambda}
		\!-\!
		\eta^{\rho\sigma} \eta^{\omega\lambda} \bigr)
	\, .
\end{equation}
The modified form of the operator in~(\ref{LichnerowiczDefinition}) is due to
the conformal rescaling in the definition of the graviton in~(\ref{MetricSplit}).

To set up the canonical quantization, we first derive the Hamiltonian form of the
gauge-invariant action~(\ref{InvariantAction}). We then extend this construction to a
one-parameter family of non-covariant multiplier gauges, obtaining the corresponding
gauge-fixed Hamiltonian that will underlie the quantized theory.

\subsection{Gauge-invariant canonical formulation}
\label{subsec: Gauge-invariant canonical formulation}

To prepare the gauge-invariant action~(\ref{InvariantAction}) for canonical analysis
it is convenient to separate temporal and spatial indices, thereby isolating the
time-derivative structure. In terms of this decomposition the action takes the form
\begin{align}
\MoveEqLeft[1.5]
S[h_{\mu\nu}]
	=
	\int\! d^{D\!}x \, a^{D-2} \biggl[
	\frac{1}{4} \Bigl( \partial_0 h_{ij} \!-\! 2 \partial_{(i} h_{j)0} 
			\!+\! \delta_{ij} \mathcal{H} h_{00} \Bigr) 
		\Bigl( \partial_0 h_{ij} \!-\! 2 \partial_{(i} h_{j)0} 
			\!+\! \delta_{ij} \mathcal{H} h_{00} \Bigr)
\nonumber \\
&
	- \frac{1}{4} \Bigl( \partial_0 h_{ii} \!-\! 2 \partial_i h_{0i} 
			\!+\! (D\!-\!1) \mathcal{H} h_{00} \Bigr) 
		\Bigl( \partial_0 h_{jj} \!-\! 2 \partial_j h_{0j} 
			\!+\! (D\!-\!1) \mathcal{H} h_{00} \Bigr)
	+ 
	\frac{1}{2} (\partial_i h_{00}) \bigl( \partial_j h_{ij} \!-\! \partial_i h_{jj} \bigr)
\nonumber \\
&	\hspace{1.cm}
	+ 
	\frac{1}{2} (\partial_i h_{ik} ) ( \partial_j h_{jk} )
	- 
	\frac{1}{2} ( \partial_i h_{ij} ) ( \partial_j h_{kk} )
	+
	\frac{1}{4} ( \partial_k h_{ii} ) ( \partial_k h_{jj} )
	-
	\frac{1}{4} ( \partial_k h_{ij} ) ( \partial_k h_{ij} )
	\biggr] \, ,
\label{DecomposedAction}
\end{align}
where the expression has been simplified using integration by parts.

The next step is to introduce the extended action, a first-order formulation that is
on-shell equivalent to~(\ref{DecomposedAction}) and well suited for identifying the
constraint structure. This is done by promoting all time derivatives in
(\ref{DecomposedAction}) to independent velocity fields,
\begin{align}
\partial_0 h_{00} \longrightarrow v_{00} \, ,
\qquad\quad
\partial_0 h_{0i} \longrightarrow v_{0i} \, ,
\qquad\quad
\partial_0 h_{ij} - 2 \partial_{(i} h_{j)0} + \delta_{ij} \mathcal{H} h_{00}
	\longrightarrow v_{ij}
	\, ,
\end{align}
and introducing Lagrange multipliers~$\pi_{00},\pi_{0i},\pi_{ij}$ to enforce equivalence with
the original dynamics. The resulting extended action is
\begin{align}
&
\mathcal{S} \bigl[ h_{00}, v_{00}, \pi_{00}, h_{0i}, v_{0i}, \pi_{0i}, h_{ij}, v_{ij}, \pi_{ij} \bigr]
	=\!\!
	\int\!\! d^{D\!}x
	\biggl\{
	a^{D-2}
	\biggl[
	\frac{1}{4} \bigl( v_{ij} v_{ij} \!-\! v_{ii} v_{jj} \bigr)
	\!+
	\frac{1}{2} (\partial_i h_{00}) \bigl( \partial_j h_{ij} \!-\! \partial_i h_{jj} \bigr)
\nonumber \\
&	\hspace{0.9cm}
	+ 
	\frac{1}{2} (\partial_i h_{ik} ) ( \partial_j h_{jk} )
	- 
	\frac{1}{2} ( \partial_i h_{ij} ) ( \partial_j h_{kk} )
	+
	\frac{1}{4} ( \partial_k h_{ii} ) ( \partial_k h_{jj} )
	-
	\frac{1}{4} ( \partial_k h_{ij} ) ( \partial_k h_{ij} )
	\biggr] 
\nonumber \\
&	\hspace{0.9cm}
	+
	\pi_{00} \bigl( \partial_0 h_{00} - v_{00} \bigr)
	+
	\pi_{0i} \bigl( \partial_0 h_{0i} - v_{0i} \bigr)
	+
	\pi_{ij} \Bigl( \partial_0 h_{ij} - 2 \partial_{(i} h_{j)0} 
		+ \delta_{ij} \mathcal{H} h_{00} - v_{ij} \Bigr)
	\biggr\}
	\, .
\end{align}

To obtain the canonical action we first solve on-shell for as many velocity components as possible, which in the case at hand are~$v_{ij}$.  
Varying the extended action with respect to $v_{ij}$ gives
\begin{equation}
\frac{\delta \mathcal{S} }{\delta v_{ij} }
	=
	\frac{ 1 }{ 2 } a^{D-2} \bigl( v_{ij} - \delta_{ij} v_{kk} \bigr) - \pi_{ij} \approx 0 \, ,
\end{equation}
which is solved by
\begin{equation}
v_{ij} \approx \overline{v}_{ij} = 
	2 a^{2-D} \Bigl( \pi_{ij} \!-\! \frac{ \delta_{ij} \pi_{kk} }{ D\!-\!2 } \Bigr) 
	\, .
\end{equation}
Here and throughout this section we use Dirac’s notation, with~$\approx$ denoting weak (on-shell) equalities and~$=$ denoting strong (off-shell) ones.  
Substituting~$\overline{v}_{ij}$ back into the extended action yields the canonical action
\begin{align}
\MoveEqLeft[3]
\mathscr{S}\bigl[ h_{00}, \pi_{00}, h_{0i}, \pi_{0i}, h_{ij},  \pi_{ij}, v_{00}, v_{0i} \bigr] 
\equiv
\mathcal{S}\bigl[ h_{00}, v_{00}, \pi_{00}, h_{0i}, v_{0i}, \pi_{0i}, h_{ij}, \overline{v}_{ij}, \pi_{ij} \bigr] 
\nonumber \\
&
	=
	\int\! d^{D\!}x \, \Bigl[
		\pi_{00} \partial_0 h_{00} 
		+ \pi_{0i} \partial_0 h_{0i} 
		+ \pi_{ij} \partial_0 h_{ij} 
		- \mathscr{H}
		- v_{00} \Phi_0
		- v_{0i} \Phi_{i} 
		\Bigr] \, .
\label{InvariantCanonicalAction}
\end{align}
Here the canonical Hamiltonian density is
\begin{align}
\mathscr{H}
	={}&
	a^{2-D} \biggl[ \pi_{ij} \pi_{ij} 
		- \frac{ \pi_{ii} \pi_{jj} }{ D\!-\!2 } \biggr]
	\!+ \pi_{ij} \Bigl( 2 \partial_{(i} h_{j)0} 
		- \delta_{ij} \mathcal{H} h_{00} \Bigr)
	\! + a^{D-2} \biggl[ 
	\frac{1}{2} (\partial_i h_{00}) \bigl( \partial_i h_{jj} \!-\! \partial_j h_{ij}  \bigr)
\nonumber \\
&
	-
	\frac{1}{2} (\partial_i h_{ik}) (\partial_j h_{jk})
	+
	\frac{1}{2} (\partial_i h_{ij}) (\partial_j h_{kk})
	-
	\frac{1}{4} (\partial_k h_{ii}) (\partial_k h_{jj})
	+
	\frac{1}{4} (\partial_k h_{ij}) (\partial_k h_{ij})
	\biggr]
	 \, ,
\end{align}
while the Lagrange multipliers $v_{00}$ and $v_{0i}$ enforce the primary constraints
\begin{equation}
\Phi_0 \equiv \pi_{00} \approx 0 \, ,
\qquad \quad
\Phi_i \equiv \pi_{0i} \approx 0 \, .
\label{PrimaryConstraints}
\end{equation}

The conservation of primary constraints produces secondary constraints~$\Psi_0$
and~$\Psi_i$,
\begin{subequations}
\begin{align}
&
\partial_0 \Phi_0
	\approx
	\frac{ a^{D-2} }{2} \bigl( \nabla^2 h_{ii} - \partial_i \partial_j h_{ij} \bigr) 
	+ 
	\mathcal{H} \pi_{ii}
	\equiv \Psi_0
	\approx 0 
	\, ,
\\
&
\partial_0 \Phi_i
	\approx
	2 \partial_j \pi_{ij} \equiv \Psi_i 
	\approx 0 
	\, .
\end{align}
\label{SecondaryConstraints}%
\end{subequations}
The conservation of secondary constraints generates no further constraints. These 
primary and secondary constraints together encode the linearized diffeomorphism symmetry of the theory.

\subsection{Gauge-fixed canonical formulation}
\label{subsec: Gauge-fixed formulation}

We now turn to the gauge-fixed formulation in the one-parameter family of non-covariant
gauges. The gauge-fixed action is obtained by adding a gauge-fixing functional to the
gauge-invariant action~(\ref{InvariantAction}),
\begin{equation}
S_\star[h_{\mu\nu}] = S[h_{\mu\nu}] + S_{\rm gf}[h_{\mu\nu}] \, .
\label{GaugeFixedAction}
\end{equation}
This addition breaks the gauge invariance of the theory. The specific choice of
gauge-fixing functional is, in principle, a matter of convenience. In what follows we adopt
a one-parameter generalization of the simple graviton gauge~\cite{Tsamis:1992xa,Woodard:2004ut} given in~(\ref{GaugeIntro}),
which, for the graviton field defined in~(\ref{MetricSplit}), reads
\begin{equation}
S_{\rm gf}[h_{\mu\nu}] = \int\! d^{D\!}x \, a^{D-2} \,
        \biggl[ - \frac{1}{2\alpha}
        \mathscr{F}^\mu \mathscr{F}_\mu \biggr]
        \, ,
\qquad \quad
\mathscr{F}_\mu = \mathscr{D}_{\mu\rho\sigma} h^{\rho\sigma}
        \, ,
\label{GF}
\end{equation}
where the associated linear differential operator is
\begin{equation}
\mathscr{D}_{\mu\rho\sigma}
        =
        \eta_{\mu(\rho}
                \Bigl[
                \partial_{\sigma)}
                +
                (D\!-\!2) \mathcal{H} \delta^0_{\sigma)}
                \Bigr]
        - \frac{1}{2} \eta_{\rho\sigma} \partial_\mu
        \, .
\label{calDdef}
\end{equation}
With this choice the gauge-fixed equations of motion take the form
\begin{equation}
\boldsymbol{D}^{\mu\nu\rho\sigma} h_{\rho\sigma} = 0 \, ,
\qquad \quad
\boldsymbol{D}^{\mu\nu\rho\sigma}
	=
	\boldsymbol{L}^{\mu\nu\rho\sigma}
	+
	\frac{1}{\alpha} \mathscr{D}_{\omega}{}^{\mu\nu}
	a^{D-2} \mathscr{D}^{\omega\rho\sigma}
	\, ,
\label{GaugeFixedKineticOperator}
\end{equation}
where the Lichnerowicz operator was defined in~(\ref{LichnerowiczDefinition}).

Setting~$\alpha\!=\!1$ reproduces the simple gauge. In the canonical formulation this choice
for the gauge-fixing functional corresponds to fixing the Lagrange multipliers in
(\ref{InvariantCanonicalAction}) to particular linear functions of the canonical variables
(see, e.g.,~\cite{Glavan:2022pmk} for an analogous discussion in the vector-field case).
Starting from the action~(\ref{GaugeFixedAction}) we now derive the corresponding
gauge-fixed canonical formulation.

We begin by performing the spatio-temporal decomposition of the gauge-fixing term,
\begin{equation}
S_{\rm gf}[h_{\mu\nu}]
	=
	\int\! d^{D\!}x \, a^{D-2}
	\biggl[
	\frac{1}{2\alpha} \mathscr{F}_0 \mathscr{F}_0
	-
	\frac{1}{2\alpha} \mathscr{F}_i \mathscr{F}_i
	\biggr]
	\, ,
\end{equation}
where
\begin{align}
&
\mathscr{F}_0 =
	- \frac{1}{2} \partial_0 h_{0 0} 
	- \frac{1}{2} \partial_0 h_{ii}
	+ \partial_i h_{0i} 
	- (D\!-\!2) \mathcal{H} h_{0 0} \, ,
\\
&
\mathscr{F}_i =
	- \partial_0 h_{0 i}
	+ \frac{1}{2} \partial_i ( h_{00}  - h_{jj} ) 
	+ \partial_j h_{ij} 
	- (D\!-\!2) \mathcal{H} h_{0 i}
	\, .
\end{align}
To pass to the canonical description in the presence of gauge fixing, we introduce an extended action by promoting specific time-derivative combinations to independent velocity fields,
\begin{align}
\partial_0 h_{00}
	+ (D\!-\!3) \mathcal{H} h_{00}
	\longrightarrow{}&
	v_{00}
	\, ,
\\
\partial_0 h_{0 i} 
	- \frac{1}{2} \partial_i ( h_{00} - h_{jj} )
	- \partial_j h_{ij} 
	+ (D\!-\!2) \mathcal{H} h_{0 i}
	\longrightarrow{}&
	v_{0i}
	\, ,
\\
\partial_0 h_{ij} - 2 \partial_{(i} h_{j)0} + \delta_{ij} \mathcal{H} h_{00}
	\longrightarrow{}& 
	v_{ij}
	\, ,
\end{align}
and introducing the corresponding Lagrange multipliers to enforce on-shell equivalence. The extended action takes the form
\begin{align}
\MoveEqLeft[3]
\mathcal{S}_\star \bigl[ h_{00}, v_{00}, \pi_{00}, h_{0i}, v_{0i}, \pi_{0i}, h_{ij}, v_{ij}, \pi_{ij} \bigr]
	=
	\int\! d^{D\!}x \, 
	\biggl\{
	a^{D-2}
	\biggl[
	\frac{1}{4} v_{ij} v_{ij}
	-
	\frac{1}{4} v_{ii} v_{jj}
\nonumber \\
&
	+ 
	\frac{1}{2} (\partial_i h_{00}) \bigl( \partial_j h_{ij} - \partial_i h_{jj} \bigr)
	+ 
	\frac{1}{2} (\partial_i h_{ik} ) ( \partial_j h_{jk} )
	- 
	\frac{1}{2} ( \partial_i h_{ij} ) ( \partial_j h_{kk} )
\nonumber \\
&
	+
	\frac{1}{4} ( \partial_k h_{ii} ) ( \partial_k h_{jj} )
	-
	\frac{1}{4} ( \partial_k h_{ij} ) ( \partial_k h_{ij} )
	+
	\frac{ 1 }{8\alpha} \bigl( v_{00} +  v_{ii} \bigr)^2
	-
	\frac{v_{0i} v_{0i}}{2\alpha}
	\biggr] 
\nonumber \\
&
	+
	\pi_{00} \Bigl( \partial_0 h_{00} + (D\!-\!3)\mathcal{H} h_{00} - v_{00} \Bigr)
	+
	\pi_{0i} \Bigl( \partial_0 h_{0i} 
	-
	\frac{1}{2} \partial_i \bigl( h_{00} - h_{jj} \bigr)
\nonumber \\
&
	-
	\partial_j h_{ij} 
	+
	(D\!-\!2) \mathcal{H} h_{0 i}
	- 
	v_{0i} \Bigr)
	+
	\pi_{ij} \Bigl( 
		\partial_0 h_{ij} 
		-
		2 \partial_{(i} h_{j)0}
		+
		\delta_{ij} \mathcal{H} h_{00} 
		-
		v_{ij} \Bigr)
	\biggr\}
	\, .
\end{align}

Varying with respect to the velocity fields yields the on-shell relations
\begin{align}
\frac{\delta \mathcal{S}_\star }{ \delta v_{00} }
	={}&
	\frac{1}{4\alpha} a^{D-2} \bigl( v_{00} + v_{ii} \bigr)
	-
	\pi_{00}
	\approx 0
	\, ,
\\
\frac{\delta \mathcal{S}_\star}{ \delta v_{0i} }
	={}&
	-
	\frac{ 1 }{ \alpha } a^{D-2} v_{0i}
	-
	\pi_{0i}
	\approx 0
	\, ,
\\
\frac{\delta \mathcal{S}_\star}{ \delta v_{ij} }
	={}&
	\frac{1}{2}
	a^{D-2}
	\bigl( v_{ij} - \delta_{ij} v_{kk} \bigr)
	+
	\frac{ \delta_{ij} }{4\alpha} a^{D-2} \bigl( v_{00} + v_{kk} \bigr)
	-
	\pi_{ij}
	\approx 0
	\, .
\end{align}
Owing to the gauge-fixing functional, it is now possible to solve for all velocity fields
on-shell,
\begin{align}
&
v_{00} \approx \overline{v}_{00}
	= 
	\frac{ 2 a^{2-D} }{ D\!-\!2 } \Bigl[
		\pi_{ii}
		- 
		\Bigl( D \!-\! 1 \!-\! 2(D\!-\!2)\alpha \Bigr) \pi_{00}
		\Bigr] 
		\, ,
\\
&
v_{0i} \approx \overline{v}_{0i}
	= 
	- \alpha a^{2-D} \pi_{0i} 
	\, ,
\\
&
v_{ij} \approx \overline{v}_{ij}
	= 
	2a^{2-D} \biggl[ 
	\pi_{ij}
	+
	\frac{\delta_{ij} ( \pi_{00} \!-\! \pi_{kk} ) }{ D\!-\!2 }
	\biggr] 
	\, .
\end{align}
Substituting these solutions as strong equalities back into the extended action gives the gauge-fixed canonical action,
\begin{align}
&
\mathscr{S}_\star 
	\bigl[ h_{00}, \pi_{00}, h_{0i}, \pi_{0i}, h_{ij}, \pi_{ij} \bigr] 
	\equiv
	\mathcal{S}_\star
	\bigl[ h_{00}, \overline{v}_{00}, \pi_{00}, h_{0i}, \overline{v}_{0i}, 
		\pi_{0i}, h_{ij}, \overline{v}_{ij}, \pi_{ij} \bigr] 
\nonumber \\
&
\hspace{2cm}
	=
	\int\! d^{D\!}x \, \Bigl[
	\pi_{00} \partial_0 h_{00}
	+ \pi_{0i} \partial_0 h_{0 i} 
	+  \pi_{ij} \partial_0 h_{ij} 
	- \mathscr{H}_\star
	\Bigr] \, .
\end{align}
where the corresponding gauge-fixed Hamiltonian is
\begin{align}
\mathscr{H}_\star
	={}&
	a^{2-D}
	\biggl[
	\pi_{ij} \pi_{ij}
	-
	\frac{ \pi_{ii} \pi_{jj} }{D\!-\!2}
	+
	\frac{ 2 \pi_{00} \pi_{ii} }{ D\!-\!2 } 
	-
	\frac{ D \!-\! 1 \!-\! 2 (D \!-\! 2) \alpha }{ D\!-\!2 } \pi_{00} \pi_{00}
	-
	\frac{\alpha}{2} \pi_{0i} \pi_{0i}
	\biggr]
\nonumber \\
&
	+
	\pi_{ij} \Bigl( 
		2 \partial_{i} h_{0j}
		-
		\delta_{ij} \mathcal{H} h_{00} 
		\Bigr)
	+
	\pi_{0i} \biggl[
		\frac{1}{2} \partial_i \bigl( h_{00} - h_{jj} \bigr)
		+
		\partial_j h_{ij}
		-
		(D\!-\!2) \mathcal{H} h_{0 i}
		\biggr]
\nonumber \\
&
	-
	(D\!-\!3)\mathcal{H} \pi_{00} h_{00}
	+
	a^{D-2}
	\biggl[
	\frac{1}{2} (\partial_i h_{00}) \bigl( \partial_i h_{jj} - \partial_j h_{ij} \bigr)
	- 
	\frac{1}{2} (\partial_i h_{ik} ) ( \partial_j h_{jk} )
\nonumber \\
&
	+ 
	\frac{1}{2} ( \partial_i h_{ij} ) ( \partial_j h_{kk} )
	-
	\frac{1}{4} ( \partial_k h_{ii} ) ( \partial_k h_{jj} )
	+
	\frac{1}{4} ( \partial_k h_{ij} ) ( \partial_k h_{ij} )
	\biggr] 
	\, .
\end{align}

The Poisson brackets are determined by the symplectic part of the canonical action and are not affected by gauge fixing:
\begin{subequations}
\begin{align}
\bigl\{ h_{00}(\eta,\vec{x}) , \pi_{00}(\eta,\vec{x}^{\,\prime}) \bigr\}
	={}&
	\delta^{D-1}( \vec{x} \!-\! \vec{x}^{\,\prime} ) \, ,
\\
\bigl\{ h_{0i}(\eta,\vec{x}) , \pi_{0j}(\eta,\vec{x}^{\,\prime}) \bigr\}
	={}&
	\delta_{ij} \delta^{D-1}( \vec{x} \!-\! \vec{x}^{\,\prime} ) \, ,
\\
\bigl\{ h_{ij}(\eta,\vec{x}) , \pi_{kl}(\eta,\vec{x}^{\,\prime}) \bigr\}
	={}&
	\delta_{i(k} \delta_{l)j} \delta^{D-1}( \vec{x} \!-\! \vec{x}^{\,\prime} ) \, ,
\end{align}
\label{PoissonBrackets}%
\end{subequations}
The equations of motion that follow from this canonical action are:
\begin{align}
\partial_0 h_{00} \approx{}&
	\frac{2 a^{2-D}}{D\!-\!2} 
		\Bigl[ \pi_{ii} - \Bigl( D \!-\! 1 \!-\! 2 (D\!-\!2) \alpha \Bigr) \pi_{00} \Bigr]
	-
	(D\!-\!3) \mathcal{H} h_{00}
	\, ,
\label{eom1}
\\
\partial_0 \pi_{00} \approx{}&
	\frac{1}{2} \partial_i \pi_{0i} 
	+
	(D\!-\!3) \mathcal{H} \pi_{00}
	+
	\mathcal{H} \pi_{ii}
	+
	\frac{a^{D-2}}{2} \bigl( \nabla^2 h_{ii} - \partial_i \partial_j h_{ij} \bigr)
	\, ,
\label{eom2}
\\
\partial_0 h_{0i} \approx{}&
	- \alpha a^{2-D} \pi_{0i}
	+
	\frac{1}{2} \partial_i \bigl( h_{00} - h_{jj} \bigr)
	+
	\partial_j h_{ij}
	-
	(D\!-\!2) \mathcal{H} h_{0i}
	\, ,
\label{eom3}
\\
\partial_0 \pi_{0i} \approx{}&
	2 \partial_j \pi_{ij}
	+
	(D\!-\!2) \mathcal{H} \pi_{0i}
	\, ,
\label{eom4}
\\
\partial_0 h_{ij} \approx{}&
	2 a^{2-D} \biggl[
		\pi_{ij} + \frac{\delta_{ij} \bigl( \pi_{00} - \pi_{kk} \bigr) }{ D \!-\! 2 }
		\biggr]
	+
	2 \partial_{(i} h_{j)0}
	-
	\delta_{ij} \mathcal{H} h_{00}
	\, ,
\label{eom5}
\\
\partial_0 \pi_{ij} \approx{}&
	\partial_{(i} \pi_{j)0}
	-
	\frac{\delta_{ij}}{2} \partial_k \pi_{0k}
	+
	\frac{a^{D-2}}{2}
	\Bigl[
		 - \bigl( \partial_i \partial_j - \delta_{ij} \nabla^2 \bigr) h_{00}
		 -
		 2 \partial_k \partial_{(i} h_{j)k}
\nonumber \\
&	\hspace{5.cm}
	 +
	 \partial_i \partial_j h_{kk}
	+
	\delta_{ij} \bigl( \partial_k \partial_l - \delta_{kl} \nabla^2 \bigr) h_{kl}
	+
	\nabla^2 h_{ij}
		\Bigr]
		\, .
\label{eom6}
\end{align}
These can be written as Hamilton equations with the help of the Poisson
brackets~(\ref{PoissonBrackets}).

We note that, when working with multiplier gauges, the gauge-fixed action does not
generate the first-class constraints~(\ref{PrimaryConstraints})
and~(\ref{SecondaryConstraints}) that are present in the gauge-invariant action.
These constraints must therefore be imposed \emph{in addition} to the gauge-fixed action:
\begin{equation}
\Phi_0 = \pi_{00} \, ,
\qquad
\Phi_i = \pi_{0i} \, ,
\qquad
\Psi_0 = \frac{ a^{D-2} }{2} \bigl( \nabla^2 h_{ii} - \partial_i \partial_j h_{ij} \bigr) 
	+ \mathcal{H} \pi_{ii} \, ,
\qquad
\Psi_i = 2 \partial_j \pi_{ij} \, .
\label{FirstClassConstraints}
\end{equation}
It is sufficient to impose these constraints at the level of initial conditions, since they
satisfy a closed system of equations of motion:
\begin{align}
\partial_0 \Phi_0 
	\approx{}&
	\Psi_0
	+ ( D\!-\!3 ) \mathcal{H} \Phi_0
	+ \frac{1}{2} \partial_i \Phi_i
	\, ,
\\
\partial_0 \Phi_i
	\approx{}&
	\Psi_i
	+ (D\!-\!2) \mathcal{H} \Phi_i 
	\, ,
\\
\partial_0 \Psi_0
	\approx{}&
	- \frac{1}{2} \partial_i \Psi_i
	+ \nabla^2 \Phi_0
	+ \mathcal{H} \Psi_0 
	- \frac{1}{2} (D\!-\!3) \mathcal{H} \partial_i \Phi_i
	\, ,
\\
\partial_0 \Psi_i
	\approx{}&
		\nabla^2 \Phi_i
		\, ,
\end{align}
These relations guarantee the conservation of the constraints. Consequently, they form a
closed first-class algebra also in the gauge-fixed canonical formulation. In the next
section we quantize this gauge-fixed canonical system, and in subsequent sections we
construct the graviton two-point function.

\section{Graviton quantization}
\label{sec: Graviton quantization}

In this section, we quantize the linearized graviton field in the multiplier gauge, starting
from the canonical formulation of the previous section. We promote the fields and their
conjugate momenta to operators, replace the classical Poisson brackets by equal-time
commutation relations, and implement the constraints as conditions on the space of states.
This establishes the operator algebra and mode decomposition needed later for the
construction of the graviton two-point function.

We thus promote the classical fields and conjugate momenta to Hermitian operators,
\begin{equation}
h_{\mu\nu} \longrightarrow \hat{h}_{\mu\nu} \, ,
\qquad \quad
\pi_{\mu\nu} \longrightarrow \hat{\pi}_{\mu\nu} \, ,
\label{Quantization}
\end{equation}
and replace their Poisson brackets by equal-time canonical commutation relations,
\begin{subequations}
\begin{align}
\bigl[ \hat{h}_{00}(\eta,\vec{x}) , \hat{\pi}_{00}(\eta,\vec{x}^{\,\prime}) \bigr]
	={}&
	i \delta^{D-1}( \vec{x} \!-\! \vec{x}^{\,\prime} ) \, ,
\\
\bigl[ \hat{h}_{0i}(\eta,\vec{x}) , \hat{\pi}_{0j}(\eta,\vec{x}^{\,\prime}) \bigr]
	={}&
	\delta_{ij} \, i \delta^{D-1}( \vec{x} \!-\! \vec{x}^{\,\prime} ) \, ,
\\
\bigl[ \hat{h}_{ij}(\eta,\vec{x}) , \hat{\pi}_{kl}(\eta,\vec{x}^{\,\prime}) \bigr]
	={}&
	\delta_{i(k} \delta_{l)j} \, i \delta^{D-1}( \vec{x} \!-\! \vec{x}^{\,\prime} ) \, .
\end{align}
\end{subequations}
These field operators satisfy the same equations of motion~(\ref{eom1})--(\ref{eom6})
as their classical counterparts. First-class constraints~(\ref{FirstClassConstraints})
are promoted to Hermitian constraint operators by substitutions in~(\ref{Quantization}).

The constraints~(\ref{FirstClassConstraints}) of the theory are implemented as a
requirement on the space of state vectors. It would be inconsistent with
the canonical commutation relations to demand that Hermitian constraint operators 
annihilate the state. For this reason, constraints are implemented at the level of matrix 
elements, by demanding that matrix elements
of polynomials of the Hermitian constraint operators vanish between physical states.
In particular, for the Gaussian states that we consider, it is sufficient here to require that 
all two-point functions of Hermitian constraint operators vanish,
\begin{equation}
\bigl\langle \Omega \bigr| \hat{\Phi}_\mu(x) 
	\hat{\Phi}_\nu(x') \bigl| \Omega \bigr\rangle
	=0
	\, ,
\qquad
\bigl\langle \Omega \bigr| \hat{\Phi}_\mu(x) 
	\hat{\Psi}_\nu(x') \bigl| \Omega \bigr\rangle
	=0
	\, ,
\qquad
\bigl\langle \Omega \bigr| \hat{\Psi}_\mu(x) 
	\hat{\Psi}_\nu(x') \bigl| \Omega \bigr\rangle
	=0
	\, ,
\label{ConstraintCorrelators}
\end{equation}
for all physical states~$\lvert\Omega\rangle$. This requirement is implemented
as a condition on the state vectors by demanding that four particular {\it non-Hermitian}
linear combinations of the constraint operators annihilate the physical state
(see~\cite{Glavan:2022pmk} for a detailed discussion in the vector field case).
The structure of the state space satisfying these conditions becomes most 
transparent after decomposing the fields into scalar, vector, and tensor components 
and transforming to momentum space. We now turn to these decompositions before 
completing the description of the quantized theory.

\subsection{Scalar-vector-tensor decomposition}
\label{subsec: Scalar-vector-tensor decomposition}

To separate physical and gauge degrees of freedom, it is convenient to 
decompose spatial tensors into their scalar, vector, and tensor parts with respect to 
spatial rotations. This is achieved by introducing transverse and longitudinal projectors,
\begin{equation}
\mathbb{P}_{ij}^{\scr T} = \delta_{ij} \!-\! \frac{\partial_i \partial_j}{\nabla^2} \, ,
\qquad \qquad
\mathbb{P}_{ij}^{\scr L} = \frac{\partial_i \partial_j}{\nabla^2} \, ,
\end{equation}
which are idempotent and mutually orthogonal,
\begin{equation}
\mathbb{P}_{ij}^{\scr T} \mathbb{P}_{jk}^{\scr T} = \mathbb{P}_{ik}^{\scr T} \, ,
\qquad \quad
\mathbb{P}_{ij}^{\scr L} \mathbb{P}_{jk}^{\scr L} = \mathbb{P}_{ik}^{\scr L} \, ,
\qquad \quad
\mathbb{P}_{ij}^{\scr T} \mathbb{P}_{jk}^{\scr L} = \mathbb{P}_{ij}^{\scr L} \mathbb{P}_{jk}^{\scr T} = 0 \, ,
\end{equation}
so that any spatial tensor can be decomposed uniquely into scalar, vector, and tensor pieces.

With these projectors, we decompose the canonical variables into their scalar, vector,
and tensor parts. The decomposition is defined by
\begin{subequations}
\begin{align}
\hat{h}_{00} 
	={}& 
	\hat{S}_1 \, , \quad
&&
\hat{h}_{0i} 
	= 
	\frac{\partial_i}{\nabla^2} \hat{S}_2
	+ 
	\hat{V}^1_i 
	\, , \quad
&&
\hat{h}_{ij} 
	=
	\mathbb{P}_{ij}^{\scr T} \hat{S}_3
	+
	\mathbb{P}_{ij}^{\scr L} \hat{S}_4
	+ 
	\frac{2}{\nabla^2} \partial_{(i} \hat{V}^2_{j)}
	+
	\hat{T}_{ij} 
	\, .
\\
\hat{\pi}_{00} 
	={}& 
	\hat{\Pi}_1 
	\, , \quad
&&
\hat{\pi}_{0i} 
	= 
	- \partial_i \hat{\Pi}_2 
	+ 
	\hat{\Pi}^1_i 
	\, , \quad
&&
\hat{\pi}_{ij} 
	=
	\frac{ \mathbb{P}_{ij}^{\scr T} \hat{\Pi}_3 }{ D\!-\!2 }
	+
	\mathbb{P}_{ij}^{\scr L} \hat{\Pi}_4
	-
	\partial_{(i} \hat{\Pi}^2_{j)}
	+ 
	\hat{\Pi}_{ij} 
	\, .
\end{align}
\label{SVTdecomposition}%
\end{subequations}
Here the vectors are transverse, 
\begin{equation}
\partial_i \hat{V}_i^I = 0 \, ,
\qquad\quad
\partial_i \hat{\Pi}_i^I = 0 \ ,
\end{equation}
and the tensors are transverse and traceless,
\begin{equation}
\partial_i \hat{T}_{ij} = 0 \, ,
\qquad\quad
\partial_i \hat{\Pi}_{ij} = 0 \ ,
\qquad\quad
\hat{T}_{ii} = 0 \, ,
\qquad\quad
\hat{\Pi}_{ii}= 0  \, .
\end{equation}
As we show below, this decomposition renders both the canonical structure and the 
dynamics block-diagonal. The transverse-traceless tensor modes already carry the physical 
graviton degrees of freedom, while the scalar and vector modes are further constrained by 
the quantum conditions~(\ref{ConstraintCorrelators}).

\paragraph{Scalar sector.}
From the definitions above, we can express the four scalar variables and their canonical
momenta in terms of the original canonical variables,
\begin{align}
&
\hat{S}_1 = \hat{h}_{00} \, ,
&&
\hat{S}_2 = \partial_i \hat{h}_{0i} \, ,
&&
\hat{S}_3 = \frac{\mathbb{P}_{ij}^{\scr T} \hat{h}_{ij} }{D\!-\!2} \, ,
&&
\hat{S}_4 = \mathbb{P}_{ij}^{\scr L} \hat{h}_{ij} \, ,
\\
&
\hat{\Pi}_1 = \hat{\pi}_{00} \, ,
&&
\hat{\Pi}_2 = - \frac{\partial_i}{\nabla^2} \hat{\pi}_{0i} \, ,
&&
\hat{\Pi}_3 = \mathbb{P}_{ij}^{\scr T} \hat{\pi}_{ij} \, ,
&&
\hat{\Pi}_4 = \mathbb{P}_{ij}^{\scr L} \hat{\pi}_{ij} \, .
\end{align}
The nonvanishing canonical commutators in this basis are
\begin{equation}
\bigl[ \hat{S}_I(\eta,\vec{x}) , \hat{\Pi}_J(\eta,\vec{x}^{\,\prime}) \bigr]
	= 
	\delta_{IJ} \delta^{D-1}(\vec{x} \!-\! \vec{x}^{\,\prime})
	\, ,
\qquad \quad
I,J = 1,2,3,4 \, .
\end{equation}
This sector contains eight first-order equations of motion, one for each scalar variable
and its conjugate momentum:
\begin{align}
\partial_0 \hat{S}_1 
	={}&
	\frac{2 a^{2-D}}{D\!-\!2} \Bigl[ 
		- \bigl( D \!-\! 1 \!-\! 2 (D\!-\!2) \alpha \bigr) \hat{\Pi}_1
		+ \hat{\Pi}_3 + \hat{\Pi}_4 \Bigr]
	-
	(D\!-\!3) \mathcal{H} \hat{S}_1
	\, ,
\\
\partial_0 \hat{\Pi}_1 
	={}&
	- \frac{1}{2} \nabla^2 \hat{\Pi}_2
	+
	\mathcal{H} \Bigl[
	(D\!-\!3) \hat{\Pi}_1 + \hat{\Pi}_3 + \hat{\Pi}_4
	\Bigr]
	+
	\frac{D\!-\!2}{2} a^{D-2} \nabla^2 \hat{S}_3
	\, ,
\\
\partial_0 \hat{S}_2
	={}&
	\alpha a^{2-D} \nabla^2 \hat{\Pi}_2
	+
	\frac{1}{2} \nabla^2 \Bigl[ \hat{S}_1 - (D\!-\!2) \hat{S}_3 + \hat{S}_4 \Bigr]
	-
	(D\!-\!2) \mathcal{H} \hat{S}_2
	\, ,
\\
\partial_0 \hat{\Pi}_2
	={}&
	(D\!-\!2) \mathcal{H} \hat{\Pi}_2
	-
	2 \hat{\Pi}_4
	\, ,
\\
\partial_0 \hat{S}_3
	={}&
	\frac{2 a^{2-D} }{ D\!-\!2 } \bigl( \hat{\Pi}_1 - \hat{\Pi}_4 \bigr)
	-
	\mathcal{H} \hat{S}_1
	\, ,
\\
\partial_0 \hat{\Pi}_3
	={}&
	\frac{D\!-\!2}{2} 
	\nabla^2 \Bigl[
	\hat{\Pi}_2
	+
	a^{D-2}
	\bigl(
	\hat{S}_1
	-
	(D \!-\! 3) \hat{S}_3
	\bigr)
	\Bigr]
	\, ,
\\
\partial_0 \hat{S}_4
	={}&
	\frac{ 2 a^{2-D} }{ D \!-\! 2 } \Bigl[
	 	\hat{\Pi}_1 - \hat{\Pi}_3 + (D\!-\!3) \hat{\Pi}_4
		\Bigr]
	+
	2 \hat{S}_2
	-
	\mathcal{H} \hat{S}_1
	\, ,
\\
\partial_0 \hat{\Pi}_4
	={}&
	- \frac{1}{2} \nabla^2 \hat{\Pi}_2
	\, .
\end{align}

\paragraph{Vector sector.}
The two transverse vectors and their conjugate momenta, expressed in terms of the
original canonical variables, are
\begin{equation}
\hat{V}_i^1 = \mathbb{P}_{ij}^{\scr T} \hat{h}_{0j} \, ,
\qquad\quad
\hat{\Pi}_i^1 = \mathbb{P}_{ij}^{\scr T} \hat{\pi}_{0j} \, ,
\qquad\quad
\hat{V}_i^2 = \mathbb{P}_{ij}^{\scr T} \partial_k \hat{h}_{jk} \, ,
\qquad\quad
\hat{\Pi}_i^2 = - 2 \mathbb{P}_{ij}^{\scr T} \frac{\partial_k}{\nabla^2} \hat{\pi}_{jk} \, ,
\end{equation}
so that the nonvanishing commutators in this sector read
\begin{equation}
\bigl[ \hat{V}_i^I(\eta, \vec{x}) , \hat{\Pi}_j^J(\eta, \vec{x}^{\,\prime} ) \bigr]
	=
	\delta_{IJ} \mathbb{P}_{ij}^{\scr T} \delta^{D-1}(\vec{x} \!-\! \vec{x}^{\,\prime} )
	\, ,
\qquad \quad
I,J=1,2 \, .
\end{equation}
The four first-order equations of motion in this sector are:
\begin{align}
\partial_0 \hat{V}^1_i 
	={}&
	- \alpha a^{2-D} \hat{\Pi}^1_i
	+
	\hat{V}^2_{i}
	-
	(D\!-\!2) \mathcal{H} \hat{V}^1_i 
	\, ,
\\
\partial_0 \hat{\Pi}^1_i 
	={}&
	-
	\nabla^2 \hat{\Pi}^2_{i}
	+
	(D\!-\!2) \mathcal{H} \hat{\Pi}^1_i 
	\, ,
\\
\partial_0 \hat{V}^2_{i}
	={}&
	- a^{2-D} \nabla^2 \hat{\Pi}^2_{i}
	+
	\nabla^2 \hat{V}_i^1
	\, ,
\\
\partial_0 \hat{\Pi}^2_i
	={}&
	- \hat{\Pi}^1_i
	\, .
\end{align}

\paragraph{Tensor sector.}
Finally, the transverse-traceless tensor modes are expressed as
\begin{equation}
\hat{T}_{ij} =
	\biggl[ \mathbb{P}_{k(i}^{\scr T} \mathbb{P}_{j)l}^{\scr T}
		- \frac{ \mathbb{P}_{ij}^{\scr T} \mathbb{P}_{kl}^{\scr T} }{ D \!-\! 2 } \biggr]
		\hat{h}_{kl} 
		\, ,
\qquad\quad
\hat{\Pi}_{ij} =
	\biggl[ \mathbb{P}_{k(i}^{\scr T} \mathbb{P}_{j)l}^{\scr T} 
		- \frac{ \mathbb{P}_{ij}^{\scr T} \mathbb{P}_{kl}^{\scr T} }{ D \!-\! 2 } \biggr]
		\hat{\pi}_{kl} 
		\, .
\end{equation}
and they satisfy the following non-vanishing commutation relation
\begin{equation}
\bigl[ \hat{T}_{ij}(\eta, \vec{x}) , \hat{\Pi}_{kl}(\eta, \vec{x}^{\,\prime}) \bigr]
	=
	\biggl[ \mathbb{P}_{i(k}^{\scr T} \mathbb{P}_{l)j}^{\scr T}
		- \frac{ \mathbb{P}_{ij}^{\scr T} \mathbb{P}_{kl}^{\scr T} }{ D \!-\! 2 } \biggr]
		\delta^{D-1}(\vec{x} \!-\! \vec{x}^{\,\prime} )
		\, .
\end{equation}
In this sector the equations of motion take the form
\begin{equation}
\partial_0 \hat{T}_{ij} =
	2 a^{2-D} \hat{\Pi}_{ij}
	\, ,
\qquad \quad
\partial_0 \hat{\Pi}_{ij} =
	\frac{ a^{D-2} }{2}
	\nabla^2 \hat{T}_{ij}
	\, .
\end{equation}

\paragraph{Constraints.}
The scalar–vector–tensor decomposition in~(\ref{SVTdecomposition}) induces an analogous
decomposition of the Hermitian constraint operators,
\begin{equation}
\hat{\Phi}_0 = \hat{K}_1 \, ,
\qquad\quad
\hat{\Phi}_i = - \partial_i \hat{K}_2 + \hat{K}_i^1 \, ,
\qquad\quad
\hat{\Psi}_0 = \hat{K}_3 \, ,
\qquad\quad
\hat{\Psi}_i = 2\partial_i \hat{K}_4 - \nabla^2 \hat{K}_i^2 \, ,
\end{equation}
so that only scalar and vector components appear. The scalar components are
\begin{equation}
\hat{K}_1 = \hat{\Pi}_1 \, ,
\qquad
\hat{K}_2 = \hat{\Pi}_2 \, ,
\qquad
\hat{K}_3 = \frac{ D\!-\!2 }{2} a^{D-2} \nabla^2 \hat{S}_3
	+ \mathcal{H} \bigl( \hat{\Pi}_3 + \hat{\Pi}_4 \bigr)
	\, ,
\qquad
\hat{K}_4 = \hat{\Pi}_4 \, ,
\end{equation}
and the vector components are
\begin{equation}
\hat{K}_i^1 = \hat{\Pi}_i^1 \, ,
\qquad \quad
\hat{K}_i^2 = \hat{\Pi}_i^2 \, .
\end{equation}
The equations of motion for these combinations split into the scalar sector,
\begin{align}
\partial_0 \hat{K}_1
	={}&
	( D\!-\!3 ) \mathcal{H} \hat{K}_1
	- \frac{1}{2} \nabla^2 \hat{K}_2
	+ \hat{K}_3
	\, ,
\\
\partial_0 \hat{K}_2
	={}&
	(D\!-\!2) \mathcal{H} \hat{K}_2
	- 2 \hat{K}_4
	\, ,
\\
\partial_0 \hat{K}_3
	={}&
	\nabla^2 \hat{K}_1
	+ \frac{1}{2} (D\!-\!3) \mathcal{H} \nabla^2 \hat{K}_2
	+ \mathcal{H} \hat{K}_3
	- \nabla^2 \hat{K}_4
	\, ,
\\
\partial_0 \hat{K}_4
	={}&
	- \frac{1}{2} \nabla^2 \hat{K}_2
	\, ,
\end{align}
and the vector sector,
\begin{equation}
\partial_0 \hat{K}_i^1
	=
	(D\!-\!2) \mathcal{H} \hat{K}_i^1
	- \nabla^2 \hat{K}_i^2
	\, ,
\qquad\quad
\partial_0 \hat{K}_i^2
	=
	- \hat{K}_i^1
	\, .
\end{equation}
These relations will be used to track the time evolution of the constraint operators
once we move to Fourier space and impose the quantum conditions~(\ref{ConstraintCorrelators})
on a mode-by-mode basis.

\subsection{Fourier space}
\label{subsec: Fourier space}

Because the background is spatially homogeneous and isotropic, it is natural to work in
Fourier space. Each comoving wavevector~$\vec{k}$ then evolves independently, and both the
canonical structure and the constraint conditions can be implemented mode by mode in the
tensor, vector, and scalar sectors.

\paragraph{Tensor sector.}
In the tensor sector we expand the transverse–traceless operators in a basis of
polarization tensors for each comoving wavevector~$\vec{k}$. The two field operators
are written as
\begin{align}
\hat{T}_{ij} (\eta,\vec{x}) ={}&
	\frac{ \sqrt{2} }{a^{\frac{D-2}{2}}} 
	 \!
	\int\! \frac{ d^{D-1}k }{ (2\pi)^{\frac{D-1}{2}} } \,
	e^{i \vec{k} \cdot \vec{x} } \!
	\sum_{\sigma=1}^{ \frac{D(D-3)}{2} } \!\!
	\varepsilon_{ij} (\sigma,\vec{k})
	\hat{\mathcal{T}}_{\sigma}(\eta,\vec{k})
	\, ,
\\
\hat{\Pi}_{ij} (\eta,\vec{x}) ={}&
	\frac{ a^{\frac{D-2}{2}} }{\sqrt{2}} \!
	\int\! \frac{ d^{D-1}k }{ (2\pi)^{\frac{D-1}{2}} } \,
	e^{i \vec{k} \cdot \vec{x} } \!
	\sum_{\sigma=1}^{ \frac{D(D-3)}{2} } \!\!
	\varepsilon_{ij} (\sigma,\vec{k})
	\hat{\mathcal{P}}_{\sigma}(\eta,\vec{k})
	\, ,
\end{align}
where the symmetric transverse–traceless polarization tensors satisfy
\begin{subequations}
\begin{align}
&
\varepsilon_{ii}(\sigma,\vec{k}) = 0 \, ,
\qquad\quad
k_i \, \varepsilon_{ij}(\sigma,\vec{k}) = 0 \, ,
\qquad\quad
\bigl[ \varepsilon_{ij}(\sigma,\vec{k}) \bigr]^*
	=
	\varepsilon_{ij}(\sigma,- \vec{k}) \, ,
\\
&
\varepsilon_{ij}^*(\sigma,\vec{k}) \varepsilon_{ij}(\sigma',\vec{k})
	=
	\delta_{\sigma\sigma'}
	\, ,
\qquad
\sum_{\sigma=1}^{ \frac{D(D-3)}{2} } 
	\varepsilon^*_{ij}(\sigma,\vec{k}) \varepsilon_{kl}(\sigma,\vec{k})
	=
	{\rm P}_{i(k}^{\scr T} (\vec{k}) {\rm P}^{\scr T}_{l)j} (\vec{k})
		- \frac{ {\rm P}^{\scr T}_{ij} (\vec{k}) {\rm P}^{\scr T}_{kl} (\vec{k}) }{D \!-\! 2}
	\, ,
\end{align}
\end{subequations}
where~${\rm P}^{\scr T}_{ij}(\vec{k}) \!=\! \delta_{ij} \!-\! k_i k_j/k^2$ is the momentum-space
transverse projector. The momentum-space operators are Hermitian in the sense that
$\hat{\mathcal{T}}_\sigma^\dag(\eta,\vec{k}) \!=\! \hat{\mathcal{T}}_\sigma(\eta,-\vec{k})$
and $\hat{\mathcal{P}}_\sigma^\dag(\eta,\vec{k}) \!=\! \hat{\mathcal{P}}_\sigma(\eta,-\vec{k})$,
and satisfy the following non-vanishing commutation relation
\begin{equation}
\bigl[ \hat{\mathcal{T}}_\sigma(\eta,\vec{k}) ,
	\hat{\mathcal{P}}_{\sigma'}(\eta,\vec{k}^{\,\prime})  \bigr]
	=
	\delta_{\sigma\sigma'} \,
	i \delta^{D-1}( \vec{k} \!+\! \vec{k}^{\,\prime} ) 
	\, .
\label{TensorMomentumCommutators}
\end{equation}
The momentum–space equations of motion are
\begin{equation}
\partial_0 \hat{\mathcal{T}}_\sigma
	=
	\hat{\mathcal{P}}_\sigma
	+
	\frac{1}{2} (D\!-\!2) \mathcal{H} \hat{\mathcal{T}}_\sigma
	\, ,
\qquad \quad
\partial_0 \hat{\mathcal{P}}_\sigma
	=
	-
	k^2 \hat{\mathcal{T}}_\sigma
	-
	\frac{1}{2} (D\!-\!2) \mathcal{H} \hat{\mathcal{P}}_\sigma
	\, .
\label{TensorMomentumEOM}
\end{equation}
Thus each tensor polarization behaves as an oscillator with a time-dependent mass;
this sector encodes the physical graviton degrees of freedom.


\paragraph{Vector sector.} 
In the vector sector, the two transverse modes for each comoving wavevector~$\vec{k}$ are expanded in a basis of polarization vectors, yielding two canonical pairs per polarization. The corresponding operator expansions are
\begin{align}
\hat{V}^I_i (\eta,\vec{x}) ={}&
	a^{- \frac{D-2}{2} } \! \!
	\int\! \frac{ d^{D-1}k }{ (2\pi)^{\frac{D-1}{2}} } \,
	e^{i \vec{k} \cdot \vec{x} } \,
	g_I \!\times\!
	\sum_{\sigma=1}^{D-2}
	\varepsilon_i(\sigma,\vec{k})
	\hat{\mathcal{V}}_{I, \sigma}(\eta,\vec{k})
	\, ,
\label{Vfourier}
\\
\hat{\Pi}^I_i (\eta,\vec{x}) ={}&
	a^{ \frac{D-2}{2} } \! \!
	\int\! \frac{ d^{D-1}k }{ (2\pi)^{\frac{D-1}{2}} } \,
	e^{i \vec{k} \cdot \vec{x} } \,
	g_I^{-1} \!\times\!
	\sum_{\sigma=1}^{D-2}
	\varepsilon_i(\sigma,\vec{k})
	\hat{\mathcal{P} }_{I, \sigma}(\eta,\vec{k})
	\, ,
\end{align}
where $I\!=\!1,2$, with coefficients $g_1\!=\!1$ and $g_2\!=\!k$ chosen so that all Fourier amplitudes carry the same dimension. The transverse polarization vectors satisfy
\begin{subequations}
\begin{align}
&
k_i \, \varepsilon_i (\sigma, \vec{k}) = 0 \, ,
&&
\varepsilon_i^*(\sigma,\vec{k}) = \varepsilon_i(\sigma,-\vec{k}) \, ,
\\
&
\varepsilon_i^* (\sigma, \vec{k}) \, \varepsilon_i(\sigma', \vec{k})
	=\delta_{\sigma\sigma'}
\, ,
&&
\sum_{\sigma=1}^{D-2} \varepsilon_i^*(\sigma,\vec{k}) \, \varepsilon_j(\sigma,\vec{k})
	=
	{\rm P}^{\scr T}_{ij} (\vec{k})
	\, .
\label{VectorPolarization}
\end{align}
\end{subequations}
Hermiticity implies 
$\hat{\mathcal{V}}_{I, \sigma}^\dag(\eta,\vec{k}) 
	\!=\! \hat{\mathcal{V}}_{I, \sigma}(\eta,-\vec{k})$ 
and 
$\hat{\mathcal{P}}_{I, \sigma}^\dag(\eta,\vec{k}) 
	\!=\! \hat{\mathcal{P}}_{I, \sigma}(\eta,-\vec{k})$.
The momentum-space commutation relations are
\begin{equation}
\bigl[ \hat{\mathcal{V}}_{I,\sigma}(\eta,\vec{k}) ,
	\hat{\mathcal{P}}_{J,\sigma'}(\eta, \vec{k}^{\,\prime}) \bigr]
	=
	\delta_{IJ} \delta_{\sigma\sigma'}
	\,
	i \delta^{D-1}(\vec{k} \!+\! \vec{k}^{\,\prime})
	\, .
\end{equation}
The equations of motion in Fourier space take the form
\begin{align}
\partial_0 \hat{\mathcal{V}}_{1,\sigma}
	={}&
	-
	\alpha \hat{\mathcal{P}}_{1,\sigma}
	+
	k \hat{\mathcal{V}}_{2,\sigma}
	-
	\frac{1}{2} (D\!-\!2) \mathcal{H} \hat{\mathcal{V}}_{1,\sigma}
	\, ,
\label{VectorModeEq1}
\\
\partial_0 \hat{\mathcal{P}}_{1,\sigma}
	={}&
	k \hat{\mathcal{P}}_{2,\sigma}
	+
	\frac{1}{2} (D\!-\!2) \mathcal{H} \hat{\mathcal{P}}_{1,\sigma}
	\, ,
\label{VectorModeEq2}
\\
\partial_0 \hat{\mathcal{V}}_{2,\sigma}
	={}&
	\hat{\mathcal{P}}_{2,\sigma}
	-
	k \hat{\mathcal{V}}_{1,\sigma}
	+
	\frac{1}{2} (D\!-\!2) \mathcal{H} \hat{\mathcal{V}}_{2,\sigma}
	\, ,
\label{VectorModeEq3}
\\
\partial_0 \hat{\mathcal{P}}_{2,\sigma}
	={}&
	- k \hat{\mathcal{P}}_{1,\sigma}
	-
	\frac{1}{2} (D\!-\!2) \mathcal{H} \hat{\mathcal{P}}_{2,\sigma}
	\, .
\label{VectorModeEq4}
\end{align}
These relations describe the coupled evolution of the two transverse vector modes for each~$\vec{k}$, whose contributions to physical correlators are further restricted by the vector constraints.

For the constraints in the vector sector we expand them in Fourier space as
\begin{equation}
\hat{K}^I_i (\eta,\vec{x}) =
	a^{\frac{D-2}{2} } \!
	\int\! \frac{ d^{D-1}k }{ (2\pi)^{\frac{D-1}{2}} } \,
	e^{i \vec{k} \cdot \vec{x} } \,
	g_I^{-1} \!\times\!
	\sum_{\sigma=1}^{D-2}
	\varepsilon_i(\sigma,\vec{k})
	\hat{\mathcal{K}}_{I, \sigma}(\eta,\vec{k})
	\, ,
\end{equation}
such that the equations of motion read
\begin{equation}
\partial_0 \hat{\mathcal{K}}_{1,\sigma}
	=
	k \hat{\mathcal{K}}_{2,\sigma}
	+ \frac{1}{2} (D\!-\!2) \mathcal{H} \hat{\mathcal{K}}_{1,\sigma}
	\, ,
\qquad\quad
\partial_0 \hat{\mathcal{K}}_{2,\sigma}
	=
	- k \hat{\mathcal{K}}_{1,\sigma}
	-
	\frac{1}{2} (D\!-\!2) \mathcal{H} \hat{\mathcal{K}}_{2,\sigma}
	\, .
\label{VecConstraintEqs}
\end{equation}
These match Eqs.~(\ref{VectorModeEq2}) and~(\ref{VectorModeEq4}), which is
expected given that by definition we 
have~$\hat{\mathcal{K}}_{I,\sigma} \!=\! \mathcal{\hat{P}}_{I,\sigma}$.


\paragraph{Scalar sector.}
In the scalar sector, the four canonical scalar pairs are expanded in Fourier modes as
\begin{align}
\hat{S}_I(\eta,\vec{x}) ={}&
	a^{-\frac{D-2}{2}} \! \!
	\int\! \frac{d^{D-1}k }{ (2\pi)^{\frac{D-1}{2}} } \, e^{i \vec{k} \cdot \vec{x} } \,
	f_I \!\times\!
	\hat{\mathcal{S}}_I(\eta,\vec{k})
	\, ,
\\
\hat{\Pi}_I(\eta,\vec{x}) ={}&
	a^{\frac{D-2}{2}} \! \!
	\int\! \frac{d^{D-1}k }{ (2\pi)^{\frac{D-1}{2}} } \, e^{i \vec{k} \cdot \vec{x} } \,
	f_I^{-1} \!\times\!
	\hat{\mathcal{P}}_I(\eta,\vec{k})
	\, ,
\end{align}
where~$I\!=\!1,2,3,4$, and where~$f_1\!=\!f_3\!=\!f_4\!=\!1$ and~$f_2\!=\!k$, chosen so that 
the momentum space operators would have the same dimension. Hermiticity of the 
momentum-space operators implies
$\hat{\mathcal{S}}_I^\dag(\eta,\vec{k}) \!=\! \hat{\mathcal{S}}_I(\eta,-\vec{k})$ and
$\hat{\mathcal{P}}_I^\dag(\eta,\vec{k}) \!=\! \hat{\mathcal{P}}_I(\eta,-\vec{k})$.
The non-vanishing commutators in momentum space are
\begin{equation}
\bigl[ \hat{\mathcal{S}}_I(\eta,\vec{k}) , \hat{\mathcal{P}}_J(\eta,\vec{k}^{\,\prime}) \bigr]
	=
	\delta_{IJ} \, i \delta^{D-1}( \vec{k} \!+\! \vec{k}^{\,\prime} )
	\, ,
\end{equation}
while the equations of motion are:
\begin{align}
\partial_0 \hat{\mathcal{S}}_1 
	={}&
	\frac{2 }{D\!-\!2} \Bigl[ 
		- \bigl( D \!-\! 1 \!-\! 2 (D\!-\!2) \alpha \bigr) \hat{\mathcal{P}}_1
		+ \hat{\mathcal{P}}_3 + \hat{\mathcal{P}}_4 \Bigr]
	-
	\frac{D \!-\! 4}{2} \mathcal{H} \hat{\mathcal{S}}_1
	\, ,
\label{ScalarEOM1}
\\
\partial_0 \hat{\mathcal{P}}_1 
	={}&
	\frac{k}{2} \hat{\mathcal{P}}_2
	+
	\mathcal{H} \Bigl[
	\frac{D\!-\!4}{2} \hat{\mathcal{P}}_1 
	+ \hat{\mathcal{P}}_3 + \hat{\mathcal{P}}_4
	\Bigr]
	-
	\frac{D\!-\!2}{2} k^2 \hat{\mathcal{S}}_3
	\, ,
\\
\partial_0 \hat{\mathcal{S}}_2
	={}&
	-
	\alpha \hat{\mathcal{P}}_2
	-
	\frac{k}{2} \Bigl[ \hat{\mathcal{S}}_1 
		- (D\!-\!2) \hat{\mathcal{S}}_3
		+ \hat{\mathcal{S}}_4 \Bigr]
	-
	\frac{D\!-\!2}{2} \mathcal{H} \hat{\mathcal{S}}_2
	\, ,
\\
\partial_0 \hat{\mathcal{P}}_2
	={}&
	-
	2 k \hat{\mathcal{P}}_4
	+
	\frac{D\!-\!2}{2} \mathcal{H} \hat{\mathcal{P}}_2
	\, ,
\\
\partial_0 \hat{\mathcal{S}}_3
	={}&
	\frac{2 }{ D\!-\!2 } \bigl( \hat{\mathcal{P}}_1 - \hat{\mathcal{P}}_4 \bigr)
	-
	\mathcal{H} \hat{\mathcal{S}}_1
	+
	\frac{D\!-\!2}{2} \mathcal{H} \hat{\mathcal{S}}_3
	\, ,
\\
\partial_0 \hat{\mathcal{P}}_3
	={}&
	-
	\frac{D\!-\!2}{2} 
	\Bigl[
	k \hat{\mathcal{P}}_2
	+
	\mathcal{H} \hat{\mathcal{P}}_3
	+
	k^2 \bigl(
	\hat{\mathcal{S}}_1
	-
	(D \!-\! 3) \hat{\mathcal{S}}_3
	\bigr)
	\Bigr]
	\, ,
\\
\partial_0 \hat{\mathcal{S}}_4
	={}&
	\frac{ 2 }{ D \!-\! 2 } \Bigl[
	 	\hat{\mathcal{P}}_1 - \hat{\mathcal{P}}_3 + (D\!-\!3)\hat{\mathcal{P}}_4
		\Bigr]
	+
	2 k \hat{\mathcal{S}}_2
	-
	\mathcal{H} \hat{\mathcal{S}}_1
	+
	\frac{D\!-\!2}{2} \mathcal{H} \hat{\mathcal{S}}_4
	\, ,
\\
\partial_0 \hat{\mathcal{P}}_4
	={}&
	\frac{k}{2} \hat{\mathcal{P}}_2
	-
	\frac{D\!-\!2}{2} \mathcal{H} \hat{\mathcal{P}}_4
	\, .
\label{ScalarEOM8}
\end{align}
For each~$\vec{k}$, the scalar sector forms a closed system of coupled first-order equations whose physical content will be determined by the scalar constraints.

We next introduce the Fourier transforms of the scalar constraint operators,
\begin{equation}
\hat{K}_{I}(\eta,\vec{x})
	=
	a^{\frac{D-2}{2}} \!\!
	\int\! \frac{d^{D-1}k }{ (2\pi)^{\frac{D-1}{2}} } \, e^{i \vec{k} \cdot \vec{x} } \,
	\ell_I \!\times\!
	\hat{\mathcal{K}}_{I}(\eta,\vec{k})
	\, ,
\qquad\quad
I = 1,2,3,4 \, ,
\label{ScalarConstraintFourier}
\end{equation}
where~$\ell_1 \!=\! \ell_4 \!=\! 1$, and $\ell_3 \!=\! 1/\ell_2 \!=\! k$.
In terms of the canonical momenta and scalar fields the constraint modes are
\begin{equation}
\hat{\mathcal{K}}_1 = \hat{\mathcal{P}}_1 \, ,
\qquad\quad
\hat{\mathcal{K}}_2 = \hat{\mathcal{P}}_2 \, ,
\qquad\quad
\hat{\mathcal{K}}_3 = 
	- \frac{ D\!-\!2 }{2} k \hat{\mathcal{S}}_3
	+ \frac{\mathcal{H}}{k} \bigl( \hat{\mathcal{P}}_3 \!+\! \hat{\mathcal{P}}_4 \bigr)
	\, ,
\qquad\quad
\hat{\mathcal{K}}_4 = \hat{\mathcal{P}}_4 \, ,
\label{scalarKs}
\end{equation}
and the corresponding equations of motion follow directly from the scalar dynamics,
\begin{align}
\partial_0 \hat{\mathcal{K}}_1
	={}&
	\frac{k}{2}  \bigl( \hat{\mathcal{K}}_2 + 2 \hat{\mathcal{K}}_3 \bigr)
	+
	\frac{1}{2} (D\!-\!4) \mathcal{H} \hat{\mathcal{K}}_1
	\, ,
\label{scalarConstraintEOM1}
\\
\partial_0 \hat{\mathcal{K}}_2
	={}&
	- 2 k \hat{\mathcal{K}}_4
	+
	\frac{1}{2} (D\!-\!2) \mathcal{H} \hat{\mathcal{K}}_2
	\, ,
\label{scalarConstraintEOM2}
\\
\partial_0 \hat{\mathcal{K}}_3
	={}&
	- k\bigl ( \hat{\mathcal{K}}_1 - \hat{\mathcal{K}}_4 \bigr)
	- \frac{1}{2} (D\!-\!3) \mathcal{H} \hat{\mathcal{K}}_2
	- \frac{1}{2} (D \!-\! 4) \mathcal{H} \hat{\mathcal{K}}_3
	\, ,
\label{scalarConstraintEOM3}
\\
\partial_0 \hat{\mathcal{K}}_4
	={}&
	\frac{k}{2} \hat{\mathcal{K}}_2
	- \frac{1}{2} (D \!-\! 2) \mathcal{H} \hat{\mathcal{K}}_4
	\, .
\label{scalarConstraintEOM4}
\end{align}

We have thus expressed all canonical variables and scalar constraint operators in terms of tensor, vector, and scalar mode operators in Fourier space, together with their linear evolution equations. In the subsequent sections we solve the mode equations for the physical tensor sector and the gauge-dependent scalar and vector sectors, impose the constraint conditions~(\ref{ConstraintCorrelators}) in momentum space, and construct the corresponding graviton two-point function. This is facilitated by the results for the
scalar fields in de Sitter, which we recall next.

\section{Scalar mode functions and propagators}
\label{sec: Scalar mode functions and two-point functions}

Before turning to the dynamics of the tensor, vector, and scalar mode operators
introduced in the previous section, and before constructing the graviton two-point
function, it is useful to recall several standard results for scalar fields in de~Sitter
space. In this section we summarize the mode functions and two-point functions of
free scalars with various effective masses. These results will serve as building
blocks for the corresponding graviton expressions that follow.

\subsection{Scalar mode functions}
\label{subsec: Scalar mode functions}

Scalar mode functions appear throughout de Sitter–space calculations, and it is useful to recall their basic properties. They satisfy the differential equation
\begin{equation}
\biggl[ \partial_0^2 + k^2 + \Bigl( \frac{1}{4} \!-\! \lambda^2 \Bigr) \mathcal{H}^2 \biggr] 
	\mathscr{U}_\lambda = 0 \, ,
\label{ModeEq}
\end{equation}
where~$\lambda$ is a parameter associated with the effective scalar field mass as
\begin{equation}
M_\lambda^2 = \biggl[ \Bigl( \frac{D\!-\!1}{2} \Bigr)^{\!2} - \lambda^2 \biggr] H^2 \, .
\end{equation}
The general solution takes the form of a linear combination
\begin{equation}
\mathscr{U}_\lambda(\eta,\vec{k}) 
	=
	A(\vec{k}) U_\lambda(\eta,k)
	+
	B(\vec{k}) U_\lambda^*(\eta,k)
	\, ,
\label{ScalarMass}
\end{equation}
where the positive-frequency Chernikov–Tagirov–Bunch–Davies (CTBD) mode function~\cite{Chernikov:1968zm,Bunch:1978yq} is
\begin{equation}
U_\lambda(\eta,k)
	=
	e^{ \frac{i\pi}{4} (2\lambda+1) }
	e^{ \frac{-ik}{2H} }
	\sqrt{ \frac{\pi}{4\mathcal{H}} } \,
	H_\lambda^{\scr (1)} \Bigl( \frac{k}{\mathcal{H}} \Bigr)
	\, ,
\end{equation}
with $H_\lambda^{\scr (1)}(z)$ the Hankel function of the first kind.  
In the limit of vanishing Hubble parameter the CTBD mode reduces to the flat-space positive-frequency mode function according to
\begin{equation}
U_\lambda(\eta,k)
	\ \overset{H \to 0}{\longsim} \
	u(\eta,k)
	\biggl[
	1
	-
	\Bigl( \frac{1}{4} \!-\! \lambda^2 \Bigr)
	\frac{iH}{2k} 
	+
	\mathcal{O}(H^2)
	\biggr]
	\, ,
\label{Uflat}
\end{equation}
where $u(\eta,k) = e^{-ik\eta}/\sqrt{2k}$ is the standard flat-space positive-frequency mode.  
These scalar mode functions will play an important role in expressing the corresponding graviton solutions and two-point functions in subsequent sections.

Recurrence relations for Hankel functions~\cite{Olver:2010,Olver:web} imply
corresponding relations among CTBD mode functions, namely
\begin{equation}
\biggl[ \partial_0 + \Bigl( \frac{1}{2} \!+\! \lambda \Bigr) \mathcal{H} \biggr] U_\lambda
	=
	-ik U_{\lambda+1}
	\, ,
\qquad \quad
\biggl[ \partial_0 + \Bigl( \frac{1}{2} \!-\! \lambda \Bigr) \mathcal{H} \biggr] U_\lambda
	=
	-ik U_{\lambda-1}
	\, ,
\label{CTBDrecurrence}
\end{equation}
which provide a convenient way of shifting mode indices up or down.  
An immediate consequence is a compact expression for the Wronskian normalization,
\begin{equation}
{\rm Re} \Bigl[ U_\lambda(\eta,k) U_{\lambda+1}^*(\eta,k) \Bigr] = \frac{1}{2k} \, .
\end{equation}
The relations~(\ref{CTBDrecurrence}), together with the homogeneous mode
equation~(\ref{ModeEq}), also yield the useful identity for a sourced mode equation,
\begin{equation}
\biggl[ \partial_0^2 + k^2 + \Bigl( \frac{1}{4} \!-\! \lambda^2 \Bigr) \mathcal{H}^2 \biggr]
	\Bigl( \frac{U_{\lambda-1}}{\mathcal{H}} \Bigr)
	=
	2 i k U_\lambda
	\, .
\label{inhomID}
\end{equation}
Another useful identity relates bilinears of CTBD modes
evaluated at different times,
\begin{equation}
	\frac{ 2 \lambda }{k^2}
	\Bigl[
	\mathcal{H} \partial_0'
	+
	\mathcal{H}' \partial_0
	+
	\mathcal{H} \mathcal{H}'
	\Bigr]
	\Bigl[
	U_{\lambda}(\eta,k) U_{\lambda}^*(\eta',k)
	\Bigr]
	=
	U_{\lambda+1}(\eta,k) U_{\lambda+1}^*(\eta',k)
	-
	U_{\lambda-1}(\eta,k) U_{\lambda-1}^*(\eta',k)
	\, .
\label{DoubleModeIdentity}
\end{equation}
Taken together, these relations streamline later computations by reducing derivative
structures and allowing systematic index shifts in products of mode functions.

\subsection{Scalar propagator}
\label{subsec: Scalar propagator}

The two-point function in de Sitter space for a scalar field with an
effective mass~(\ref{ScalarMass}) satisfies
\begin{equation}
\Bigl( {\dalembertian} - M_\lambda^2 \Bigr)
	i \bigl[ \tensor*[^{\tt a \!}]{\Delta}{^{\tt \! b }} \bigr]_\lambda(x;x')
	=
	{\tt S}^{\tt ab} \frac{ i\delta^D(x\!-\!x') }{ \sqrt{-g} }
	\, ,
	\qquad \quad
	{\tt S}^{\tt ab} = {\rm diag}(1,-1) \, ,
\label{ScalarPropagatorEOM}
\end{equation}
where~${\dalembertian}\!\equiv\! g^{\mu\nu} \nabla_\mu \nabla_\nu$ is the 
covariant d'Alembertian operator, and
where ${\tt a,b=\pm}$ are the Schwinger--Keldysh polarity labels appropriate for nonequilibrium
quantum field theory (see, e.g.,~\cite{Berges:2004yj,NoneqLectures}).  
The positive-frequency Wightman function admits the mode-sum representation
\begin{equation}
i \bigl[ \tensor*[^{\scr \!-\!}]{\Delta}{^{\scr \!+\!}} \bigr]_{\lambda}(x;x')
        =
        (aa')^{- \frac{D-2}{2}} \!\!
        \int\! \frac{ d^{D-1} k }{ (2\pi)^{D-1} } \,
        e^{i \vec{k} \cdot (\vec{x} - \vec{x}^{\,\prime} ) } \,
        \mathscr{U}_\lambda(\eta,k) \mathscr{U}_\lambda^*(\eta',k)
        \, ,
\label{WightmanModeSum}
\end{equation}
where $\mathscr{U}_\lambda$ satisfies Eq.~(\ref{ModeEq}), and where henceforth
$a \equiv a(\eta)$ and $a' \equiv a(\eta')$, with analogous conventions for all primed and
unprimed quantities inside two-point functions.
Implicit in~(\ref{WightmanModeSum}) is the usual $i\varepsilon$ prescription,
implemented as $\eta \to \eta - i\varepsilon/2$ and 
$\eta' \to \eta' + i\varepsilon/2$, that renders the result an analytic function whose distributional limit defines
the two-point function.

The complex conjugate of~(\ref{WightmanModeSum}) is the negative-frequency Wightman function,~$i \bigl[ \tensor*[^{\scr \!+\!}]{\Delta}{^{\scr \!-\! }} \bigr]_{\lambda}(x;x')
        \!=\!
        \bigl\{ i \bigl[ \tensor*[^{\scr \!-\!}]{\Delta}{^{\scr \!+\!}} \bigr]_{\lambda}(x;x') \bigr\}^* $.
All remaining Schwinger--Keldysh two-point functions follow from the Wightman ones.
The Feynman propagator, i.e.\ the time-ordered two-point function, is
\begin{equation}
i \bigl[ \tensor*[^{\scr \!+\!}]{\Delta}{^{\scr \! +\! }} \bigr]_{\lambda}(x;x')
        =
        \theta(\eta\!-\!\eta') \,
        i \bigl[ \tensor*[^{\scr \!-\!}]{\Delta}{^{\scr \!+\!}} \bigr]_{\lambda}(x;x')
        +
        \theta(\eta'\!-\!\eta) \,
        i \bigl[ \tensor*[^{\scr \!+\!}]{\Delta}{^{\scr \!-\!}} \bigr]_{\lambda}(x;x')
        \, ,
\end{equation}
while its complex conjugate,~$i \bigl[ \tensor*[^{\scr \!-\!}]{\Delta}{^{\scr \!-\!}} \bigr]_{\lambda}(x;x') \!=\!
        \bigl\{
        i \bigl[ \tensor*[^{\scr \!+\!}]{\Delta}{^{\scr \!+\!}} \bigr]_{\lambda}(x;x')
        \bigr\}^*$,
is the Dyson propagator.  
These scalar two-point functions, evaluated for various values of~$\lambda$, will
serve as essential building blocks in the construction of the graviton propagator.

The mode-sum representation in Eq.~(\ref{WightmanModeSum}) is finite for the CTBD choice of mode function when the effective mass-squared parameter~$M_\lambda^2$ is positive. For massless or tachyonic values of the effective mass-squared, however, the mode sum develops an infrared divergence~\cite{Allen:1987tz}. In such cases the Bogolyubov coefficients appearing in the general solution of Eq.~(\ref{ScalarMass}) must be adjusted so that the resulting mode sum is infrared convergent. 
The precise implementation of this procedure is not essential for our purposes; to leading order it can be modeled by introducing an infrared cutoff~$k_0 \!\ll\! H_0$, which regulates the contribution of super-horizon modes.

With this prescription, the scalar two-point functions in de Sitter space take the following form:
\begin{equation}
i \bigl[ \tensor*[^{\tt a \!}]{\Delta}{^{\tt \! b}} \bigr]_{\lambda}(x;x')
	= \mathcal{F}_{\lambda} \bigl( y_{\tt ab} \bigr) 
	+ \Delta \mathcal{F}_{\lambda} \bigl( y_{\tt ab},u,v \bigr) \, ,
\label{ScalarPropagator}
\end{equation}
The de Sitter invariance of the first contribution is manifested through its dependence on the
de Sitter-invariant length function
\begin{equation}
y_{\scr -+} =
	H^2 aa' \Bigl[ \| \vec{x} \!-\! \vec{x}^{\,\prime}  \|^2 
		- \bigl( \eta\!-\!\eta' \!-\! i \varepsilon \bigr)^{\!2} \Bigr]
	\, ,
\qquad
y_{\scr ++} =
	H^2 aa' \Bigl[ \| \vec{x} \!-\! \vec{x}^{\,\prime}  \|^2 
		- \bigl( | \eta\!-\!\eta' | \!-\! i \varepsilon \bigr)^{\!2} \Bigr]
	\, ,
\end{equation}
with~$y_{\scr +-} \!=\! y_{\scr -+}^*$ and~$y_{\scr --} \!=\! y_{\scr ++}^*$.
In contrast, the de Sitter–breaking part is characterized by its dependence on
\begin{equation}
u = \ln(aa') \, ,
\qquad \quad
v = \ln(a/a') \, .
\end{equation}
Henceforth, we omit explicitly indicating the Schwinger--Keldysh indices on~$y$
that determine its~$i\varepsilon$-prescription, as they will be clear from context.

The de Sitter invariant part of~(\ref{ScalarPropagator}) is expressed in terms of a hypergeometric function~\cite{Onemli:2002hr},
\begin{equation}
\mathcal{F}_{\lambda}(y) = \frac{ H^{D-2} }{ (4\pi)^{\frac{D}{2}} }
	\frac{ \Gamma\bigl( \frac{D-1}{2} \!+\! \lambda \bigr) \, 
		\Gamma\bigl( \frac{D-1}{2} \!-\! \lambda \bigr) }
			{ \Gamma\bigl( \frac{D}{2} \bigr) } \,
		{}_2F_1 \biggl( \! \Bigl\{ \frac{D\!-\!1}{2} \!+\! \lambda , 
				\frac{D\!-\!1}{2} \!-\! \lambda \Bigr\} ,
			\Bigl\{ \frac{D}{2} \Bigr\} , 1 \!-\! \frac{y}{4} \biggr)
			\, ,
\label{PropagatorFunction}
\end{equation}
which has a useful power-series representation,
\begin{align}
\MoveEqLeft[1]
\mathcal{F}_\lambda(y)
	=
	\frac{ H^{D-2} \, 
		\Gamma\bigl( \frac{D-2}{2} \bigr) }{ (4\pi)^{ \frac{D}{2} } }
	\biggl\{
	\Bigl( \frac{y}{4} \Bigr)^{ \! -\frac{D-2}{2}} 
	+ \frac{\Gamma\bigl( \frac{4-D}{2} \bigr) }
		{ \Gamma\bigl( \frac{1}{2} \!+\! \lambda \bigr) \, \Gamma\bigl( \frac{1}{2} \!-\! \lambda \bigr) } \times
\nonumber \\
&
	\times \! 
	\sum_{n=0}^{\infty} \biggl[
	\frac{ \Gamma\bigl( \frac{3}{2} \!+\! \lambda \!+\! n \bigr) \, \Gamma\bigl( \frac{3}{2} \!-\! \lambda \!+\! n \bigr) }
		{ \Gamma\bigl( \frac{6-D}{2} \!+\! n \bigr) \, (n\!+\!1)! }
		\Bigl( \frac{y}{4} \Bigr)^{\! n- \frac{D-4}{2} }
	-
	\frac{ \Gamma\bigl( \frac{D-1}{2} \!+\! \lambda \!+\! n \bigr) \, \Gamma\bigl( \frac{D-1}{2} \!-\! \lambda \!+\! n \bigr) }
		{ \Gamma\bigl( \frac{D}{2} \!+\! n \bigr) \, n! }
		\Bigl( \frac{y}{4} \Bigr)^{\! n }
	\biggr]
	\biggr\} 
	\, .
\label{Fseries}
\end{align}
The de Sitter–breaking contribution, which arises for tachyonic effective 
masses~\cite{Janssen:2008px} and also in the massless case~\cite{Onemli:2002hr}, 
is given by the multiple series
\begin{align}
\MoveEqLeft[9]
\Delta \mathcal{F}_{\lambda}(y,u,v)
	=
	\frac{H^{D-2}}{ (4\pi)^{\frac{D}{2}} }
	\frac{ \Gamma(\lambda) \, \Gamma(2\lambda) }
		{ \Gamma\bigl( \frac{D-1}{2} \bigr) \, \Gamma\bigl( \frac{1}{2} \!+\! \lambda \bigr) }
	\!\!
	\sum_{n=0}^{ \left\lfloor \lambda - \frac{D-1}{2} \right\rfloor} 
	\sum_{k=0}^{n}
	\sum_{\ell=0}^{n-k}
	c_{nk\ell}
\nonumber \\
&
	\times
		\frac{ e^{ ( \lambda - \frac{D-1}{2} - n ) u} }
			{ \bigl( \lambda \!-\! \frac{D-1}{2} \!-\! n \bigr) }
		\Bigl[ y \!+\! 4 \sh^2\Bigl( \frac{v}{2} \Bigr) \Bigr]^k
		\ch\bigl[ (n \!-\! k \!-\! 2\ell) v \bigr]
	\, ,
\end{align}
with coefficients
\begin{align}
c_{nk\ell}
	=
	\frac{ \bigl( - \frac{1}{4} \bigr)^{\!n} }{ k! \, \ell! \, (n \!-\! k \!-\! \ell)! }
	\frac{ \Gamma\bigl( \frac{D-1}{2} \bigr) \, \Gamma^2(1\!-\!\lambda) }
		{ \Gamma\bigl( \frac{D-1}{2} \!+\! k \bigr) \,
			\Gamma(\ell \!+\! 1 \!-\! \lambda) \,
			\Gamma(n \!-\! k \!-\! \ell \!+\! 1 \!-\! \lambda) }
	\, .
\end{align}

The recurrence relations~(\ref{CTBDrecurrence}) between mode functions imply
the following reflection identities for the scalar two-point functions:
\begin{subequations}
\begin{align}
&
\biggl[ \partial_0 \!+\! \Bigl( \frac{D \!-\! 1}{2} \!+\! \lambda \Bigr) \mathcal{H} \biggr]
	i \bigl[ \tensor*[^{\tt a\!}]{\Delta}{^{\tt\! b}} \bigr]_{\lambda}(x;x')
	=
	-
	\biggl[ \partial_0' \!+\! \Bigl( \frac{D \!-\! 3}{2} \!-\! \lambda \Bigr) \mathcal{H}' \biggr]
	i \bigl[ \tensor*[^{\tt a\!}]{\Delta}{^{\tt\! b}} \bigr]_{\lambda+1}(x;x')
	\, ,
\label{ReflectionIdentity1}
\\
&
\biggl[ \partial_0 \!+\! \Bigl( \frac{D \!-\! 1}{2} \!-\! \lambda \Bigr) \mathcal{H} \biggr]
	i \bigl[ \tensor*[^{\tt a\!}]{\Delta}{^{\tt\! b}} \bigr]_{\lambda}(x;x')
	=
	-
	\biggl[ \partial_0' \!+\! \Bigl( \frac{D \!-\! 3}{2} \!+\! \lambda \Bigr) \mathcal{H}' \biggr]
	i \bigl[ \tensor*[^{\tt a\!}]{\Delta}{^{\tt\! b}} \bigr]_{\lambda-1}(x;x')
	\, .
\label{ReflectionIdentity2}
\end{align}
\label{ReflectionIds}%
\end{subequations}
In addition, the mode-function identity~(\ref{DoubleModeIdentity}) possesses a useful
position-space analogue,
\begin{equation}
	\frac{ 2 \lambda }{\nabla^2}
	\Bigl[
	\mathcal{H} \partial_0'
	+
	\mathcal{H}' \partial_0
	+
	(D\!-\!1) \mathcal{H} \mathcal{H}'
	\Bigr]
	i \bigl[ \tensor*[^{\tt a\!}]{\Delta}{^{\tt\! b}} \bigr]_{\lambda} (x;x')
	=
	i \bigl[ \tensor*[^{\tt a\!}]{\Delta}{^{\tt\! b}} \bigr]_{\lambda-1} (x;x')
	-
	i \bigl[ \tensor*[^{\tt a\!}]{\Delta}{^{\tt\! b}} \bigr]_{\lambda+1} (x;x')
	\, ,
\label{LaplaceIdentity}
\end{equation}
which enables the evaluation of the inverse Laplace operator acting on a particular
combination of temporal derivatives of a scalar two-point function.

\section{Graviton field dynamics and state}
\label{sec: Dynamics and state}

In this section we solve the momentum–space equations of motion:
(\ref{TensorMomentumEOM}) for the tensor sector,
(\ref{VectorModeEq1})–(\ref{VectorModeEq4}) for the vector sector,
and (\ref{ScalarEOM1})–(\ref{ScalarEOM8}) for the scalar sector, together with the equations of motion for the constraint operators in the latter two sectors.
The solutions are expressed in terms of the scalar mode functions presented in the previous section, which provide the time–dependent structure common to all three sectors.
We also specify the quantum state for each sector, thereby completing the definition of the  framework necessary for computing the graviton two-point function in the subsequent section.

\subsection{Tensor sector}
\label{subsec: Tensor sector}

The two tensor equations of motion combine into a single homogeneous second–order
equation that determines the field,
\begin{align}
\biggl[
	\partial_0^2
	+
	k^2
	+
	\Bigl( \frac{1}{4} \!-\! \nu^2 \Bigr)
		\mathcal{H}^2
	\biggr]
	\hat{\mathcal{T}}_{\sigma}
	={}&
	0
	\, .
\label{SecondOrderTensor}
\end{align}
with
\begin{equation}
\nu = \frac{D \!-\! 1}{2} \, .
\end{equation}
The corresponding momentum field is then obtained from
\begin{equation}
\hat{\mathcal{P}}_{\sigma} 
	=
	\biggl[ \partial_0 + \Bigl( \frac{1}{2} \!-\!  \nu \Bigr) \mathcal{H}
	\biggr] \hat{\mathcal{T}}_{\sigma}
	\, .
\end{equation}
We can recognize the second–order equation~(\ref{SecondOrderTensor}) as the scalar 
mode equation~(\ref{ModeEq}), while the associated first–order equation is
essentially a recurrence relation~(\ref{CTBDrecurrence}). Therefore, the solutions for the
tensor sector field operators are
\begin{align}
\hat{\mathcal{T}}_\sigma(\eta,\vec{k})
	={}&
	U_\nu(\eta,k) \, \hat{b}_{\scr T}(\sigma,\vec{k})
	+
	U_\nu^*(\eta,k) \, \hat{b}_{\scr T}^\dag(\sigma,-\vec{k})
	\, ,
\\
\hat{\mathcal{P}}_\sigma(\eta,\vec{k})
	={}&
	-
	ik U_{\nu-1}(\eta,k) \, \hat{b}_{\scr T}(\sigma,\vec{k})
	+
	ik U_{\nu-1}^*(\eta,k) \, \hat{b}_{\scr T}^\dag(\sigma,-\vec{k})
	\, .
\end{align}

The time-independent operators in the solutions above are constants of integration. 
Their commutation relations follow from~(\ref{TensorMomentumCommutators}), and 
they obey the standard commutation relations for creation and annihilation operators,
\begin{equation}
\bigl[ \hat{b}_{\scr T}(\sigma,\vec{k}) , \hat{b}_{\scr T}^\dag(\sigma', \vec{k}^{\,\prime}) \bigr]
	=
	\delta_{\sigma\sigma'} \,
	\delta^{D-1}(\vec{k} \!-\! \vec{k}^{\,\prime}) \, .
\end{equation}
They therefore provide the natural operators with which to define the Fock space
of tensor-sector states, whose vacuum is the CTBD analogue specified by
\begin{equation}
\hat{b}_{\scr T}(\sigma, \vec{k}) \bigl| \Omega \bigr\rangle = 0 \, ,
\qquad \quad
\forall \vec{k}, \sigma \, .
\end{equation}
This is the tensor-sector state assumed in the computation of two-point functions.

\subsection{Vector sector}
\label{subsec: Vector sector}

In the vector sector the two constraint equations~(\ref{VecConstraintEqs}) decouple
from the dynamical ones, so it is convenient to solve them first. They combine into
a single homogeneous second–order equation for the first constraint operator,
\begin{equation}
\biggl[ \partial_0^2
	+ k^2
	+ \Bigl( \frac{1}{4} \!-\! \nu^2 \Bigr) \mathcal{H}^2
	\biggr] \hat{\mathcal{K}}_{1,\sigma}
	= 0
	\, ,
\end{equation}
and its solution determines the second constraint through
\begin{equation}
\hat{\mathcal{K}}_{2,\sigma}
	=
	\frac{1}{k}
	\biggl[
	\partial_0
	+
	\Bigl( \frac{1}{2} \!-\! \nu \Bigr) \mathcal{H}
	\biggr]
	\hat{\mathcal{K}}_{1,\sigma}
	\, .
\end{equation}
Using the results collected in Sec.~\ref{subsec: Scalar mode functions}, we can
immediately write down the corresponding mode expansions,
\begin{align}
\hat{\mathcal{K}}_{1, \sigma}(\eta,\vec{k})
	={}&
	k U_\nu(\eta,k) \hat{b}_{\scr P}(\sigma,\vec{k})
	+
	k U_\nu^*(\eta,k) \hat{b}_{\scr P}^\dag(\sigma,-\vec{k})
	\, ,
\\
\hat{\mathcal{K}}_{2, \sigma}(\eta,\vec{k})
	={}&
	- ik U_{\nu-1}(\eta,k) \hat{b}_{\scr P}(\sigma,\vec{k})
	+
	ik U_{\nu-1}^*(\eta,k) \hat{b}_{\scr P}^\dag(\sigma,-\vec{k})
	\, .
\end{align}
These solutions for cthe onstraints then immediately determine the two canonical momenta 
of the
vector sector,~$\hat{\mathcal{P}}_{1,\sigma} \!=\! \hat{\mathcal{K}}_{1,\sigma}$ and
$\hat{\mathcal{P}}_{2,\sigma} \!=\! \hat{\mathcal{K}}_{2,\sigma}$.

The remaining two equations~(\ref{VectorModeEq1}) and~(\ref{VectorModeEq3})
of the vector sector combine into a single second–order equation sourced by the solutions
for the constraints,
\begin{equation}
\biggl[ \partial_0^2 + k^2 + \Bigl( \frac{1}{4} \!-\! \nu^2 \Bigr) \mathcal{H}^2 \biggr] \hat{\mathcal{V}}_{2,\sigma}
	= 
	- (1\!-\!\alpha) k \hat{\mathcal{K}}_{1,\sigma}
	\, ,
\end{equation}
while the remaining field is determined from
\begin{equation}
\hat{\mathcal{V}}_{1,\sigma}
	=
	-
	\frac{1}{k}
	\biggl[
	\partial_0
	+
	\Bigl( \frac{1}{2} \!-\! \nu \Bigr) \mathcal{H}
	\biggr]
	\hat{\mathcal{V}}_{2,\sigma}
	+
	\frac{ \hat{\mathcal{K}}_{2,\sigma} }{k}
	\, .
\end{equation}
The corresponding solutions take the form
\begin{align}
&\!\!\!
\hat{\mathcal{V}}_{2,\sigma}(\eta, \vec{k} )
	=
	U_\nu(\eta,k) \hat{b}_{\scr H}(\sigma,\vec{k})
	+
	U_{\nu}^*(\eta,k) \hat{b}_{\scr H}^\dag(\sigma, - \vec{k})
	+
	v_2(\eta,k) \hat{b}_{\scr P}(\sigma,\vec{k})
	+
	v_2^*(\eta,k) \hat{b}_{\scr P}^\dag(\sigma,-\vec{k})
	\, ,
\label{V2sol}
\\
&\!\!\!
\hat{\mathcal{V}}_{1,\sigma}(\eta, \vec{k} )
	=
	i U_{\nu-1}(\eta,k) \hat{b}_{\scr H}(\sigma,\vec{k})
	-
	i U_{\nu-1}^*(\eta,k) \hat{b}_{\scr H}^\dag(\sigma, - \vec{k})
	+
	i v_1(\eta,k) \hat{b}_{\scr P}(\sigma,\vec{k})
	-
	i v_1^*(\eta,k) \hat{b}_{\scr P}^\dag(\sigma,-\vec{k})
	\, ,
\label{V1sol}
\end{align}
where the particular mode functions satisfy
\begin{align}
\biggl[ \partial_0^2 + k^2 + \Bigl( \frac{1}{4} \!-\! \nu^2 \Bigr) \mathcal{H}^2 \biggr] v_2
	={}& 
	- (1\!-\!\alpha) k^2 U_\nu
	\, ,
\\
v_1
	={}&
	\frac{i}{k}
	\biggl[
	\partial_0
	+
	\Bigl( \frac{1}{2} \!-\! \nu \Bigr) \mathcal{H}
	\biggr] v_2
	-
	U_{\nu-1}
	\, .
\end{align}

The particular mode functions are obtained by using the identity in~(\ref{inhomID})
together with the recurrence relations~(\ref{CTBDrecurrence}),
\begin{align}
v_2 ={}&
	\frac{(1\!-\!\alpha) i k }{2\mathcal{H}} 
		\biggl[
		U_{\nu-1}
		-
		\frac{\mathcal{H}}{H} U_\nu
		\biggr]
	+
	\frac{ 1 \!-\! (1\!-\!\alpha) \nu}{2} 
	U_\nu
	\, ,
\label{v1sol}
\\
v_1 ={}&
	\frac{(1\!-\!\alpha) ik }{2\mathcal{H}} 
	\biggl[
	U_{\nu}
	-
	\frac{\mathcal{H}}{H} U_{\nu-1}
	\biggr]
	-
	\frac{ 1\!-\!(1\!-\!\alpha) \nu}{2} 
	U_{\nu-1}
	\, .
\label{v2sol}
\end{align}
The homogeneous components of these solutions are fixed by imposing the
Wronskian-like normalization condition
\begin{equation}
{\rm Re} \Bigl( U_\nu^* v_1 + U_{\nu-1}^* v_2 \Bigr) = 0 \, ,
\end{equation}
and by requiring a regular flat-space limit inferred from~(\ref{Uflat}),
\begin{align}
&
v_2 \xrightarrow{H_0 \to 0}
	\frac{1}{4}
	\Bigl[
	(1\!+\!\alpha)
	- 2 (1\!-\!\alpha) i k (\eta \!-\! \eta_0)
	\Bigr]
	u(\eta,k)
	\, ,
\\
&
v_1 \xrightarrow{H_0 \to 0}
	\frac{1}{4}
	\Bigl[
	-
	(1\!+\!\alpha)
	- 2(1\!-\!\alpha) i k (\eta \!-\! \eta_0)
	\Bigr]
	u(\eta,k)
	\, .
\end{align}

Having computed the field operators of the vector sector, we can now infer the
commutation relations for the time-independent operators. The only nonvanishing
ones are
\begin{equation}
\bigl[ \hat{b}_{\scr H}(\sigma, \vec{k}) , 
	\hat{b}^\dag_{\scr P}(\sigma', \vec{k}^{\,\prime}) \bigr]
	=
\bigl[ \hat{b}_{\scr P}(\sigma, \vec{k}) , 
	\hat{b}^\dag_{\scr H}(\sigma', \vec{k}^{\,\prime}) \bigr]
	=
	\delta_{\sigma\sigma'} \,
	\delta^{D-1}(\vec{k} \!-\! \vec{k}^{\,\prime})
	\, ,
\label{VectorCommutators}
\end{equation}
which shows that these operators are not of canonical creation/annihilation type.
This does not pose a problem; on the contrary, their form is convenient for constructing
the representation of the state space~\cite{Glavan:2022pmk}.

Any physical state must be annihilated by a time-independent, non-Hermitian linear
combination of the vector-sector constraints. The choice of this linear combination is
not unique, and is instead dictated by convenience. A convenient choice is obtained by
appealing to the flat-space limit in which manifest Lorentz invariance requires the
state to be annihilated by the positive-frequency part of~$\hat{\mathcal{K}}_{1,\sigma}$.
Using the flat-space limit of the scalar mode functions~(\ref{Uflat}), this condition
is readily seen to be
\begin{equation}
\hat{b}_{\scr P}(\sigma, \vec{k}) \bigl| \Omega \bigr\rangle = 0 \, ,
\qquad \quad
\forall \vec{k}, \sigma \, .
\label{VectorSubsidiary}
\end{equation}
This is precisely the choice made in Gupta's quantization of the graviton
field~\cite{Gupta:1952zz}, based on the method originally developed for the
electromagnetic field~\cite{Gupta:1949rh,Bleuler:1950cy}, and it generalizes
naturally to de Sitter space.

The construction of the vector sector space of states is completed by defining the
vacuum-like state~$\bigl| \Omega \bigr\rangle$ to satisfy
\begin{equation}
\hat{b}_{\scr H}(\sigma, \vec{k}) \bigl| \Omega \bigr\rangle =  0 \, ,
\qquad \quad
\forall \vec{k}, \sigma \, .
\label{OtherVectorStateCondition}
\end{equation}
The remainder of the state space is then generated by acting with the two remaining
time-independent operators. This construction yields an indefinite inner-product
space; however, this is not a physical feature, since the physical states are
required to satisfy the subsidiary condition~(\ref{VectorSubsidiary}), which
selects a subspace with positive-definite inner product. In computing two-point
functions we will assume a state obeying both~(\ref{VectorSubsidiary}) and
(\ref{OtherVectorStateCondition}).

\subsection{Scalar sector}
\label{subsec: Scalar sector}

The four constraint equations~(\ref{scalarConstraintEOM1})--(\ref{scalarConstraintEOM4})
decouple from the remaining four equations of motion in the scalar sector. They combine
into two homogeneous second–order equations,
\begin{subequations}
\begin{align}
\biggl[ \partial_0^2 + k^2
	+
	\Bigl( \frac{1}{4} \!-\! (\nu\!-\!1)^2 \Bigr) \mathcal{H}^2
	\biggr]
	\hat{\mathcal{K}}_1
	={}&
	0
	\, ,
\\
\biggl[ \partial_0^2 + k^2
	+
	\Bigl( \frac{1}{4} \!-\! \nu^2 \Bigr) \mathcal{H}^2
	\biggr]
	\hat{\mathcal{K}}_2
	={}&
	0
	\, ,
\end{align}
\label{ScalarConstraintSecondOrder}%
\end{subequations}
which determine two of the fields, while the remaining two are obtained from
\begin{subequations}
\begin{align}
\hat{\mathcal{K}}_3
	={}&
	\frac{1}{k} \biggl[ \partial_0
	+ \Bigl( \frac{1}{2} \!-\! (\nu\!-\!1) \Bigr) \mathcal{H} \biggr] \hat{\mathcal{K}}_1
	-
	\frac{ \hat{\mathcal{K}}_2 }{2}
	\, ,
\\
\hat{\mathcal{K}}_4
	={}&
	- \frac{1}{2k} \biggl[ \partial_0 +
	\Bigl( \frac{1}{2} \!-\! \nu \Bigr) \mathcal{H} 
	\biggr] \hat{\mathcal{K}}_2
	\, .
\end{align}
\label{ScalarConstraintOther}%
\end{subequations}
The two second--order equations~(\ref{ScalarConstraintSecondOrder}) 
are in the form of the scalar mode 
equation~(\ref{ModeEq}), while Eqs.~(\ref{ScalarConstraintOther})
are in the form of recurrence relations~(\ref{CTBDrecurrence}),
so that we can immediately write the solutions:
\begin{align}
\hat{\mathcal{K}}_1(\eta,\vec{k})
	={}&
	i k U_{\nu-1}(\eta,k) \hat{c}_1(\vec{k})
	-
	i k U_{\nu-1}^*(\eta,k) \hat{c}_1^\dag(-\vec{k})
	\, ,
\label{K1solution}
\\
\hat{\mathcal{K}}_2(\eta,\vec{k})
	={}&
	k U_{\nu}(\eta,k) \hat{c}_2(\vec{k})
	+
	k U_{\nu}^*(\eta,k) \hat{c}_2^\dag(-\vec{k})
	\, ,
\label{K2solution}
\\
\hat{\mathcal{K}}_3(\eta,\vec{k})
	={}&
	k U_{\nu-2}(\eta,k) \hat{c}_1(\vec{k})
	+
	k U_{\nu-2}^*(\eta,k) \hat{c}_1^\dag(-\vec{k})
	-
	\frac{k}{2} U_{\nu}(\eta,k) \hat{c}_2(\vec{k})
	-
	\frac{k}{2} U_{\nu}^*(\eta,k) \hat{c}_2^\dag(-\vec{k})
	\, ,
\label{K3solution}
\\
\hat{\mathcal{K}}_4(\eta,\vec{k})
	={}&
	\frac{ik}{2} U_{\nu-1}(\eta,k) \hat{c}_2(\vec{k})
	- \frac{i k}{2} U_{\nu-1}^*(\eta,k) \hat{c}_2^\dag(-\vec{k})
	\, .
\label{K4solution}
\end{align}
According to their definition~(\ref{scalarKs}), these expressions automatically
determine three of the three canonical momenta,
\begin{equation}
\hat{\mathcal{P}}_1 = \hat{\mathcal{K}}_1 \, ,
\qquad \quad
\hat{\mathcal{P}}_2 = \hat{\mathcal{K}}_2 \, ,
\qquad \quad
\hat{\mathcal{P}}_4 = \hat{\mathcal{K}}_4 \, ,
\end{equation}
while the remaining equation we can consider to determine~$\hat{\mathcal{S}}_3$
in terms of~$\hat{\mathcal{P}}_3$,
\begin{equation}
\hat{\mathcal{S}}_3
 =
 	\frac{- 2}{(D\!-\!2)k}
 	\biggl[
 	\hat{\mathcal{K}}_3
 	-
	\frac{\mathcal{H}}{k} \bigl( \hat{\mathcal{P}}_3 + \hat{\mathcal{K}}_4 \bigr)
	\biggr]
	\, .
\label{S3intermediate}
\end{equation}

From the remaining four equations we can form two second–order equations, sourced by
the previously obtained constraint solutions,
\begin{align}
\biggl[ \partial_0^2 + k^2
	+
	\Bigl( \frac{1}{4} \!-\! (\nu \!-\! 2)^2 \Bigr) \mathcal{H}^2
	\biggr]
	\hat{\mathcal{S}}_1
	={}&
	- 2(1\!-\!\alpha) k \bigl( \hat{\mathcal{K}}_2 + 2 \hat{\mathcal{K}}_3 \bigr)
	\, ,
\label{S1eq}
\\
\biggl[ \partial_0^2 + k^2
	+
	\Bigl( \frac{1}{4} \!-\! (\nu \!-\! 1)^2 \Bigr) \mathcal{H}^2
	\biggr]
	\hat{\mathcal{S}}_2
	={}&
	- 2 (1\!-\!\alpha) k \bigl( \hat{\mathcal{K}}_4 - \hat{\mathcal{K}}_1 \bigr)
	\, .
\label{S2eq}
\end{align}
Once these are solved the third canonical momentum can be written as
\begin{equation}
\hat{\mathcal{P}}_3
	=
	\frac{D\!-\!2}{2} \biggl[
	\partial_0 
	+
	\Bigl( \frac{1}{2} \!+\! (\nu \!-\! 2) \Bigr) \mathcal{H}
	\biggr] \hat{\mathcal{S}}_1 
	+
	\bigl[ D \!-\! 1 \!-\! 2 (D\!-\!2) \alpha \bigr] \hat{\mathcal{K}}_1
	-
	\hat{\mathcal{K}}_4
	\, ,
\end{equation}
which simultaneously fixes~$\hat{\mathcal{S}}_3$ through~(\ref{S3intermediate}).
This in turn fixes the last scalar,
\begin{equation}
\hat{\mathcal{S}}_4 = 
	- \frac{2}{k} \biggl[ \partial_0 + \Bigl( \frac{1}{2} \!+\! (\nu\!-\!1) \Bigr) \mathcal{H} \biggr] \hat{\mathcal{S}}_2
	- \hat{\mathcal{S}}_1
	+ (D\!-\!2) \hat{\mathcal{S}}_3
	- \frac{2\alpha}{k} \hat{\mathcal{P}}_2
	\, .
\end{equation}

The solutions for the five remaining fields are parametrized by the initial conditions of the
two second–order equations~(\ref{S1eq}) and~(\ref{S2eq}),
\begin{align}
&
\hat{\mathcal{S}}_1(\eta,\vec{k})
	=
	U_{\nu-2}(\eta,k) \hat{d}_1(\vec{k})
	+
	U_{\nu-2}^*(\eta,k) \hat{d}_1^\dag(-\vec{k})
	+
	w_1(\eta,k) \hat{c}_1(\vec{k})
	+
	w_1^*(\eta,k) \hat{c}_1^\dag(-\vec{k})
	\, ,
\\
&
\hat{\mathcal{S}}_2(\eta,\vec{k})
	=
	i U_{\nu-1}(\eta,k) \hat{d}_2(\vec{k})
	-
	i U_{\nu-1}^*(\eta,k) \hat{d}_2^\dag(-\vec{k})
\nonumber \\
&	\hspace{2cm}
	+
	i w_2(\eta,k) \Bigl[ 2  \hat{c}_1(\vec{k}) - \hat{c}_2(\vec{k}) \Bigr]
	-
	i w_2^*(\eta,k) \Bigl[ 2 \hat{c}_1^\dag(-\vec{k}) -\hat{c}_2^\dag(-\vec{k}) \Bigr]
	\, .
\\
&
\hat{\mathcal{P}}_3(\eta,\vec{k})
	=
	\frac{D\!-\!2}{2}
	\Bigl[
	- i k U_{\nu-1}(\eta,k) \hat{d}_1(\vec{k})
	+
	i k U_{\nu-1}^*(\eta,k) \hat{d}_1^\dag(-\vec{k})
	-
	i k w_3(\eta,k) \hat{c}_1(\vec{k})
\nonumber \\
&	\hspace{2cm}
	+
	i k w_3^*(\eta,k) \hat{c}_1^\dag(-\vec{k})
	\Bigr]
	-
	\frac{ik}{2} U_{\nu-1}(\eta,k) \hat{c}_2(\vec{k})
	+
	\frac{i k}{2} U_{\nu-1}^*(\eta,k) \hat{c}_2^\dag(-\vec{k}) 
	\, ,
\\
&
\hat{\mathcal{S}}_3(\eta,\vec{k})
 	=
	-
	\frac{i \mathcal{H}}{k}
	\Bigl[
	U_{\nu-1}(\eta,k) \hat{d}_1(\vec{k})
	-
	U_{\nu-1}^*(\eta,k) \hat{d}_1^\dag(-\vec{k})
	\Bigr]
	+
	\frac{ 1}{D\!-\!2}
 	\Bigl[
	U_{\nu}(\eta,k) \hat{c}_2(\vec{k})
	+
	U_{\nu}^*(\eta,k) \hat{c}_2^\dag(-\vec{k})
	\Bigr]
\nonumber \\
&	\hspace{0.4cm}
	-
	\biggl[
	\frac{2}{D\!-\!2}
 	U_{\nu-2}(\eta,k)
	+
	\frac{ i \mathcal{H}}{k} w_3(\eta,k)
	\biggr]
	\hat{c}_1(\vec{k})
	-
	\biggl[
	\frac{2}{D\!-\!2}
	U_{\nu-2}^*(\eta,k)
	-
	\frac{ i \mathcal{H}}{k} w_3^*(\eta,k) 
	\biggr]
	\hat{c}_1^\dag(-\vec{k})
	\, ,
\\
&
\hat{\mathcal{S}}_4(\eta,\vec{k})
	=
	-
	\biggl[ 
	U_{\nu-2}(\eta,k)
	+
	\frac{(D\!-\!2) i \mathcal{H}}{k}
	U_{\nu-1}(\eta,k)
	\biggr]
	\hat{d}_1(\vec{k})
	-
	\biggl[
	U_{\nu-2}^*(\eta,k)
\nonumber \\
&	\hspace{1.9cm}
	-
	\frac{(D\!-\!2) i \mathcal{H}}{k}
	U_{\nu-1}^*(\eta,k)
	\biggr]
	\hat{d}_1^\dag(-\vec{k})
	- 
	2U_{\nu}(\eta,k) \hat{d}_2(\vec{k})
	-
	2 U_{\nu}^*(\eta,k) \hat{d}_2^\dag(-\vec{k})
\nonumber \\
&	\hspace{1.9cm}
	-
	2 w_4(\eta,k)
	\Bigl[ 2  \hat{c}_1(\vec{k}) - \hat{c}_2(\vec{k}) \Bigr]
	-
	2 w_4^*(\eta,k)
	\Bigl[ 2 \hat{c}_1^\dag(-\vec{k}) -\hat{c}_2^\dag(-\vec{k}) \Bigr]
\nonumber \\
&	\hspace{1.9cm}
	-
	\biggl[ 
	w_1(\eta,k)
	+
	\frac{ (D\!-\!2) i \mathcal{H}}{k} w_3(\eta,k)
	+
 	2U_{\nu-2}(\eta,k)
	-
	2 (1 \!-\! 2\alpha) U_{\nu}(\eta,k)
	\biggr]
	\hat{c}_1(\vec{k})
\nonumber \\
&	\hspace{1.9cm}
	-
	\biggl[
	w_1^*(\eta,k)
	-
	\frac{ (D\!-\!2) i \mathcal{H}}{k} w_3^*(\eta,k) 
	+
	2U_{\nu-2}^*(\eta,k)
	-
	2(1 \!-\! 2\alpha) U_{\nu}^*(\eta,k)
	\biggr]
	\hat{c}_1^\dag(-\vec{k})
	\, .
\end{align}
and expressed in terms of the particular mode functions~$w_1$--$w_4$.
The first two particular mode functions satisfy sourced second--order equations,
\begin{align}
\biggl[ \partial_0^2 + k^2 + \Bigl( \frac{1}{4} \!-\! (\nu\!-\!2) \Bigr) \mathcal{H}^2 \biggr]
	w_1
	={}&
	-
	4(1\!-\!\alpha) k^2 U_{\nu-2} \, ,
\\
\biggl[ \partial_0^2 + k^2 + \Bigl( \frac{1}{4} \!-\! (\nu\!-\!1) \Bigr) \mathcal{H}^2 \biggr]
	w_2
	={}&
	(1\!-\!\alpha) k^2 U_{\nu-1}
	\, .
\end{align}
whose solutions determine the two remaining particular mode functions,
\begin{align}
w_3 ={}& 
	\frac{i}{k} \biggl[ \partial_0 + \Bigl( \frac{1}{2} \!+\! (\nu\!-\!2) \Bigr) \mathcal{H} \biggr]
	w_1
	+
	2\biggl[ \frac{D \!-\! 3}{D\!-\!2} \!-\! 2 (1 \!-\! \alpha) \biggr] U_{\nu-1}
	\, ,
\\
w_4 ={}& 
	\frac{i}{k}
	\biggl[ \partial_0
	+
	\Bigl( \frac{1}{2} \!+\! (\nu \!-\! 1) \Bigr)  \mathcal{H} 
	\biggr] w_2
	+
	\frac{1 \!-\! 2\alpha}{2} U_\nu
	\, .
\end{align}

The second–order equations for the particular mode functions are solved using the
identity~(\ref{inhomID}), and the resulting expressions are simplified through the
recurrence relations~(\ref{CTBDrecurrence}):
\begin{align}
w_1 ={}&
	\frac{2(1\!-\!\alpha) ik}{\mathcal{H}} 
		\biggl[
		U_{\nu-3}
		-
		\frac{\mathcal{H}}{H} U_{\nu-2}
		\biggr]
	-
	\biggl[
	\frac{D \!-\! 3}{D\!-\!2} 
	-
	2(1\!-\!\alpha) ( 3 \!-\! \nu ) 
	\biggr]
	U_{\nu-2}
	\, ,
\label{w1solution}
\\
w_2 ={}&
	- \frac{ (1\!-\!\alpha) ik }{2\mathcal{H}}  
		\biggl[
		U_{\nu-2}
		-
		\frac{\mathcal{H}}{H} U_{\nu-1}
		\biggr]
	-
	\frac{1}{4}
	\biggl[
	\frac{D \!-\! 3}{D\!-\!2} 
	+
	2(1\!-\!\alpha) ( 1 \!-\! \nu ) 
	\biggr]
	U_{\nu-1}
	=
	- \frac{1}{4} w_3
	\, ,
\label{w2solution}
\\
w_4 ={}&
	- \frac{ (1\!-\!\alpha) ik }{2\mathcal{H}}  
		\biggl[
		U_{\nu-1}
		-
		\frac{\mathcal{H}}{H} U_{\nu}
		\biggr]
	-
	\frac{1}{4}
	\biggl[
	\frac{3D \!-\! 7}{D\!-\!2} 
	-
	2(1\!-\!\alpha) ( 1 \!+\! \nu ) 
	\biggr]
	U_{\nu}
	\, .
\label{w4solution}
\end{align}
The homogeneous parts of these solutions are fixed by requiring a regular
flat-space limit, and by imposing the normalization condition
\begin{equation}
	{\rm Re}
	\Bigl(
	U_{\nu-1}^* w_1 - 4 U_{\nu-2}^* w_2
	\Bigr)
	=
	0 \, .
\end{equation}
Naturally, different choices of integration constants are possible; these choices
manifest themselves in the commutation relations obtained next. As expected,
however, the final two-point functions remain invariant under such redefinitions.

Inverting the field-operator solutions to extract the corresponding initial-condition
operators is a tedious exercise, but it yields the full algebra of commutators.
Among these, the non-vanishing ones are
\begin{align}
&
\bigl[ \hat{d}_1(\vec{k}) , \hat{c}_1^\dag(\vec{k}^{\,\prime}) \bigr]
	=
	\bigl[ \hat{c}_1(\vec{k}) , \hat{d}_1^\dag(\vec{k}^{\,\prime}) \bigr]
	=
	- \delta^{D-1}(\vec{k} \!-\! \vec{k}^{\,\prime})
	\, ,
\\
&
\bigl[ \hat{d}_2(\vec{k}) , \hat{c}_1^\dag(\vec{k}^{\,\prime}) \bigr]
	=
	\bigl[ \hat{c}_1(\vec{k}) , \hat{d}_2^\dag(\vec{k}^{\,\prime}) \bigr]
	=
	\frac{1}{2} \delta^{D-1}(\vec{k} \!-\! \vec{k}^{\,\prime})
	\, ,
\\
&
\bigl[ \hat{d}_2(\vec{k}) , \hat{c}_2^\dag(\vec{k}^{\,\prime}) \bigr]
	=
	\bigl[ \hat{c}_2(\vec{k}) , \hat{d}_2^\dag(\vec{k}^{\,\prime}) \bigr]
	=
	\delta^{D-1}(\vec{k} \!-\! \vec{k}^{\,\prime})
	\, ,
\\
&
\bigl[ \hat{d}_2(\vec{k}) , \hat{d}_2^\dag(\vec{k}^{\,\prime}) \bigr]
	=
	-
	\alpha \,
	\delta^{D-1}(\vec{k} \!-\! \vec{k}^{\,\prime})
	\, .
\end{align}
It is convenient to redefine the momentum-space operators so that these commutation
relations become block diagonal and take the same structure as in the vector
sector~(\ref{VectorCommutators}). This is achieved by replacing~$\hat{d}_1$ and
$\hat{d}_2$ with
\begin{equation}
\hat{e}_1(\vec{k}) = - \hat{d}_1(\vec{k}) \, ,
\qquad\quad
\hat{e}_2(\vec{k}) = \hat{d}_2(\vec{k}) 
	+ \frac{1}{2} \hat{d}_1(\vec{k}) + \frac{\alpha}{2} \hat{c}_2(\vec{k})
\, .
\label{MomentumOperatorTrans}
\end{equation}
In terms of these new operators, the nonvanishing commutators become
\begin{align}
&
\bigl[ \hat{e}_1(\vec{k}) , \hat{c}_1^\dag(\vec{k}^{\,\prime}) \bigr]
	=
	\bigl[ \hat{c}_1(\vec{k}) , \hat{e}_1^\dag(\vec{k}^{\,\prime}) \bigr]
	=
	\delta^{D-1}(\vec{k} \!-\! \vec{k}^{\,\prime} )
	\, ,
\nonumber \\
&
\bigl[ \hat{e}_2(\vec{k}) , \hat{c}_2^\dag(\vec{k}^{\,\prime}) \bigr]
	=
	\bigl[ \hat{c}_2(\vec{k}) , \hat{e}_2^\dag(\vec{k}^{\,\prime}) \bigr]
	=
	\delta^{D-1}(\vec{k} \!-\! \vec{k}^{\,\prime} )
	\, .
\end{align}

The vanishing of the two-point functions of the scalar-sector constraints 
is ensured by requiring that two independent non-Hermitian linear combinations of the 
four Hermitian constraints~(\ref{K1solution})--(\ref{K4solution}) annihilate the physical 
state. As in the vector sector, the choice of these combinations is not unique.  
A convenient choice is obtained by examining the flat-space 
limit. In Minkowski space, Lorentz invariance of the graviton two-point function is 
maintained provided the positive-frequency parts of the first two scalar constraints,
\begin{align}
{\tt Minkowski:}
\quad
\hat{K}_1^{\scr (+)} = \int\! \frac{d^{D-1}k}{(2\pi)^{\frac{D-1}{2}}} \,
	u(\eta,k) \hat{c}_1(\vec{k})
\, , \qquad
\hat{K}_2^{\scr (+)} = \int\! \frac{d^{D-1}k}{(2\pi)^{\frac{D-1}{2}}} \,
	k^{-1} u(\eta,k) \hat{c}_2(\vec{k}) \, ,
\end{align}
are required to annihilate the physical state,
\begin{equation}
{\tt Minkowski:}
\quad\qquad
\hat{K}_1^{\scr (+)}(\vec{x}) \bigl| \Omega \bigr\rangle = 0 \, ,
\qquad\quad
\hat{K}_2^{\scr (+)}(\vec{x}) \bigl| \Omega \bigr\rangle = 0 \, ,
\qquad\quad
\forall \vec{x} \, ,
\label{ScalarFlatStateCondition}
\end{equation}
in accordance with the Gupta–Bleuler construction~\cite{Gupta:1952zz}.  
Taking this limit therefore implies that the constraints are implemented in a 
manifestly Lorentz-invariant manner whenever
\begin{equation}
\hat{c}_1(\vec{k}) \bigl| \Omega \bigr\rangle = 0 \, ,
\qquad \quad
\hat{c}_2(\vec{k}) \bigl| \Omega \bigr\rangle = 0 \, ,
\qquad \quad
\forall \vec{k} \, .
\label{ScalarFirstCon}
\end{equation}
We adopt this same condition in de Sitter space.

To complete the construction of the indefinite-metric state space, one must also 
specify the vacuum-like state.  
Following the vector-sector analysis, we define it by
\begin{equation}
\hat{e}_1(\vec{k}) \bigl| \Omega \bigr\rangle = 0 \, ,
\qquad \quad
\hat{e}_2(\vec{k}) \bigl| \Omega \bigr\rangle = 0 \, ,
\qquad \quad
\forall \vec{k} \, ,
\label{ScalarOtherCondition}
\end{equation}
so that all other states are generated by acting with the conjugate operators.  
Using the transformation~(\ref{MomentumOperatorTrans}), this condition is equivalent to
\begin{equation}
\hat{d}_1(\vec{k}) \bigl| \Omega \bigr\rangle = 0 \, ,
\qquad\quad
\hat{d}_2(\vec{k}) \bigl| \Omega \bigr\rangle = 0 \, ,
\label{ScalarSecondCon}
\end{equation}
and we assume that the state satisfies both~(\ref{ScalarFirstCon}) and 
(\ref{ScalarSecondCon}) in the computation of the two-point function in the next section.

\section{Graviton two-point function}
\label{sec: Two-point function}

In this section we present the positive-frequency Wightman two-point function
\begin{equation}
i \bigl[ \tensor*[_{\mu\nu}^{\scr - }]{\Delta}{_{\rho\sigma}^{\scr + }} \bigr] (x;x') =
\bigl\langle \Omega \bigr| \hat{h}_{\mu\nu}(x) \hat{h}_{\rho\sigma}(x') 
	\bigl| \Omega \bigr\rangle
\label{2ptDefinition}
\end{equation}
for the graviton in de Sitter space, evaluated in the one-parameter gauge
specified in~(\ref{GF}).

\subsection{Generalities}
\label{subsec: Generalities}

Apart from the Wightman function~(\ref{2ptDefinition}), perturbative computations
in nonequilibrium quantum field theory (see e.g.~\cite{Berges:2004yj,NoneqLectures})
require three additional two-point functions. The first is the negative-frequency
Wightman function, which is simply the complex conjugate of~(\ref{2ptDefinition}),
$\,i \bigl[ \tensor*[_{\mu\nu}^{\scr + }]{\Delta}{_{\rho\sigma}^{\scr - }} \bigr] (x;x')
= \bigl\{ i \bigl[ \tensor*[_{\rho\sigma}^{\scr - }]{\Delta}{_{\mu\nu}^{\scr + }} \bigr] (x';x) \bigr\}^{*}$.
The second is the Feynman propagator, defined as the expectation value of the
time-ordered product of field operators, and expressible in terms of Wightman
functions and temporal step functions,
\begin{equation}
i \bigl[ \tensor*[_{\mu\nu}^{\scr + }]{\Delta}{_{\rho\sigma}^{\scr + }} \bigr] (x;x')
	=
	\theta(\eta \!-\! \eta') \,
	i \bigl[ \tensor*[_{\mu\nu}^{\scr - }]{\Delta}{_{\rho\sigma}^{\scr + }} \bigr] (x;x')
	+
	\theta(\eta' \!-\! \eta) \,
	i \bigl[ \tensor*[_{\mu\nu}^{\scr + }]{\Delta}{_{\rho\sigma}^{\scr - }} \bigr] (x;x')
	\, .
\end{equation}
Its complex conjugate defines the Dyson propagator,
$\,i \bigl[ \tensor*[_{\mu\nu}^{\scr - }]{\Delta}{_{\rho\sigma}^{\scr - }} \bigr] (x;x')
= \bigl\{ i \bigl[ \tensor*[_{\rho\sigma}^{\scr + }]{\Delta}{_{\mu\nu}^{\scr + }} \bigr] (x';x) \bigr\}^{*}$.

Operator equations of motion~(\ref{eom1})--(\ref{eom6}) 
can be written in a more compact second-order form,
\begin{equation}
\boldsymbol{D}^{\omega\lambda\mu\nu}  \hat{h}_{\mu\nu}= 0 \, ,
\label{covariantEOM}
\end{equation}
where the kinetic operator was given in~(\ref{GaugeFixedKineticOperator}).
Consequently, the same equation of motion is inherited by two-point functions,
\begin{equation}
\boldsymbol{D}^{\omega\lambda\mu\nu} \,
	i \bigl[ \tensor*[_{\mu\nu}^{\tt a}]{\Delta}{_{\rho\sigma}^{\tt b}} \bigr] (x;x') = 
	{\tt S}^{\tt ab} 
	\delta^{(\omega}_\rho \delta^{\lambda)}_\sigma
	\frac{i \delta^D(x\!-\!x')}{\sqrt{-g}}
	\, .
\label{2ptCovariantEOM}
\end{equation}

In addition to this equation, the two-point function must satisfy a number of
state-independent subsidiary conditions~(\ref{ConstraintCorrelators}). These are
essentially equivalent to the Ward identities, which are known to hold in the
simple-gauge limit~$\alpha\!=\!1$~\cite{Tsamis:1992zt}. The primary constraint
operators can be written in a covariantized form,
\begin{equation}
\hat{\Phi}_\omega = 
	\frac{a^{D-2} }{\alpha}
	\mathscr{D}_\omega{}^{\mu\nu} \hat{h}_{\mu\nu} \, ,
\end{equation}
with the operator on the right-hand side defined in~(\ref{calDdef}). Consequently,
the first condition in~(\ref{ConstraintCorrelators}) can be expressed as two linear
derivative operators acting on the propagator,
\begin{equation}
\mathscr{D}_\omega{}^{\mu\nu}
\mathscr{D}'_\lambda{}^{\rho\sigma} \,
i \bigl[ \tensor*[_{\mu\nu}^{\tt a \,}]{\Delta}{_{\rho\sigma}^{\tt \, b}} \bigr](x;x')
	=
	- \alpha \,
	{\tt S}^{\tt ab}
	g_{\omega\lambda}
	\frac{i \delta^D(x \!-\! x') }{ \sqrt{-g} }
	\, .
\label{doubleWard}
\end{equation}
The remaining conditions in~(\ref{ConstraintCorrelators}) are then satisfied
automatically, provided that the equation of motion~(\ref{covariantEOM}) and the
subsidiary condition in~(\ref{doubleWard}) hold. Equations~(\ref{2ptCovariantEOM})
and~(\ref{doubleWard}) therefore provide powerful consistency checks for the result
derived at the end of this section, and the verification is given in the Appendix.

\subsection{Evaluating mode sums}
\label{subsec: Evaluating mode sums}

The components of the graviton two-point function can be expressed in terms
of the two-point functions of the scalar, vector, and tensor sectors,
\begin{align}
i \bigl[ \tensor*[_{00}^{\scr - }]{\Delta}{_{00}^{\scr + }} \bigr] (x;x')
	={}&
	\bigl\langle \Omega \bigr| \hat{S}_1(x) \hat{S}_1(x') \bigl| \Omega \bigr\rangle
	\, ,
\\
i \bigl[ \tensor*[_{0i}^{\scr - }]{\Delta}{_{00}^{\scr + }} \bigr] (x;x')
	={}&
	\frac{\partial_i}{\nabla^2} 
	\bigl\langle \Omega \bigr| 
	\hat{S}_2(x) \hat{S}_1(x') 
	\bigl| \Omega \bigr\rangle
	\, ,
\\
i \bigl[ \tensor*[_{ij}^{\scr - }]{\Delta}{_{00}^{\scr + }} \bigr] (x;x')
	={}&
	\mathbb{P}_{ij}^{\scr T} 
	\bigl\langle \Omega \bigr| 
	\hat{S}_3 (x) \hat{S}_1(x')
	\bigl| \Omega \bigr\rangle
	+
	\mathbb{P}_{ij}^{\scr L} 
	\bigl\langle \Omega \bigr| 
	\hat{S}_4(x) \hat{S}_1(x')
	\bigl| \Omega \bigr\rangle
	\, ,
\\
i \bigl[ \tensor*[_{0i}^{\scr - }]{\Delta}{_{0k}^{\scr + }} \bigr] (x;x')
	={}&
	\frac{\partial_i}{\nabla^2} 
	\frac{\partial'_k}{\nabla'^2} 
	\bigl\langle \Omega \bigr| 
	\hat{S}_2(x) \hat{S}_2(x')
	\bigl| \Omega \bigr\rangle
	+
	\bigl\langle \Omega \bigr| 
	\hat{V}^1_i(x) \hat{V}^1_k(x')
	\bigl| \Omega \bigr\rangle
	\, ,
\\
i \bigl[ \tensor*[_{ij}^{\scr - }]{\Delta}{_{0k}^{\scr + }} \bigr] (x;x')
	={}&
	\mathbb{P}_{ij}^{\scr T} 
	\frac{\partial_k'}{\nabla'^2} 
	\bigl\langle \Omega \bigr| 
	\hat{S}_3(x) \hat{S}_2(x')
	\bigl| \Omega \bigr\rangle
	+
	\mathbb{P}_{ij}^{\scr L} 
	\frac{\partial_k'}{\nabla'^2} 
	\bigl\langle \Omega \bigr| 
	\hat{S}_4(x)
	\hat{S}_2(x')
	\bigl| \Omega \bigr\rangle
\nonumber \\
&
	+
	\frac{2}{\nabla^2} \delta_{(i}^m \partial_{j)}
	\bigl\langle \Omega \bigr| 
	\hat{V}^2_{m}(x)
	\hat{V}^1_k(x')
	\bigl| \Omega \bigr\rangle
	\, ,
\\
i \bigl[ \tensor*[_{ij}^{\scr - }]{\Delta}{_{kl}^{\scr + }} \bigr] (x;x')
	={}&
	\mathbb{P}_{ij}^{\scr T}
	\mathbb{P}_{kl}'^{\scr T}
	\bigl\langle \Omega \bigr| 
	\hat{S}_3(x) \hat{S}_3(x')
	\bigl| \Omega \bigr\rangle
	+
	\mathbb{P}_{ij}^{\scr T}
	\mathbb{P}_{kl}'^{\scr L}
	\bigl\langle \Omega \bigr| 
	\hat{S}_3(x) \hat{S}_4(x')
	\bigl| \Omega \bigr\rangle
\nonumber \\
&
	+
	\mathbb{P}_{ij}^{\scr L}
	\mathbb{P}_{kl}'^{\scr T}
	\bigl\langle \Omega \bigr| 
	\hat{S}_4(x) \hat{S}_3(x')
	\bigl| \Omega \bigr\rangle
	+
	\mathbb{P}_{ij}^{\scr L}
	\mathbb{P}_{kl}'^{\scr L}
	\bigl\langle \Omega \bigr| 
	\hat{S}_4(x) \hat{S}_4(x')
	\bigl| \Omega \bigr\rangle
\nonumber \\
&
	+
	\frac{2}{\nabla^2} \delta_{(i}^m \partial_{j)}
	\frac{2}{\nabla'^2}  \delta_{(k}^n \partial'_{l)}
	\bigl\langle \Omega \bigr| 
	\hat{V}^2_{m}(x) \hat{V}^2_{n}(x')
	\bigl| \Omega \bigr\rangle
	+
	\bigl\langle \Omega \bigr| 
	\hat{T}_{ij}(x) \hat{T}_{kl}(x')
	\bigl| \Omega \bigr\rangle
	\, .
\end{align}
The remaining components follow from
$
i \bigl[ \tensor*[_{\mu\nu}^{\scr - }]{\Delta}{_{\rho\sigma}^{\scr + }} \bigr] (x;x')
	\!=\!
	\bigl\{ 
	i \bigl[ \tensor*[_{\rho\sigma}^{\scr - }]{\Delta}{_{\mu\nu}^{\scr + }} \bigr] (x';x) 
	\bigr\}^*
$,
as implied by the definition~(\ref{2ptDefinition}). We first compute all sectorial
two-point functions and then assemble them to obtain the full graviton two-point
function.

\subsubsection{Tensor sector two-point functions}
\label{subsubsec: Tensor sector two-point functions}

There is only a single two-point function in the tensor sector,
which is readily evaluated using the results of Sec.~\ref{subsec: Tensor sector},
\begin{equation}
\bigl\langle \Omega \bigr| 
	\hat{T}_{ij}(x) \hat{T}_{kl}(x')
	\bigl| \Omega \bigr\rangle
	=
	2 \biggl[ \mathbb{P}^{\scr T}_{k(i} \mathbb{P}^{\scr T}_{j)l} 
		- \frac{ \mathbb{P}^{\scr T}_{ij} \mathbb{P}^{\scr T}_{kl}}{D \!-\! 2} \biggr]
	 i \bigl[ \tensor*[^{\scr \!-\! }]{\Delta}{^{\scr \!+\!}} \bigr]_\nu(x;x')
	 \, ,
\end{equation}
where we have recognized the mode-sum representation~(\ref{WightmanModeSum}) 
of the scalar two-point function. 
This expression contains inverse Laplacians inside the transverse projection operators, 
which act on the scalar two-point function, i.e.~the scalar two-point function is integrated against
the Green's function of the Laplace operator. Although this action can be worked out 
explicitly, the resulting expressions are rather cumbersome~\cite{Domazet:2024dil}. 
Fortunately, in the present gauge the inverse Laplacians either cancel or simplify once 
the contributions of all sectors are assembled into the graviton two-point function.
Thus no explicit evaluation of inverse Laplacians is required here or in the vector 
and scalar sectors.

\subsubsection{Vector sector two-point functions}
\label{subsubsec: Vector sector two-point functions}

Two-point functions in the vector sector can all be written as mode sums,
\begin{equation}
\bigl\langle \Omega \bigr| 
	\hat{V}^I_i(x) \hat{V}^J_j(x')
	\bigl| \Omega \bigr\rangle
	=
	\mathbb{P}_{ij}^{\scr T} \,
	(aa')^{-\frac{D-2}{2} } \!\!
	\int\! \frac{ d^{D-1}k }{ (2\pi)^{D-1} } \,
	e^{i \vec{k} \cdot ( \vec{x} - \vec{x}^{\,\prime } ) } \,
	Y_{IJ}(\eta,\eta',k)
	\, .
\end{equation}
where the integrands are
\begin{align}
Y_{11}(\eta,\eta',k)
	={}&
	U_{\nu-1}(\eta,k) 
	v_1^*(\eta',k)
	+
	v_1(\eta,k) 
	U_{\nu-1}^*(\eta',k)
	\, ,
\\
Y_{21}(\eta,\eta',k)	
	={}&
	- i k
	\Bigl[
	U_{\nu}(\eta,k) v_1^*(\eta',k)
	+
	v_2(\eta,k) U_{\nu-1}^*(\eta',k)
	\Bigr]
	\, ,
\\
Y_{22}(\eta,\eta',k)
	={}&
	k^2
	\Bigl[
	U_\nu(\eta,k) v_2^*(\eta',k) 
	+
	v_2(\eta,k) U_\nu^*(\eta',k) 
	\Bigr]
	\, ,
\end{align}
with $Y_{21}(\eta,\eta',k) = \bigl[ Y_{12}(\eta',\eta,k) \bigr]^*$.  
These expressions follow from substituting the momentum-space solutions
(\ref{V2sol}) and~(\ref{V1sol}) into the position-space operators~(\ref{Vfourier}),
evaluating the expectation value using the state conditions
(\ref{VectorSubsidiary}) and~(\ref{OtherVectorStateCondition}), applying the
commutation relations~(\ref{VectorCommutators}), and finally making use of the
polarization sum~(\ref{VectorPolarization}) to identify the transverse projector.

Substituting the explicit particular solutions~(\ref{v1sol}) and~(\ref{v2sol}), and
using the recurrence relations~(\ref{CTBDrecurrence}) wherever appropriate,
yields compact expressions for the vector sector two-point functions,
in terms of operators acting on scalar two-point functions:
\begin{align}
&
\bigl\langle \Omega \bigr| 
	\hat{V}^1_i(x) \hat{V}^1_j(x')
	\bigl| \Omega \bigr\rangle
	=
	\mathbb{P}_{ij}^{\scr T}
	\biggl\{
	- 1 - \frac{ 1\!-\!\alpha }{2\mathcal{H}\mathcal{H}' } 
		\Bigl[
		\mathcal{H}' \partial_0 
		+
		\mathcal{H} \partial'_0 
		+
		(D \!-\! 3) \mathcal{H}\mathcal{H}'
		\Bigr]
	\biggr\}
	i \bigl[ \tensor*[^{\scr \!-\!}]{\Delta}{^{\scr \!+\!}} \bigr]_{\nu-1}(x;x')
	\, ,
\\
&
\bigl\langle \Omega \bigr| 
	\hat{V}^2_i(x) \hat{V}^1_j(x')
	\bigl| \Omega \bigr\rangle
	=
	\frac{ 1\!-\!\alpha }{2 \mathcal{H} \mathcal{H}'}
	\mathbb{P}_{ij}^{\scr T} \nabla^2
	\biggl\{
	\mathcal{H} \,
	i \bigl[ \tensor*[^{\scr \!-\!}]{\Delta}{^{\scr \!+\!}} \bigr]_{\nu}(x;x')
	-
	\mathcal{H}' \,
	i \bigl[ \tensor*[^{\scr \!-\!}]{\Delta}{^{\scr \!+\!}} \bigr]_{\nu-1}(x;x')
	\biggr\}
	\, ,
\\
&
\bigl\langle \Omega \bigr| 
	\hat{V}^2_i(x) \hat{V}^2_j(x')
	\bigl| \Omega \bigr\rangle
	=
	\mathbb{P}_{ij}^{\scr T} 
	\biggl\{
	- 1
	+
	\frac{ 1\!-\!\alpha }{2\mathcal{H}\mathcal{H}'} 
		\Bigl[
			\mathcal{H}' \partial_0 
			+
			\mathcal{H} \partial'_0 
			+
			(D \!-\!1 ) \mathcal{H}\mathcal{H}' 
		\Bigr]
		\biggr\}
		\nabla^2 
	i \bigl[ \tensor*[^{\scr \!-\!}]{\Delta}{^{\scr \!+\!}} \bigr]_{\nu}(x;x')
	\, .
\end{align}
%

\subsubsection{Scalar sector two-point functions}
\label{subsubsec: Scalar sector two-point functions}

Scalar-sector correlators are again most conveniently expressed as mode integrals.
Using the results of Sec.~\ref{subsec: Scalar sector},
for each pair of fields $\hat{S}_I$ and $\hat{S}_J$ we may write
\begin{equation}
\bigl\langle \Omega \bigr| \hat{S}_I(x) \hat{S}_J(x') \bigl| \Omega \bigr\rangle
	=
	(aa')^{-\frac{D-2}{2} } \!\!
	\int\! \frac{ d^{D-1}k }{ (2\pi)^{D-1} } \,
	e^{i \vec{k} \cdot ( \vec{x} - \vec{x}^{\,\prime } ) } \,
	Z_{IJ}(\eta,\eta',k)
\end{equation}
where the integrands take the form
\begin{align}
Z_{11}(\eta,\eta',k) ={}&
	-
	U_{\nu-2}(\eta,k) 
	w_1^*(\eta',k)
	-
	w_1(\eta,k) 
	U_{\nu-2}^*(\eta',k)
	\, ,
\\
Z_{21}(\eta,\eta',k) ={}&
	\frac{i k}{2}
	\Bigl[
	U_{\nu-1}(\eta,k) w_1^*(\eta',k)
	-
	4 w_2(\eta,k) U_{\nu-2}^*(\eta',k)
	\Bigr]
	\, ,
\\
Z_{31}(\eta,\eta',k) ={}&
		\frac{ i \mathcal{H} }{ k } \Bigl[
			w_3(\eta,k) U_{\nu-2}^*(\eta',k)
			+
			U_{\nu-1}(\eta,k) w_1^*(\eta',k)
			\Bigr] 
	+
	\frac{2 }{D\!-\!2} 
	U_{\nu-2}(\eta,k) U_{\nu-2}^*(\eta',k)
			\, ,
\\
Z_{41}(\eta,\eta',k) ={}&
	\Bigl[ 
	w_1(\eta,k)
	+
	4 w_4(\eta,k)
	\Bigr]
	U_{\nu-2}^*(\eta',k)
	+
	\Bigl[ 
	U_{\nu-2}(\eta,k)
	- 
	U_{\nu}(\eta,k)
	\Bigr]
	w_1^*(\eta',k)
\nonumber \\
&
	+
	\frac{(D\!-\!2) i \mathcal{H}}{k} 
	\Bigl[
	U_{\nu-1}(\eta,k) w_1^*(\eta',k)
	+
	w_3(\eta,k) U_{\nu-2}^*(\eta',k)
	\Bigr]
\nonumber \\
&
	+
	2 \Bigl[
 	U_{\nu-2}(\eta,k)
	-
	(1 \!-\! 2\alpha) U_{\nu}(\eta,k)
	\Bigr]
	U_{\nu-2}^*(\eta',k)
	\, ,
\\
Z_{22}(\eta,\eta',k) ={}&
	- \alpha k^2 U_{\nu-1}(\eta,k) U_{\nu-1}^*(\eta',k)
	\, ,
\\
Z_{32}(\eta,\eta',k) ={}&
	\frac{\mathcal{H}}{2}
	\Bigl[
	4 U_{\nu-1}(\eta,k) w_2^*(\eta',k) 
	-
	w_3(\eta,k) U_{\nu-1}^*(\eta',k)
	\Bigr]
\nonumber \\
&
	+
 	\frac{ ik }{D\!-\!2}
 	\Bigl[
 	U_{\nu-2}(\eta,k)
 	-
 	U_{\nu}(\eta,k)
 	\Bigr]
 	U_{\nu-1}^*(\eta',k)
 	\, ,
\\
Z_{42}(\eta,\eta',k) ={}&
	\frac{ik}{2} \Bigl[
	w_1(\eta,k) U_{\nu-1}^*(\eta',k)
	-
	4 U_{\nu-2}(\eta,k) w_2^*(\eta',k)
	\Bigr]
	-
	\frac{(D\!-\!2)\mathcal{H}}{2}
	\Bigl[ 
	w_3(\eta,k) U_{\nu-1}^*(\eta',k)
\nonumber \\
&
	-
	4 U_{\nu-1}(\eta,k) w_2^*(\eta',k) 
	\Bigr]
	+
	i k \Bigl[
 	U_{\nu-2}(\eta,k)
	-
	U_{\nu}(\eta,k) 
	\Bigr]
	U_{\nu-1}^*(\eta',k)
	\, ,
\\
Z_{33}(\eta,\eta',k) ={}&
	-
	\frac{ \mathcal{H} \mathcal{H}' }{k^2}
	\Bigl[
	U_{\nu-1}(\eta,k)
	w_3^*(\eta',k) 
	+
	w_3(\eta,k)
	U_{\nu-1}^*(\eta',k) 
	\Bigr]
\nonumber \\
&
	+
	\frac{2i}{(D\!-\!2)k}
	\Bigl[
	\mathcal{H}'
 	U_{\nu-2}(\eta,k)
	U_{\nu-1}^*(\eta',k) 
	-
	\mathcal{H}
	U_{\nu-1}(\eta,k)
	U_{\nu-2}^*(\eta',k)
	\Bigr]
	\, ,
\\
Z_{43}(\eta,\eta',k) ={}&
	-
	\frac{ (D\!-\!2) \mathcal{H}\mathcal{H}'}{k^2}
	\Bigl[
	U_{\nu-1}(\eta,k) w_3^*(\eta',k)
	+
	w_3(\eta,k) U_{\nu-1}^*(\eta',k)
	\Bigr]
\\
&
	+
	\frac{ i \mathcal{H}' }{k} 
	\Bigl[ w_1(\eta,k)  + 4 w_4(\eta,k)\Bigr]
	U_{\nu-1}^*(\eta',k)
	-
	\frac{ i\mathcal{H}'}{k} 
	\Bigl[ U_{\nu}(\eta,k) - U_{\nu-2}(\eta,k) \Bigr]
	w_3^*(\eta',k)
\nonumber \\
&
	-
 	\frac{2i \mathcal{H} }{k}
 	U_{\nu-1}(\eta,k) U_{\nu-2}^*(\eta',k)
	+
 	\frac{2i \mathcal{H}' }{k}
 	\Bigl[
	U_{\nu-2}(\eta,k)
	-
	(1 \!-\! 2\alpha) U_{\nu}(\eta,k)
	\Bigr]
	U_{\nu-1}^*(\eta',k) 
\nonumber \\
&
	-
 	\frac{2}{D\!-\!2}
	\Bigl[
	U_{\nu}(\eta,k) U_{\nu}^*(\eta',k)
	-
	U_{\nu}(\eta,k) U_{\nu-2}^*(\eta',k)
	+
	U_{\nu-2}(\eta,k) U_{\nu-2}^*(\eta',k)
	\Bigr]
	\, 
	\nonumber 
\\
Z_{44}(\eta,\eta',k)
	={}&
	-
	4 U_{\nu}(\eta,k) w_4^*(\eta',k)
	-
	4 w_4(\eta,k) U_{\nu}^*(\eta',k)
	-
	4 \alpha U_{\nu}(\eta,k) U_{\nu}^*(\eta',k) 
\nonumber \\
&
	+
	\biggl[ 
	U_{\nu}(\eta,k)
	-
	U_{\nu-2}(\eta,k)
	-
	\frac{(D\!-\!2) i \mathcal{H}}{k}
	U_{\nu-1}(\eta,k)
	\biggr]
	\!\times\!
	\biggl[
	w_1^*(\eta',k)
\nonumber \\
&	\hspace{0.8cm}
	-
	\frac{ (D\!-\!2) i \mathcal{H}' }{k} w_3^*(\eta',k)
	+
	4 w_4^*(\eta',k)
	+
	2U_{\nu-2}^*(\eta',k)
	-
	2(1 \!-\! 2\alpha) U_{\nu}^*(\eta',k)
	\biggr]
\nonumber \\
&
	+
	\biggl[
	w_1(\eta,k)
	+
	\frac{ (D\!-\!2) i \mathcal{H}}{k} w_3(\eta,k)
	+
	4 w_4(\eta,k)
	+
 	2U_{\nu-2}(\eta,k)
	-
	2 (1 \!-\! 2\alpha) U_{\nu}(\eta,k)
	\biggr]
\nonumber \\
&	\hspace{1cm}
	\times\!
	\biggl[
	U_{\nu}^*(\eta',k)
	-
	U_{\nu-2}^*(\eta',k)
	+
	\frac{(D\!-\!2) i \mathcal{H}' }{k}
	U_{\nu-1}^*(\eta',k)
	\biggr]
	\, .
\end{align}
Finally, the remaining combinations follow from the symmetry 
relation~$ Z_{IJ}(\eta,\eta',k) \!=\! \bigl[ Z_{JI}(\eta',\eta,k) \bigr]^*$.

Using the explicit solutions for the scalar-sector particular mode functions given in~(\ref{w1solution})--(\ref{w4solution}), together with the recurrence relations~(\ref{CTBDrecurrence}), we can express all scalar-sector two-point functions directly in terms of scalar Wightman functions. The resulting expressions are:
\begin{align}
\bigl\langle \Omega \bigr| \hat{S}_1(x) \hat{S}_1(x') \bigl| \Omega \bigr\rangle
	={}&
	\biggl\{
	\frac{2(D\!-\!3)}{D\!-\!2} 
	+
	\frac{ 2(1\!-\!\alpha) }{ \mathcal{H} \mathcal{H}' }
	\Bigl[ \mathcal{H}' \partial_0 + \mathcal{H} \partial'_0
		+ (D\!-\!3) \mathcal{H} \mathcal{H}' \Bigr]
	\biggr\}
	i \bigl[ \tensor*[^{\scr \!-\!}]{\Delta}{^{\scr \!+\! }} \bigr]_{\nu-2}(x;x')
	\, ,
\\
\bigl\langle \Omega \bigr| \hat{S}_2(x) \hat{S}_1(x') \bigl| \Omega \bigr\rangle
	={}&
	\frac{1\!-\!\alpha}{ \mathcal{H} \mathcal{H}' } \nabla^2 
	\Bigl\{
		\mathcal{H}' \,
		i \bigl[ \tensor*[^{\scr \!-\!}]{\Delta}{^{\scr \!+\!}} \bigr]_{\nu-2} (x;x')
		-
		\mathcal{H} \,
		i \bigl[ \tensor*[^{\scr \!-\!}]{\Delta}{^{\scr \!+\!}} \bigr]_{\nu-1} (x;x')
	\Bigr\}
	\, ,
\\
\bigl\langle \Omega \bigr| \hat{S}_3(x) \hat{S}_1(x') \bigl| \Omega \bigr\rangle
	={}&
	\bigl\langle \Omega \bigr| \hat{S}_4(x) \hat{S}_1(x') \bigl| \Omega \bigr\rangle
	=
	\frac{2}{D\!-\!2} \,
	i \bigl[ \tensor*[^{\scr \!-\!}]{\Delta}{^{\scr \!+\!}} \bigr]_{\nu-2} (x;x')
\nonumber \\
&
	-
 	\frac{2 (1\!-\!\alpha)}{\mathcal{H}'}
	\Bigl\{
	\mathcal{H}' \,
	i \bigl[ \tensor*[^{\scr \!-\!}]{\Delta}{^{\scr \!+\!}} \bigr]_{\nu-2} (x;x')
	-
	\mathcal{H} \,
	i \bigl[ \tensor*[^{\scr \!-\!}]{\Delta}{^{\scr \!+\!}} \bigr]_{\nu-1} (x;x')
	\Bigr\}
	 \, ,
\nonumber \\
\bigl\langle \Omega \bigr| \hat{S}_2(x) \hat{S}_2(x') \bigl| \Omega \bigr\rangle
	={}&
	\alpha \nabla^2 
	i \bigl[ \tensor*[^{\scr \!-\!}]{\Delta}{^{\scr \!+\!}} \bigr]_{\nu-1} (x;x')
	\, ,
\\
\bigl\langle \Omega \bigr| \hat{S}_3(x) \hat{S}_2(x') \bigl| \Omega \bigr\rangle
	={}&
		\frac{1\!-\!\alpha}{\mathcal{H}'}
		\Bigl[
		\mathcal{H}' \partial_0 
		+
		\mathcal{H} \partial'_0 
		+ 
		(D\!-\!1)\mathcal{H}\mathcal{H}'
		\Bigr]
		i \bigl[ \tensor*[^{\scr \!-\!}]{\Delta}{^{\scr \!+\!}} \bigr]_{\nu-1} (x;x')
		\, ,
\\
\bigl\langle \Omega \bigr| \hat{S}_4(x) \hat{S}_2(x') \bigl| \Omega \bigr\rangle
	={}&
	\frac{1\!-\!\alpha}{ \mathcal{H} \mathcal{H}' }
	\nabla^2 \Bigl\{
	\mathcal{H}' \,
		i \bigl[ \tensor*[^{\scr \!-\!}]{\Delta}{^{\scr \!+\!}} \bigr]_{\nu-1} (x;x')
	- \mathcal{H} \,
		i \bigl[ \tensor*[^{\scr \!-\!}]{\Delta}{^{\scr \!+\!}} \bigr]_{\nu-2} (x;x')
		\Bigr\}
\nonumber \\
&
	+
	\frac{(D\!-\!2) (1\!-\!\alpha) }{\mathcal{H}'}
		\Bigl[
		\mathcal{H}' \partial_0 
		+
		\mathcal{H} \partial_0' 
		+
		(D\!-\!1) \mathcal{H} \mathcal{H}' \Bigr] 
		i \bigl[ \tensor*[^{\scr \!-\!}]{\Delta}{^{\scr \!+\!}} \bigr]_{\nu-1} (x;x')
		\, ,
\\
\bigl\langle \Omega \bigr| \hat{S}_3(x) \hat{S}_3(x') \bigl| \Omega \bigr\rangle
	={}&
	\frac{2}{D\!-\!2}
	\Bigl[ 
	1
	\!-\!
	(D\!-\!2)(1\!-\!\alpha)
	\Bigr]
	\frac{1}{\nabla^2}
	\Bigl[ \mathcal{H}' \partial_0 +\mathcal{H} \partial_0' 
		+ (D\!-\!1) \mathcal{H} \mathcal{H}' \Bigr] 
	i \bigl[ \tensor*[^{\scr \!-\!}]{\Delta}{^{\scr \!+\!}} \bigr]_{\nu-1} (x;x')
	\, ,
\\
\bigl\langle \Omega \bigr| \hat{S}_4(x) \hat{S}_3(x') \bigl| \Omega \bigr\rangle
	={}&
	-
 	\frac{2}{D\!-\!2} \,
	i \bigl[ \tensor*[^{\scr \!-\!}]{\Delta}{^{\scr \!+\!}} \bigr]_{\nu} (x;x')
	+
	\frac{ 2 }{D\!-\!2}
	\Bigl[ 1 \!-\! (D\!-\!2) (1\!-\!\alpha) \Bigr]
\nonumber \\
&
	\times
	\frac{ 1 }{\nabla^2}
	\Bigl[
	\mathcal{H}' \partial_0 
	+
	\mathcal{H} \partial'_0 
	+
	(D\!-\!1) \mathcal{H} \mathcal{H}'
	\Bigr]
	i \bigl[ \tensor*[^{\scr \!-\!}]{\Delta}{^{\scr \!+\!}} \bigr]_{\nu-1} (x;x')
	\, ,
\\
\bigl\langle \Omega \bigr| \hat{S}_4(x) \hat{S}_4(x') \bigl| \Omega \bigr\rangle
	={}&
	\frac{2}{D\!-\!2} 
	\Bigl[ 1 - (D\!-\!2)(1\!-\!\alpha) \Bigr]
	\frac{1}{\nabla^2}
	\Bigl[
	\mathcal{H} \partial_0' 
	+
	\mathcal{H}' \partial_0 
	+
	(D\!-\!1) \mathcal{H}\mathcal{H}'
	\Bigr]
	i \bigl[ \tensor*[^{\scr \!-\!}]{\Delta}{^{\scr \!+\!}} \bigr]_{\nu-1} (x;x')
\nonumber \\
&
	+
	2 \biggl(
	\frac{D \!-\! 3}{D\!-\!2}
	-
	\frac{ (1\!-\!\alpha) }{\mathcal{H} \mathcal{H}'}
	\Bigl[
	\mathcal{H} \partial_0'
	+
	\mathcal{H}' \partial_0 
	+
	(D \!-\! 1) \mathcal{H} \mathcal{H}'
	\Bigr]
	\biggl)
	i \bigl[ \tensor*[^{\scr \!-\!}]{\Delta}{^{\scr \!+\!}} \bigr]_{\nu} (x;x')
	\, .
\end{align}
%

\subsubsection{Consolidating graviton two-point function}
\label{subsubsec: Consolidating graviton two-point function}

Combining the results for sectorial two-point functions, we obtain the results 
for the full components of the graviton two-point function,
\begin{align}
i \bigl[ \tensor*[_{00}^{\scr - }]{\Delta}{_{00}^{\scr + }} \bigr] (x;x')
	={}&
	2 \biggl\{
	\frac{D\!-\!3}{D\!-\!2} 
	+ \frac{ 1\!-\!\alpha }{ \mathcal{H} \mathcal{H}' }
		\Bigl[ \mathcal{H}' \partial_0 + \mathcal{H} \partial'_0
			+ (D\!-\!3) \mathcal{H} \mathcal{H}' \Bigr] 
	\biggr\}
	i \bigl[ \tensor*[^{\scr \!-\!}]{\Delta}{^{\scr \!+\! }} \bigr]_{\nu-2}(x;x')
	\, ,
\label{0000Component}
\\
i \bigl[ \tensor*[_{0i}^{\scr - }]{\Delta}{_{00}^{\scr + }} \bigr] (x;x')
	={}&
	\frac{ 1\!-\!\alpha }{ \mathcal{H} \mathcal{H}' }
	\partial_i
	\Bigl\{
		\mathcal{H}' \,
		i \bigl[ \tensor*[^{\scr \!-\!}]{\Delta}{^{\scr \!+\!}} \bigr]_{\nu-2} (x;x')
		-
		\mathcal{H} \,
		i \bigl[ \tensor*[^{\scr \!-\!}]{\Delta}{^{\scr \!+\!}} \bigr]_{\nu-1} (x;x')
	\Bigr\}
	\, ,
\label{0i00Component}
\\
i \bigl[ \tensor*[_{ij}^{\scr - }]{\Delta}{_{00}^{\scr + }} \bigr] (x;x')
	={}&
	\frac{2 \delta_{ij}}{D\!-\!2} \,
	i \bigl[ \tensor*[^{\scr \!-\!}]{\Delta}{^{\scr \!+\!}} \bigr]_{\nu-2} (x;x')
	+
	\frac{ 2 \delta_{ij}(1\!-\!\alpha) }{ \mathcal{H}' } 
	\Bigl\{
	\mathcal{H} \,
	i \bigl[ \tensor*[^{\scr \!-\!}]{\Delta}{^{\scr \!+\!}} \bigr]_{\nu-1} (x;x')
	-
	\mathcal{H}'
	i \bigl[ \tensor*[^{\scr \!-\!}]{\Delta}{^{\scr \!+\!}} \bigr]_{\nu-2} (x;x')
	\Bigr\}
	\, ,
\label{ij00Component}
\\
i \bigl[ \tensor*[_{0i}^{\scr - }]{\Delta}{_{0k}^{\scr + }} \bigr] (x;x')
	={}&
	\biggl\{
	-
	\alpha \delta_{ik}
	-
	\frac{ 1\!-\!\alpha }{2\mathcal{H}\mathcal{H}' } 
		\mathbb{P}_{ik}^{\scr T} \Bigl[
		\mathcal{H}' \partial_0 
		+
		\mathcal{H} \partial'_0 
		+
		(D \!-\! 1) \mathcal{H}\mathcal{H}'
		\Bigr]
	\biggr\}
	i \bigl[ \tensor*[^{\scr \!-\!}]{\Delta}{^{\scr \!+\!}} \bigr]_{\nu-1}(x;x')
	\, ,
\label{0i0kComponent}
\\
i \bigl[ \tensor*[_{ij}^{\scr - }]{\Delta}{_{0k}^{\scr + }} \bigr] (x;x')
	={}&
	\frac{1\!-\!\alpha}{\mathcal{H}'}
	\frac{ \delta_{ij} \partial_k'}{\nabla^2} 
		\Bigl[
		\mathcal{H}' \partial_0 
		+
		\mathcal{H} \partial'_0 
		+ 
		(D\!-\!1)\mathcal{H}\mathcal{H}'
		\Bigr]
		i \bigl[ \tensor*[^{\scr \!-\!}]{\Delta}{^{\scr \!+\!}} \bigr]_{\nu-1} (x;x')
\nonumber \\
&
	+
	\frac{(D\!-\!3)(1\!-\!\alpha)}{\mathcal{H}'}
	\frac{\partial_i \partial_j \partial_k'}{ \nabla^4 }
		\Bigl[
		\mathcal{H}' \partial_0 
		+
		\mathcal{H} \partial_0' 
		+
		(D\!-\!1) \mathcal{H} \mathcal{H}' \Bigr] 
		i \bigl[ \tensor*[^{\scr \!-\!}]{\Delta}{^{\scr \!+\!}} \bigr]_{\nu-1} (x;x')
\nonumber \\
&
	+
	\frac{ 1\!-\!\alpha }{ \mathcal{H}'}
	\frac{\partial_i \partial_j \partial_k'}{\nabla^2}
	\Bigl\{
	i \bigl[ \tensor*[^{\scr \!-\!}]{\Delta}{^{\scr \!+\!}} \bigr]_{\nu}(x;x')
	- 
	i \bigl[ \tensor*[^{\scr \!-\!}]{\Delta}{^{\scr \!+\!}} \bigr]_{\nu-2} (x;x')
	\Bigr\}
\nonumber \\
&
	+
	\frac{ 1\!-\!\alpha }{ \mathcal{H} \mathcal{H}'}
	\delta_{k(i} \partial_{j)}
	\Bigl\{
	\mathcal{H} \,
	i \bigl[ \tensor*[^{\scr \!-\!}]{\Delta}{^{\scr \!+\!}} \bigr]_{\nu}(x;x')
	-
	\mathcal{H}' \,
	i \bigl[ \tensor*[^{\scr \!-\!}]{\Delta}{^{\scr \!+\!}} \bigr]_{\nu-1}(x;x')
	\Bigr\}
	\, ,
\label{ij0kComponent}
\\
i \bigl[ \tensor*[_{ij}^{\scr - }]{\Delta}{_{kl}^{\scr + }} \bigr] (x;x')
	={}&
	\frac{ 2(1\!-\!\alpha) }{\mathcal{H}\mathcal{H}'} 
	\frac{ \partial_{(i} \delta_{j)(k} \partial_{l)}' }{\nabla^2}
		\Bigl[
			\mathcal{H}' \partial_0 
			+
			\mathcal{H} \partial'_0 
			+
			(D \!-\!1 ) \mathcal{H}\mathcal{H}' \Bigr]
	i \bigl[ \tensor*[^{\scr \!-\!}]{\Delta}{^{\scr \!+\!}} \bigr]_{\nu}(x;x')
\nonumber \\
&
	+
	\frac{2}{D\!-\!2}
	\Bigl[ 
	1
	\!-\!
	(D\!-\!2)(1\!-\!\alpha)
	\Bigr]
	\frac{ \delta_{ij} \delta_{kl} }{\nabla^2}
	\Bigl[ \mathcal{H}' \partial_0 +\mathcal{H} \partial_0' 
		+ (D\!-\!1) \mathcal{H} \mathcal{H}' \Bigr] 
	i \bigl[ \tensor*[^{\scr \!-\!}]{\Delta}{^{\scr \!+\!}} \bigr]_{\nu-1} (x;x')
\nonumber \\
&	\hspace{0.5cm}
	+
	2 \Bigl[
	\delta_{k(i} \delta_{j)l}
	-
	\frac{ \delta_{ij} \delta_{kl} }{D \!-\! 2} 
	\Bigr]
	i \bigl[ \tensor*[^{\scr \!-\! }]{\Delta}{^{\scr \!+\!}} \bigr]_\nu(x;x')
	\, .
\label{ijklComponent}
\end{align}
Using the identity in~(\ref{LaplaceIdentity}), all inverse Laplacians appearing in components~(\ref{ij0kComponent}) and~(\ref{ijklComponent}) can be removed, giving
\begin{align}
i \bigl[ \tensor*[_{ij}^{\scr - }]{\Delta}{_{0k}^{\scr + }} \bigr] (x;x')
	={}&
	\frac{1\!-\!\alpha}{ (D\!-\!3) \mathcal{H}'}
	\delta_{ij} \partial_k'
	\Bigl\{
	i \bigl[ \tensor*[^{\scr \!-\!}]{\Delta}{^{\scr \!+\!}} \bigr]_{\nu-2} (x;x')
	-
	i \bigl[ \tensor*[^{\scr \!-\!}]{\Delta}{^{\scr \!+\!}} \bigr]_{\nu} (x;x')
	\Bigr\}
\nonumber \\
&
	+
	\frac{ 1\!-\!\alpha }{ \mathcal{H} \mathcal{H}'}
	\delta_{k(i} \partial_{j)}
	\Bigl\{
	\mathcal{H} \,
	i \bigl[ \tensor*[^{\scr \!-\!}]{\Delta}{^{\scr \!+\!}} \bigr]_{\nu}(x;x')
	-
	\mathcal{H}' \,
	i \bigl[ \tensor*[^{\scr \!-\!}]{\Delta}{^{\scr \!+\!}} \bigr]_{\nu-1}(x;x')
	\Bigr\}
	\, ,
\\
i \bigl[ \tensor*[_{ij}^{\scr - }]{\Delta}{_{kl}^{\scr + }} \bigr] (x;x')
	={}&
	2 \biggl[
	\delta_{k(i} \delta_{j)l}
	-
	\frac{ \delta_{ij} \delta_{kl} }{D \!-\! 3}
	\biggr]
	i \bigl[ \tensor*[^{\scr \!-\! }]{\Delta}{^{\scr \!+\!}} \bigr]_\nu(x;x')
	+
	\frac{2 \delta_{ij} \delta_{kl} }{(D\!-\!2) (D\!-\!3) }
	i \bigl[ \tensor*[^{\scr \!-\!}]{\Delta}{^{\scr \!+\!}} \bigr]_{\nu-2} (x;x')
\nonumber \\
&
	+
	\frac{ 2(1\!-\!\alpha) }{ (D\!-\!1) \mathcal{H}\mathcal{H}'}
	\partial_{(i} \delta_{j)(k} \partial_{l)}'
	\Bigl\{
	i \bigl[ \tensor*[^{\scr \!-\!}]{\Delta}{^{\scr \!+\!}} \bigr]_{\nu-1} (x;x')
	-
	i \bigl[ \tensor*[^{\scr \!-\!}]{\Delta}{^{\scr \!+\!}} \bigr]_{\nu+1} (x;x')
	\Bigr\}
\nonumber \\
&
	+
	\frac{2 (1\!-\!\alpha) }{ (D\!-\!3) }
	\delta_{ij} \delta_{kl}
	\Bigl\{
	i \bigl[ \tensor*[^{\scr \!-\!}]{\Delta}{^{\scr \!+\!}} \bigr]_{\nu} (x;x')
	-
	i \bigl[ \tensor*[^{\scr \!-\!}]{\Delta}{^{\scr \!+\!}} \bigr]_{\nu-2} (x;x')
	\Bigr\}
\end{align}
Thus, the graviton two-point function can be written entirely in terms of scalar propagators with different effective masses, acted on by at most two derivatives. The remaining two-point functions follow directly after assigning the appropriate Schwinger--Keldysh polarities to the scalar propagators.

The entire set of two-point function components can be written in a significantly more compact form,
\begin{equation}
i \bigl[ \tensor*[_{\mu\nu}^{\tt a}]{\Delta}{_{\rho\sigma}^{\tt b}} \bigr](x;x')
	=
	i \bigl[ \tensor*[_{\mu\nu}^{\tt a \,}]{\Upsilon}{_{\rho\sigma}^{\, \tt b}} \bigr](x;x')
	+
	(1 \!-\! \alpha) \! \times\!
	i \bigl[ \tensor*[_{\mu\nu}^{\tt a \,}]{\Theta}{_{\rho\sigma}^{\tt \, b}} \bigr](x;x')
	\, ,
\label{SplitPropagator}
\end{equation}
where the first contribution corresponds to the simple gauge graviton 
propagator~\cite{Tsamis:1992xa,Woodard:2004ut}, obtained by setting $\alpha\!=\!1$,
\begin{align}
\MoveEqLeft[4]
	i \bigl[ \tensor*[_{\mu\nu}^{\tt a \,}]{\Upsilon}{_{\rho\sigma}^{\, \tt b}} \bigr](x;x')
	=
	2\Bigl[ \overline{\eta}_{\rho(\mu} \overline{\eta}_{\nu)\sigma}
		- \frac{ \overline{\eta}_{\mu\nu} \overline{\eta}_{\rho\sigma} }{ D \!-\! 3 } \Bigr]
	i \bigl[ \tensor*[^{\tt a\! }]{\Delta}{^{\tt \! b}} \bigr]_{\nu} (x;x')
	-
	4 \delta^0_{(\mu} \overline{\eta}_{\nu)(\rho} \delta^0_{\sigma)} \,
	i \bigl[ \tensor*[^{\tt a \!}]{\Delta}{^{\tt \! b}} \bigr]_{\nu-1} (x;x')
\nonumber \\
&
	+
	\frac{2}{(D \!-\! 2)(D \!-\! 3)}
	\Bigl[ \overline{\eta}_{\mu\nu} \!+\! (D\!-\!3) \delta_\mu^0 \delta_\nu^0 \Bigr]
	\Bigl[ \overline{\eta}_{\rho\sigma} \!+\! (D\!-\!3) \delta_\rho^0 \delta_\sigma^0 \Bigr]
	i \bigl[ \tensor*[^{\tt a \!}]{\Delta}{^{\tt \! b}} \bigr]_{\nu-2} (x;x')
	\, .
\label{UpsilonPropagator}
\end{align}
The second contribution in~(\ref{SplitPropagator}) inherits its structure from the
linearized gauge transformation~(\ref{GaugeTransformation}),\footnote{This form can also
be written more compactly using covariant derivatives:
\begin{equation*}
i \bigl[ \tensor*[_{\mu\nu}^{\tt a \,}]{\Theta}{_{\rho\sigma}^{\tt \, b}} \bigr](x;x')
	=
	\frac{ 1 }{ H^2 (aa')^2 }
	\nabla'_{\rho)} \nabla_{(\mu} 
	i \bigl[ \tensor*[_{\nu)}^{\tt a\, }]{\Xi}{_{(\sigma}^{\tt \, b}} \bigr](x;x')
	\, .
\end{equation*}
}
\begin{equation}
i \bigl[ \tensor*[_{\mu\nu}^{\tt a \,}]{\Theta}{_{\rho\sigma}^{\tt \, b}} \bigr](x;x')
	=
	\Bigl[
	\delta^\omega_{(\mu} \partial_{\nu)}
	-
	\eta_{\mu\nu} \delta^\omega_0 \mathcal{H}
	\Bigr]
	\Bigl[
	\delta^\lambda_{(\rho} \partial_{\sigma)}'
	-
	\eta_{\rho\sigma} \delta^\lambda_0 \mathcal{H}'
	\Bigr]
	\frac{ i \bigl[ \tensor*[_{\omega}^{\tt a \,}]{\Xi}{_{\lambda}^{\, \tt b}} \bigr](x;x') }
	{ H^2(aa')^2 }
	\, ,
\label{ThetaSolution}
\end{equation}
where the vector two-point function on which the derivatives act takes the form
\begin{align}
\MoveEqLeft[11]
i \bigl[ \tensor*[_{\omega}^{\tt a \, }]{\Xi}{_{\lambda}^{\, \tt b}} \bigr](x;x')
	=
	aa'
	\biggl[
	\frac{ 2 \delta_\omega^0 \delta_\lambda^0 }{ D \!-\! 3 } 
		\Bigl\{
		i \bigl[ \tensor*[^{\tt a \!}]{\Delta}{^{\tt \! b}} \bigr]_{\nu}(x;x')
		-
		i \bigl[ \tensor*[^{\tt a \!}]{\Delta}{^{\tt \! b}} \bigr]_{\nu-2}(x;x')
		\Bigr\}
\nonumber \\
&
	- \frac{ 2 \overline{\eta}_{\omega\lambda} }{ D \!-\! 1 }
		\Bigl\{
		i \bigl[ \tensor*[^{\tt a \!}]{\Delta}{^{\tt \! b}} \bigr]_{\nu+1}(x;x')
		-
		i \bigl[ \tensor*[^{\tt a \!}]{\Delta}{^{\tt \! b}} \bigr]_{\nu-1}(x;x')
		\Bigr\}
	\biggr]
	\, .
\label{XiPropagator}
\end{align}
Verifying that these expressions correctly reproduce the components
(\ref{0000Component})–(\ref{ijklComponent}) requires acting with the derivatives,
applying the reflection identities~(\ref{ReflectionIds}), and using the identity~(\ref{LaplaceIdentity}).  
The relations~(\ref{SplitPropagator})–(\ref{XiPropagator}) constitute the main results of
this work.

\section{Discussion}
\label{sec: Discussion}

In this work we have employed canonical methods to construct a graviton propagator in 
de Sitter space for a one-parameter family of simple non-covariant gauges defined in
(\ref{GaugeIntro}).  A key advantage of the resulting expression, given at the end of
Sec.~\ref{sec: Two-point function}, is its tractable and compact structure: the full
propagator is expressed in terms of a small number of scalar propagators with at most
two derivatives acting on them.  Moreover, all scalar propagators that appear admit
power-series representations that terminate in $D\!=\!4$, as follows from
(\ref{Fseries}).  This makes the expression particularly convenient for practical loop
computations and for transparent analyses of gauge dependence in proposed observables,
such as in~\cite{Glavan:2024elz}.

It is instructive to compare our result with the two-parameter family of gauges 
introduced in~\cite{Glavan:2019msf}, defined by the gauge-fixing action
\begin{equation}
S_{\rm gf}[h_{\mu\nu}] \!=\!\! \int\! d^{D}x \, a^{D-2} \,
	\biggl[ - \frac{ \eta^{\mu\nu} \mathscr{F}_\mu \mathscr{F}_\nu }{2(1\!+\!\delta\alpha)} 
	 \biggr] \, ,
\qquad
\mathscr{F}_\mu = \eta^{\rho\sigma} \Bigl[ \partial_\rho h_{\sigma\mu} 
	+ (D\!-\!2) \mathcal{H} \delta_{(\rho}^0 h_{\sigma)\mu}
	- \frac{1}{2} (1 \!+\! \delta\beta )\partial_\mu h_{\rho\sigma} \Bigr] \, .
\end{equation}
The propagator in that work was obtained perturbatively, to linear order in the 
infinitesimal variations $\delta\alpha$ and $\delta\beta$.  
Our closed-form result exactly reproduces the $\delta\beta\!=\!0$ limit of
\cite{Glavan:2019msf}, demonstrating that the dependence of the propagator on 
$\alpha$ in this two-parameter family is in fact \emph{exactly linear}.  
This extends and sharpens the understanding of the structure of this gauge family well
beyond the infinitesimal regime.

We have also verified explicitly that the propagator constructed here satisfies the
correct equation of motion and obeys the full Ward--Takahashi identity, as shown in the
Appendix.  In addition, the propagator reduces to the standard Lorentz-invariant form
in the flat-space limit.  These checks confirm the internal consistency of the
construction and show that the one-parameter generalization preserves the expected
gauge structure.

The simplicity of the resulting propagator and the presence of a free gauge-fixing parameter
 make it a promising tool for a variety of
applications, including loop computations, analyses of infrared behaviour, and systematic
studies of gauge dependence in cosmological graviton observables.  Indeed, one such 
application utilizing this propagator---demonstrating the gauge independence of the observable 
introduced in~\cite{Glavan:2024elz}---is already in progress~\cite{Glavan:2026pug}.

\section*{Acknowledgements}
\label{sec: Acknowledgements}
\addcontentsline{toc}{section}{\protect\numberline{}Acknowledgments}

This work was supported by project 24-13079S of the Czech Science Foundation 
(GA\v{C}R).

\appendix

\section{Checking propagator solution}
\label{app: Checking propagator solution}

\subsection{Flat space limit}
\label{subsec: Flat space limit}

In the flat-space limit,~$H\!\to\!0$, the de Sitter space scalar field propagators with various
effective masses give
\begin{align}
i \bigl[ \tensor*[^{ \tt a\!}]{\Delta}{^{\! \tt b }} \bigr]_\lambda (x;x')
	\xrightarrow{H \to 0}
	\frac{ \Gamma\bigl( \frac{D-2}{2} \bigr) }
		{ 4 \pi^{\frac{D}{2}} \bigl( \Delta x^2_{\tt ab} \bigr)^{\! \frac{D-2}{2}} }
	\equiv
	i \Delta^{\tt \! ab}(x\!-\!x')
	\, ,
\end{align}
while the particular difference of the two scalar propagators gives
\begin{equation}
\frac{1}{H^2} \Bigl\{
	i \bigl[ \tensor*[^{ \tt a\!}]{\Delta}{^{\! \tt b }} \bigr]_{\lambda+1}(x;x')
		-
		i \bigl[ \tensor*[^{ \tt a\!}]{\Delta}{^{\! \tt b }} \bigr]_{\lambda-1}(x;x')
		\Bigr\}
	\xrightarrow{H \to 0}
	- 4 \lambda \partial^{-2} i \Delta^{\tt \! ab}(x\!-\!x')
	\, .
\end{equation}
These relations allow us to take the flat-space limit of the propagator
in~(\ref{SplitPropagator}), obtaining
\begin{equation}
i \bigl[ \tensor*[_{\mu\nu}^{\tt a \,}]{\Delta}{_{\rho\sigma}^{\tt \, b}} \bigr](x;x')
	=
	2\biggl[
	\eta_{\rho(\mu} \eta_{\nu)\sigma}
	-
	\frac{ \eta_{\mu\nu} \eta_{\rho\sigma} }{ D\!-\!2 }
	-
	2 (1\!-\!\alpha) 
		\frac{\partial_{(\mu} \eta_{\nu)(\rho} \partial_{\sigma)} }{\partial^2} 
	\biggr]
	i \Delta^{\tt \! ab}(x\!-\!x')
	\, .
\end{equation}
This expression matches Capper’s graviton propagator in~\cite{Capper:1979ej} for
$\beta\!=\!1$, as expected.

\subsection{Ward-Takahashi identity}
\label{subapp: Ward-Takahashi identity}

To verify the double Ward–Takahashi identity~(\ref{doubleWard}), we first act the
operator~(\ref{calDdef}) on each of the two terms in the propagator
(\ref{SplitPropagator}). Acting this operator on~(\ref{UpsilonPropagator}) and
performing the required contractions gives
\begin{align}
\MoveEqLeft[3]
\mathscr{D}_\omega{}^{\mu\nu} \,
i \bigl[ \tensor*[_{\mu\nu}^{\tt a \,}]{\Upsilon}{_{\rho\sigma}^{\tt \, b}} \bigr](x;x')
	=
	2 \Bigl[
	\overline{\eta}_{\omega(\rho} \overline{\partial}_{\sigma)}
	+
	\frac{ 1  }{ D \!-\! 3 } \overline{\eta}_{\rho\sigma} \delta_\omega^0 \partial_0
	\Bigr]
	i \bigl[ \tensor*[^{\tt a\! }]{\Delta}{^{\tt \! b}} \bigr]_{\nu} (x;x')
\nonumber \\
&
	-
	2 \Bigl\{
	\delta^0_{\omega} \delta^0_{(\rho} \overline{\partial}_{\sigma)}
	-
	\overline{\eta}_{\omega(\rho} \delta^0_{\sigma)} 
	\Bigl[
	\partial_0
	+
	(D\!-\!2) \mathcal{H} 
	\Bigr]
	\Bigr\}
	i \bigl[ \tensor*[^{\tt a\! }]{\Delta}{^{\tt \! b}} \bigr]_{\nu-1} (x;x')
\nonumber \\
&
	-
	\frac{2 \delta_\omega^0  }{D \!-\! 3}
	\Bigl[ \overline{\eta}_{\rho\sigma} \!+\! (D\!-\!3) \delta_\rho^0 \delta_\sigma^0 \Bigr]
	\Bigl[
	\partial_0
	+
	(D\!-\!3)  \mathcal{H}
	\Bigr]
	i \bigl[ \tensor*[^{\tt a\! }]{\Delta}{^{\tt \! b}} \bigr]_{\nu-2} (x;x')
	\, .
\label{Dupsilon}
\end{align}
Using the reflection identities for spatial and temporal derivatives~(\ref{ReflectionIds}),
we convert all unprimed derivatives into primed ones and simplify the expression to
\begin{equation}
\mathscr{D}_\omega{}^{\mu\nu} \,
i \bigl[ \tensor*[_{\mu\nu}^{\tt a \,}]{\Upsilon}{_{\rho\sigma}^{\tt \, b}} \bigr](x;x')
	=
	-
	2 \overline{\eta}_{\omega(\rho} \partial_{\sigma)}'
	i \bigl[ \tensor*[^{\tt a\! }]{\Delta}{^{\tt \! b}} \bigr]_{\nu} (x;x')
	+
	2 \delta^0_{\omega}  \Bigl[
	\delta^0_{(\rho} \partial_{\sigma)}' 
		\!-\! \eta_{\rho\sigma} \mathcal{H}'
	\Bigr]
	i \bigl[ \tensor*[^{\tt a\! }]{\Delta}{^{\tt \! b}} \bigr]_{\nu-1} (x;x')
	\, .
\label{singleDupsilon}
\end{equation}
This agrees with Eqs.~(35a) and (35b) of~\cite{Tsamis:1992zt} once the different
normalizations are taken into account.

Acting the same operator on the remaining part of the two-point function
in~(\ref{SplitPropagator}) yields
\begin{align}
\MoveEqLeft[3]
\mathscr{D}_\omega{}^{\mu\nu}  \,
i \bigl[ \tensor*[_{\mu\nu}^{\tt a \,}]{\Theta}{_{\rho\sigma}^{\tt \, b}} \bigr](x;x')
	=
	\frac{1}{ (D\!-\!3) \mathcal{H} \mathcal{H}' } \delta_\omega^0
	\Bigl[
	\delta^0_{(\rho} \partial_{\sigma)}'
	-
	\overline{\eta}_{\rho\sigma} \mathcal{H}'
	\Bigr]
	\Bigl[
	\partial^2 - (D\!-\!4) \mathcal{H} \partial_0
	\Bigr]
\nonumber \\
&
	\times\!
	\Bigl\{ i \bigl[ \tensor*[^{\tt a\! }]{\Delta}{^{\tt \! b}} \bigr]_{\nu} (x;x')
		- i \bigl[ \tensor*[^{\tt a\! }]{\Delta}{^{\tt \! b}} \bigr]_{\nu-2} (x;x')
			\Bigr\}
	- \frac{1}{(D\!-\!1) \mathcal{H} \mathcal{H}' }
	\overline{\eta}_{\omega(\rho} \Bigl[
		\partial'_{\sigma)}
		-
		\delta_{\sigma)}^0 \mathcal{H}'
		\Bigr] 
\nonumber \\
&	\hspace{1cm}
	\times\!
	\Bigl[ \partial^2 
		- (D\!-\!4) \mathcal{H} \partial_0 + (D\!-\!2) \mathcal{H}^2 \Bigr]
	\Bigl\{ i \bigl[ \tensor*[^{\tt a\! }]{\Delta}{^{\tt \! b}} \bigr]_{\nu+1} (x;x')
		- i \bigl[ \tensor*[^{\tt a\! }]{\Delta}{^{\tt \! b}} \bigr]_{\nu-1} (x;x') 
		\Bigr\}
	\, .
\end{align}
Applying the equations of motion for the scalar propagators
(\ref{ScalarPropagatorEOM}) to eliminate the unprimed derivatives and using the
reflection identities~(\ref{ReflectionIds}) to convert the remaining ones into primed
derivatives, we find that this expression evaluates to minus the right-hand side of
Eq.~(\ref{singleDupsilon}):
\begin{equation}
\mathscr{D}_\omega{}^{\mu\nu}  \,
i \bigl[ \tensor*[_{\mu\nu}^{\tt a \,}]{\Theta}{_{\rho\sigma}^{\tt \, b}} \bigr](x;x')
	=
	-
	\mathscr{D}_\omega{}^{\mu\nu} \,
	i \bigl[ \tensor*[_{\mu\nu}^{\tt a \,}]{\Upsilon}{_{\rho\sigma}^{\tt \, b}} \bigr](x;x')
	\, .
\label{DTheta}
\end{equation}
Consequently, for the full two-point function we obtain
\begin{equation}
\mathscr{D}_\omega{}^{\mu\nu} \,
i \bigl[ \tensor*[_{\mu\nu}^{\tt a \,}]{\Delta}{_{\rho\sigma}^{\tt \, b}} \bigr](x;x')
	=
	\alpha \times
	\mathscr{D}_\omega{}^{\mu\nu}
	i \bigl[ \tensor*[_{\mu\nu}^{\tt a \,}]{\Upsilon}{_{\rho\sigma}^{\tt \, b}} \bigr](x;x')
	\, .
\label{Ddelta}
\end{equation}
Acting with the second linear differential operator on this expression gives
\begin{align}
\mathscr{D}_\omega{}^{\mu\nu}
\mathscr{D}'_\lambda{}^{\rho\sigma} \,
i \bigl[ \tensor*[_{\mu\nu}^{\tt a \,}]{\Delta}{_{\rho\sigma}^{\tt \, b}} \bigr](x;x')
	={}&
	\alpha
	\biggl\{
	-
	\overline{\eta}_{\omega\lambda}
	\Bigl[
		\partial'^2 \!-\! (D\!-\!2) \mathcal{H}' \partial_0'
		\Bigr]
		i \bigl[ \tensor*[^{\tt a\! }]{\Delta}{^{\tt \! b}} \bigr]_{\nu} (x;x')
\nonumber \\
&	\hspace{0.5cm}
	+
	\delta_\omega^0 \delta_\lambda^0
		\Bigl[ 
			\partial'^2 
			\!-\!
			(D\!-\!2) \mathcal{H}' \partial'_0 
			\!-\! 
			(D\!-\!2) \mathcal{H}'^2 
			\Bigr]
			i \bigl[ \tensor*[^{\tt a\! }]{\Delta}{^{\tt \! b}} \bigr]_{\nu-1} (x;x')
	\biggr\}
	\, .
\end{align}
Finally, applying the equations of motion for the scalar propagators
(\ref{ScalarPropagatorEOM}) produces the double Ward–Takahashi identity
(\ref{doubleWard}).

\subsection{Equation of motion}
\label{subapp: Equation of motion}

The simple gauge propagator~(\ref{UpsilonPropagator}) 
satisfies the equation of motion~(\ref{2ptCovariantEOM})
with~$\alpha\!=\!1$~\cite{Tsamis:1992xa},
\begin{equation}
\Bigl[ \boldsymbol{L}^{\omega\lambda\mu\nu} 
	+
	\mathscr{D}^{\alpha\omega\lambda} a^{D-2} \mathscr{D}_{\alpha}{}^{\mu\nu} \Bigr]
	i \bigl[ \tensor*[_{\mu\nu}^{\tt a \,}]{\Upsilon}{_{\rho\sigma}^{\tt \, b}} \bigr](x;x')
	=
	\delta^{(\omega}_\rho \delta^{\lambda)}_\sigma \, 
	\frac{i \delta^D(x \!-\! x')}{\sqrt{-g}}
	\, .
\end{equation}
Together with~(\ref{DTheta}), this implies that the part of the full 
propagator~(\ref{SplitPropagator}) proportional to~$(1\!-\!\alpha)$
must be a homogeneous solution of the Lichnerowicz operator,
\begin{equation}
\boldsymbol{L}^{\omega\lambda\mu\nu}
	i \bigl[ \tensor*[_{\omega\lambda}^{\tt a \, }]{\Theta}{_{\rho\sigma}^{\tt \, b}} \bigr](x;x')
	=
	0
	\, ,
\end{equation}
This indeed follows from the
fact that the Lichnerowicz operator annihilates the tensor structure appearing
in~(\ref{ThetaSolution}),
\begin{equation}
\boldsymbol{L}^{\omega\lambda\mu\nu} 
	\Bigl[
	\delta^\alpha_{(\mu} \partial_{\nu)}
	-
	\eta_{\mu\nu} \delta^\alpha_0 \mathcal{H}
	\Bigr]
	=
	0
	\, ,
\end{equation}
and ensures that the full propagator satisfies the equation of motion
(\ref{2ptCovariantEOM}) in the one-parameter simple gauge.

\end{document}